%% file: ms.tex
\documentclass{jfm}
\usepackage{epstopdf, epsfig}
\pdfoutput=1
\usepackage{xcolor}
\usepackage{multirow}
\usepackage{graphicx}
\usepackage{subcaption}
\usepackage{siunitx}
\usepackage{float}
\usepackage{gensymb}
\usepackage{amssymb}
\usepackage{bm}
\usepackage{wrapfig}
\usepackage{tikz}
\usepackage{makecell}
\usepackage{mathtools}

\usepackage{overpic}
\usepackage{hyperref}

\usepackage{amsmath}
\usepackage[version=4]{mhchem}
\usepackage{longtable,tabularx}

\shorttitle{Linear modal instabilities around post-stall swept finite-aspect ratio wings} %for header on odd pages
\shortauthor{A. Burtsev et al.} %for header on even pages

\title{Linear modal instabilities around post-stall swept finite-aspect ratio wings at low Reynolds numbers}

\author
 {
  Anton Burtsev\aff{1},
  Wei He\aff{1}
  \corresp{\email{wei.he@liverpool.ac.uk}},
  Shelby Hayostek\aff{2},
  Kai Zhang\aff{3,4},
  Vassilios Theofilis\aff{1,5},
  Kunihiko Taira\aff{4},
  \and
  Michael Amitay\aff{2}
  }

\affiliation
{
\aff{1}
Department of Mechanical, Materials and Aerospace Engineering, University of Liverpool, Brownlow Hill, England L69 3GH, United Kingdom
\aff{2}
Department of Mechanical, Aeronautical, and Nuclear Engineering, Rensselaer Polytechnic
Institute, Troy, NY 12180, USA
\aff{3}
Department of Mechanical and Aerospace Engineering, Rutgers University, Piscataway, NJ 08854, USA
\aff{4}
Department of Mechanical and Aerospace Engineering, University of California, Los Angeles, CA 90095, USA
\aff{5}
Escola Politecnica, Universidade S\~ao Paulo, Avda. Prof. Mello Moraes 2231, CEP 5508-900, S\~ao Paulo-SP, Brasil
}
\begin{document}

\maketitle

\begin{abstract}
Linear modal instabilities of flow over finite-span untapered wings have been investigated numerically at Reynolds number 400, at a range of angles of attack and sweep on two wings having aspect ratios 4 and 8. Base flows have been generated by direct numerical simulation, marching the unsteady incompressible three-dimensional Navier-Stokes equations to a steady state, or using selective frequency damping to obtain stationary linearly unstable flows. Unstable three-dimensional linear global modes of swept wings have been identified for the first time using spectral-element time-stepping solvers. The effect of the wing geometry and flow parameters on these modes has been examined in detail. An increase of the angle of attack was found to destabilize the flow, while an increase of the sweep angle had the opposite effect. On unswept wings, TriGlobal analysis revealed that the most unstable global mode peaks in the midspan region of the wake; the peak of the mode structure moves towards the tip as sweep is increased. Data-driven analysis was then employed to study the effects of wing geometry and flow conditions on the nonlinear wake. On unswept wings, the dominant mode at low angles of attack is a Kelvin-Helmholtz-like instability, qualitatively analogous with global modes of infinite-span wings under same conditions. At higher angles of attack and moderate sweep angles, the dominant mode is a structure denominated the {\em interaction mode}. At high sweep angles, this mode evolves into elongated streamwise vortices on higher aspect ratio wings, while on shorter wings it becomes indistinguishable from tip-vortex instability.
\end{abstract}

\begin{keywords}
%Authors should not enter keywords on the manuscript, as these must be chosen by the author during the online submission process and will then be added during the typesetting process (see http://journals.cambridge.org/data/\linebreak[3]relatedlink/jfm-\linebreak[3]keywords.pdf for the full list)
\end{keywords}

\input{sections/introduction.tex}

\input{sections/theory.tex}

\input{sections/numerical.tex}

\input{sections/results.tex}

%\section{Discussion}
%\textcolor{red}{To be added later}

%\newpage
\section{Summary}
\label{sec:conclusion}

Linear modal three-dimensional (TriGlobal) instability analysis of laminar three-dimensional separated flows over finite aspect ratio, constant-chord wings have been performed at $Re=400$, two aspect ratios and a range of angles of attack and sweep. Monitoring the unsteady, three-dimensional base flows, the following observations were made, as the angle of sweep increased. When $0^\circ \leqslant \Lambda < 10^\circ$ the three distinct regions reported by \cite{zhang_2020} were also observed, namely the tip vortex, wake and the interaction regions with braid-like vortices. For $15^\circ \leqslant \Lambda < 25^\circ$ the braid-like vortices of the interaction region become dominant and absorb the tip vortex. Finally, at $25^\circ \le \Lambda \le 30^\circ$ hair-pin vortices form in the wake, tip stall and "ram's horn" vortices are present on the wing. The overall effect of an increasing sweep is flow stabilisation.

Linear TriGlobal instability analysis, performed at conditions where steady or stationary unstable base flows could be computed, revealed the leading eigenmodes of this class of flows for the first time. This analysis gave insight to the formation of the unstable wake, which is found to be caused by a unstable global wake mode at half-span for an unswept wing. Changes in the aspect ratio were seen to have little effect on the shape of the leading global flow eigenmode. As sweep increases from $\Lambda=0^\circ$ to $10^\circ$ the structures of the unstable mode peak closer to the wing tip and at $\Lambda=15^\circ$ the mode evolves into a vortical instability. At the maximum considered sweep of $\Lambda=30^\circ$ the leading mode is stable and takes the form of a vortical instability growing further away from the wing.

Data-driven analysis employing POD, and to a lesser extent DMD, was subsequently conducted during the nonlinear saturation stage, with the objective of classifying the dominant structures of the separated flow and assessing the effects of wing geometry. On the larger aspect ratio wing, $sAR=4$, at low angles of attack the dominant wake mode takes the form of Kelvin-Helmholtz instability. For higher $\alpha$, this mode exhibits more complex three-dimensional features, changing into a structure denominated the interaction mode. The interaction mode consists of both periodic structures, located at half-span of the wing associated with the wake, as well as vortical structures, that correspond to the location of braid-like vortices in the wing wake, recently reported by \cite{zhang_2020}. The interaction mode is also found to be associated with unsteady motion of the reattachment line, observed near the trailing edge in direct numerical simulation. The spatial characteristics of the interaction mode change with sweep, reflecting the changes of the underlying base flow. The mode eventually splits into separate structures on either side of the symmetric wing at $\Lambda\approx15^\circ$. At even higher angles of sweep the dominant structure unravelled in the data-driven analysis takes the form of elongated streamwise vortices on the larger $sAR=4$ wing, which are associated with the recently discovered multiple streamwise-aligned vortices that form on the finite aspect ratio wing at higher sweep angles \citep{zhang_2020b}. On the shorter, $sAR=2$, wing the tip effects lead to the dominant mode becoming indistinguishable from a tip-vortex instability.

The wake structures observed in \citet{zhang_2020} have been analysed and classified into two classes of modes, the interaction mode (IM) and the wake wode (WM), having distinct energy content at a range of angles of attack and sweep. The frequency content and the spatial structures of the interaction mode could be linked to the oscillatory motion of reattachment line observed in large-scale simulations. The distinct frequency peaks of the lift coefficient signal observed by \citet{zhang_2020} have also been identified in the frequency spectra of the interaction and wake modes, leaving little doubt that the IM and WM are the major contributors to lift on finite aspect ratio wings. Finally, the stabilisation effects due to sweep and tip vortex discussed by \citet{zhang_2020,zhang_2020b} were quantified over a range of moderate sweep angles on the basis of the relative importance between the spanwise velocity component and the velocity induced by the tip vortex over at fixed values of the wing aspect ratio and angle of attack. The balance of these stabilisation mechanisms at $\Lambda=5^\circ$ was found to lead to a quasi-two-dimensional structure of the wake close to the wing.

Overall, the effect of sweep on the structure of the leading global flow eigenmodes, as well as of the modes computed in the data-driven analysis is relatively mild, up to and including $\Lambda=10^\circ$, and mainly affects the spanwise location of the peak of the mode structures. 
%This result is analogous to observations made on the effect of sweep in turbulent flow at orders-of-magnitude higher Reynolds numbers. \tb{Miki, could you please add a couple of references to typical high-Re works here?}. 
On the other hand, higher sweep angle results, up to $\Lambda=30^\circ$ examined herein, show that flow over the $sAR=4$ wing is dominated by the strength of the spanwise velocity on the suction side of the wing, which prevails over that of the tip-induced spanwise velocity. Conversely, the flow over the shorter $sAR=2$ wing at the same conditions is dominated by tip effects, with the tip-induced velocity countering the spanwise flow.

Future work may address the effect of the last remaining unexplored parameter, namely wing taper, on the present findings. However, the low-Reynolds number results reported here establish a basis for understanding flow dynamics and instabilities on finite three-dimensional constant-chord wings at low Reynolds numbers, as a first step towards understanding turbulent flow at higher Reynolds numbers and providing theoretically-founded, physics-driven flow control strategies.

\vskip 1em
\noindent{\bf Acknowledgements\bf{.}}\\
Support of AFOSR Grant FA9550-17-1-0222 with Dr. Gregg Abate and Dr. Douglas Smith as Program Officers is gratefully acknowledged. The authors also acknowledge computational time made available on the UK supercomputing facility ARCHER via the UKTC Grant EP/R029326/1 and on the DoD Copper supercomputer, via project AFVAW10102F62 with Dr. Nicholas Bisek as Principal Investigator. 
\vskip 1em
\noindent{\bf Declaration of Interests\bf{.}}\\
The authors report no conflict of interest.

\input{sections/appendix.tex}

%-------------------------------

\bibliographystyle{jfm}
\bibliography{jfm-instructions}

\end{document}

%% file: sections/introduction.tex
%\newpage
\section{Introduction}
Our present concern is with linear global instability mechanisms associated with unsteadiness of laminar three-dimensional separated flows over finite aspect ratio, constant-chord unswept and swept wings at low Reynolds numbers. To-date the vast majority of instability studies has focused on simplified models of laminar separation with no spanwise base flow component, as encountered in flows over two-dimensional profiles, or spanwise homogeneous flow over infinite-span three-dimensional wings, that have been used as a proxy to understand fundamental mechanisms of separation in practical fixed- or rotary-wing applications. However, either of these approximations fails to address the essential three-dimensionality of the flow field \citep{WygnanskiEtAl2011,WygnanskiEtAl2014} and the implications of linear instability of three-dimensional separated flow on the ensuing unsteadiness on a finite-span swept wing. On finite aspect ratio wings work exists that focuses on the mid-span wake \citep[e.g.][]{Manolesos_2014} or the vicinity of the tip vortex \citep[e.g.][]{Edstrand_2016} and has improved the understanding of specific phenomena, namely the formation of stall-cells and tip-vortex instability, respectively. However, presently there exists limited knowledge on linear instability mechanisms associated with three-dimensional separation {\em per se} on the wing surface, or a deep understanding of the complex vortex dynamics arising from this instability on a finite-span wing, as a function of the angles of attack $(\alpha)$ and sweep $(\Lambda)$. In fact, there is a void in the literature that employs three-dimensional global (TriGlobal) linear instability analysis appropriate for the fully inhomogeneous three-dimensional flow field around a finite aspect ratio wing. The present work aims to close this knowledge gap by documenting modal instability mechanisms and their evolution on different wing geometries and flow conditions. 

A short review of existing literature on the subject sets the scene for the work performed herein. Laminar separation studies have extensively employed adverse pressure gradients to generate nominally two-dimensional separation bubbles on a flat plate geometry. Although such bubbles were known to be structurally unstable \citep[e.g.][]{Dallmann_1988}, \citet{Theofilis_2000} showed that the physical mechanism leading to unsteadiness and three-dimensionalisation of a nominally two-dimensional laminar separation bubble, as well as to breakdown of the associated two-dimensional vortex, arises from self-excitation of a previously unknown stationary three-dimensional global mode.
Soon after the latter work, global linear stability theory was applied to understand instability mechanisms in two-dimensional airfoils \citep{Theofilis_2002} and three-dimensional unswept wings of infinite span \citep{Kitsios_2009}. 
\citet{Rodriguez_2010} revisited the adverse pressure gradient boundary layer to study structural changes experienced by the laminar separation bubble in a flat-plate boundary layer due to the presence of the unstable stationary three-dimensional global mode and established a criterion of $\sim 7.5\%$ of backflow as a necessary condition for self-excitation of the nominally two-dimensional flow. Furthermore, linear superposition of the global mode discovered by \citet{Theofilis_2000} upon the two-dimensional laminar separation bubble revealed the well-known three-dimensional U-separation pattern \citep{HornungPerry,Perry_1987,delery2013topology}, while the surface streamlines topology induced by the global mode resembled the characteristic cellular structures known as stall cells (SC), that are observed experimentally to form on stalled wings. 

%------------------------------------------------------------------------
% 2.5D and Stall Cells
%------------------------------------------------------------------------
Stall cells consisting of pairs of counter rotating vortices on the suction side of a wing arise at angles of attack near maximum lift and have been observed experimentally for a long time \citep{Moss_1968,Bippes_1980,Winkelman_1980,Weihs_Katz_1983,Bippes_1984,Schewe_2001,Broeren_2001}. Application of the appropriate for the spanwise homogeneous nature of the base flow BiGlobal analysis concept attributed the appearance of stall cells to self-excitation of the stationary global mode of the laminar separation bubble formed on a stalled airfoil \citep{Rodriguez_2011}. However that mechanism did not explain the existence of stall cells at low angles of attack far away from stall, as experimentally observed by \cite{Elimelech_2010}. 

Steady massively separated spanwise homogeneous flow over stalled wings was revisited by \cite{HeJFM} using global linear modal and nonmodal stability tools. Flow over three different NACA airfoils (symmetric 0009 and 0015 as well as cambered 4415) was analysed at $150 \leq Re \leq 300$ and $10^\circ \leq \alpha \leq 20^\circ$. Two different mechanisms were identified by linear modal instability analysis: a travelling Kelvin-Helmholtz (K-H) mode dominating the flow at large spanwise periodicity length and a three-dimensional stationary mode most active as the spanwise periodicity length becomes smaller. On all three airfoils considered, nonmodal analysis showed that linear optimal perturbations evolve into travelling K-H modes. Furthermore, the travelling K-H mode was the first to become unstable as Reynolds number or angle of attack were increased, with the flow becoming two-dimensionally unsteady prior to becoming three-dimensionally unstable. Secondary instability analysis of the time-periodic base flow ensuing linear amplification of the K-H mode, via temporal Floquet theory, revealed two secondary amplified modes with spanwise wavelengths of approximately 0.6 and 2 chords. These modes are reminiscent of the classic Mode A and B instabilities of the circular cylinder \citep{Barkley_Henderson_1996, Williamson_1996} although, unlike the cylinder, the short-wavelength perturbation was the first to become linearly unstable over all three airfoils considered. 

The work of \cite{HeJFM} showed that stall-cell like streamline patterns on the wing surface arise from linear amplification of this short-wavelength secondary instability. By contrast to the primary-instability based scenario proposed by \citet{Rodriguez_2011}, the mechanism discovered by \citet{HeJFM} could explain the emergence of stall cells at lower angles of attack. \citet{Zhang_2016} extended the analysis of \cite{HeJFM} to study instability of unsteady flow over a NACA 0012 spanwise periodic wing at higher Reynolds numbers, $400 \leq Re \leq 1000$ at $\alpha=16^\circ$. At $Re=800$ and 1000 these authors identified two oscillatory unstable modes corresponding to near-wake and far-wake instabilities, alongside a stationary unstable mode, while only one unstable mode was found at the lower $Re=400$ and 600. Ground-proximity effects on the stability of separated flow over NACA 4415 at low Reynolds numbers were studied using BiGlobal theory with consideration of both flat \citep{He_2019b} and wavy ground surfaces \citep{He_2019}. Finally, \citet{Rossi_2018} considered incompressible flow over a NACA 0010 airfoil and a narrow ellipse of same thickness at a large angle of attack of $30^\circ$ ($100 \leq Re \leq 3000$) documenting multiple bifurcations. 

BiGlobal analysis of spanwise homogeneous flow over infinite-span wings has been extended to study compressibility effects and Reynolds numbers more representative of flight conditions by \citet{Plante_2019}, who investigated cellular patterns on infinite swept wings in subsonic stall (NACA 4412, $Re=3.5 \times 10^5$, Mach $=$ 0.2) and transonic buffet (OALT25, $Re=3 \times 10^6$, Mach $\approx$ 0.73) conditions. The most interesting result of this work has been that stall cells at the conditions examined were found to be 
stationary for unswept wings, but convected in the spanwise direction at a speed proportional to the sweep angle. 
The aforementioned efforts have certainly enriched understanding of instability mechanisms of spanwise homogeneous flow over wings of infinite span. However, BiGlobal analysis cannot be applied to address instability of the fully three-dimensional vortical patterns arising in finite aspect ratio wing flows. 

%------------------------------------------------------------------------
% 3D
%------------------------------------------------------------------------
Before discussing application of the appropriate linear TriGlobal modal analysis to shed light upon instabilities on finite aspect ratio wings, a brief review of experimental and numerical work of the latter configuration is discussed. Early experimental studies on finite aspect ratio ($AR$) wings are summarised in \cite{Boiko_1996}. More recently aerodynamic performance of small aspect ratio $(AR=0.5 - 2)$ wings that can be found in small unmanned aerial vehicles has been studied experimentally \citep{Torres_2004} and computationally \citep{Cosyn_2006}. \citet{Taira_2009} used three-dimensional direct numerical simulation (DNS) to study impulsively translated flat-plate wings ($AR=1-4$) of different planforms at a wide range of angles of attack and Reynolds numbers $(300 \leq Re \leq 500)$. These authors found the aspect ratio, angle of attack and Reynolds number to have a large influence on the stability of the wake profile and the force experienced by the finite wing with the flow reaching a stable steady state, a periodic cycle or aperiodic shedding. The three-dimensional nature of the flow was highlighted, and tip effects were found to stabilize the flow and exhibit nonlinear interaction of the shedding vortices. Even at larger aspect ratio of 4 the flow did not reach two-dimensional von K{\'a}rm{\'a}n vortex shedding due to the emergence of stall cell-like patterns. The effects of trapezoidal rather than rectangular planform \citep{Huang_2015}, and larger aspect ratio wings \citep{Son_2017} have been considered in more recent publications.

\citet{HeTCFD} performed linear global instability analysis using spatial BiGlobal eigenvalue problem and linear PSE-3D disturbance equations in the wake of a low aspect ratio three-dimensional wing of elliptic planform constructed using the Eppler E387 airfoil at $Re=1750$. 
Two types of disturbances were identified: symmetric perturbations corresponding to instability of the vortex sheet connecting the trailing vortices and antisymmetric perturbations peaking at the vortex sheet and also in the neighbourhood of the trailing vortex cores. A comparison with local stability theory has demonstrated that local analysis has limited ability to accurately capture the physics of the instability.

\cite{Edstrand_2018} carried out spatial and temporal stability analysis of a wake and trailing vortex region behind a NACA 0012 finite wing at $Re=1000$, $\alpha=5^{\circ}$ and $AR=1.25$, documenting seven unstable modes with the wake instability dominating in both temporal and spatial analyses. Unlike many stability analysis works focusing only on the vicinity of the tip vortex, the full half-span of the wing was considered. Although BiGlobal stability analysis was employed, streamwise rather than spanwise homogeneity was used exploiting the absence of large scale separation at the low angle of attack. This allowed capturing three-dimensional modes with structures in the tip and the wake regions. Subsequent work of \citet{Edstrand_2018b} on the same geometry employed parabolised stability analysis to guide the design on active flow control for tip vortex. A subdominant fifth instability mode was found to counter-rotate with the tip vortex. Flow control based on this fifth mode and forcing introduced at the trailing edge rather than the wing tip attenuated the tip vortex. Dynamic mode decomposition of the controlled flow was used to assess the effectiveness of such control.

\citet{Navrose_2019} conducted nonmodal stability analysis of a trailing vortex system over a flat plate and NACA 0012 wing at a range of angles of attack, aspect ratios and Reynolds numbers. Unlike in earlier studies their analysis included the tip vortex and flow over the wing. It was shown that the linear optimal perturbation is located near the wing surface which advects into the tip vortex region during its evolution, which agrees with the findings of \citet{Edstrand_2018b}. The displacement of the vortex core due to evolution of the optimal perturbation was proposed as a possible mechanism behind trailing vortex meandering. The results of this study point out the need to consider the entire complex three-dimensional flow over the wing rather than focusing on a specific isolated region in order to capture the full mechanics of the flow. These studies have demonstrated that addressing the three-dimensionality of finite wing wake through stability analysis allows for enhanced understanding of the underlaying physical mechanisms. However, the relatively low angles of attack considered in these studies meant that the underlying base flows had a relatively simple vortical structure.

In the general context of vortex dynamics, a large body of experimental and large-scale numerical simulation work exists on separated flows over finite aspect ratio wings. There are studies analysing complex vortex dynamics of finite wings under unsteady manoeuvres including translation and rotation \citep{Kim_2010, Jones_2016}, surging and plunging \citep{Calderon_2014,Manchini_2015}, pitching \citep{Jantzen_2014,Son_2017, Smith_2020}, and flapping \citep{Dong_2006, Medina_2015}. These works focused on the analysis of large scale flow structures such as leading edge vortices~\citep{Gursul_2007,Eldredge_2019} which can augment unsteady vortical lift and offer opportunities for flow control~\citep{Gursul_2014}.

%Practical applications often employ more sophisticated wing planforms than constant chord unswept wings that were analysed in most works discussed above. The effects of wing sweep are of particular interest since separated flow over finite swept wings has little in common with straight wing flow. 
On three-dimensional swept wings in particular, the presence of significant spanwise flow leads to three-dimensional flow structures like the {"\em ram's horn"} vortex \citep{Black_1956}. As soon as local stall appears on a swept wing spanwise boundary layer flow alters the stall characteristics of sections of attached flow along the span \citep{Harper_1964}. \citet{Yen_2007} have experimentally studied the surface flow patterns and wake structures of a NACA 0012 with a sweep angle, $\Lambda$ of $15^\circ$ $(AR=5, Re=2785)$ at a wide range of angles of attack: from fully attached flow all the way to bluff body like behaviour at $\alpha=45^\circ$ and above. Further experimental investigations at higher Reynolds numbers performed over a range of sweep angles were used to classify the flows into seven boundary layer flow regimes that were found to be closely related to aerodynamic performance \citep{Yen_2009}. However, the effect of the spanwise velocity component to laminar three-dimensional flow separation and vortex dynamics, emphatically demonstrated by \citet{WygnanskiEtAl2011,WygnanskiEtAl2014} to be central to the prediction of turbulent boundary layer properties on a swept wing, remained unexplored.

In the framework of our present combined theoretical/numerical and experimental efforts, \citet{zhang_2020} employed direct numerical simulation to analyse the development of three-dimensional separated flow over unswept finite wings at a range of angles of attack $(Re=400, 1 < AR < 6)$. The formation of three-dimensional structures in the separated flow was discussed in detail, showing that vorticity is introduced to the flow from the wing surface in a predominantly two-dimensional manner. The vortex sheet from the wing tip rolls up around the free end to form the tip vortex which at first is weak with its effects spatially confined. As the flow around the tip separates, the tip effects extend farther in the spanwise direction, generating three-dimensionality in the wake. It was shown that the tip-vortex induced downwash keeps the wake stable at low aspect ratio while at higher aspect ratios unsteady vortical flow emerges and vortices are shed forming closed loops. At $AR>4$ tip effects slow down shedding process near the tip, which desynchronizes from the two-dimensional shedding over the midspan region, giving rise to vortex dislocation. The interactions of the tip vortex with the unsteady wake structures at high angles of attack lead to noticeable tip vortex undulations. Force element analysis was used to identify wake structures responsible for the generation of lift and drag forces.

Subsequently, \citet{zhang_2020b} addressed swept wing flows at the same conditions. Several stabilisation mechanisms additional to those found in \citet{zhang_2020} were reported for swept wings. At small aspect ratios and low sweep angles the tip vortex downwash effects still stabilise the wake while the weakening of the downwash with increasing span allows the formation of unsteady vortex shedding. For higher sweep angles the source of three-dimensionality was shown to transition from the tip of the wing to midspan where a pair of symmetric vortical structures is formed. Their mutually induced downward velocity stabilising the wake. Therefore, three-dimensional midspan effects leading to formation of the stationary vortical structures allow steady flow formation at higher aspect ratios which would not be feasible on unswept wings. At higher aspect ratios the midspan effects weaken near the tip leading to unsteady vortex shedding in the wing tip region. Finally, for high aspect ratio highly swept wings steady flow featuring repetitive formation of the streamwise aligned finger-like vortices along the span ensues. \cite{Hayostek_2020} conducted an experimental study of unswept, cantilevered, low aspect ratio NACA 0015 wings $(Re=600$ and $ 1000)$ under wall effects closely integrated with numerical simulations. Secondary vortical structures along the span were observed, arising from the interaction between the tip and root regions and the separated flow over the wing. A global stability analysis at $AR=2$ indicated that the least stable mode, while having structures in the tip and horseshoe vortex regions is most pronounced in the wake. 

Despite the substantial improvement of understanding of complex vortical structures arising in separated flows over finite aspect ratio wings that our experimental and large-scale computational efforts have offered, several key questions remain open and motivate the present work. The origin of the unsteadiness of the trailing edge separation line, observed in the simulations of \citet{zhang_2020} and those performed herein, remains unexplained and the conjecture that this unsteadiness arises on account of a presently unknown flow eigenmode needs to be examined. Further, the frequency content and spatial structure of this, and possibly other modes existing in the flow both during the linear regime and at nonlinear saturation needs to be documented and classified, e.g. in terms of the respective energy content. From an aerodynamic point of view, it is necessary to understand the origin of the concentration of frequencies in the lift coefficient spectrum, reported in \citet{zhang_2020}, at particular nondimensional frequencies; again, it can be conjectured that a flow eigenmode with that frequency is responsible for this observation. Last but not least, the stabilization effects observed by \citet{zhang_2020b} on swept wings, their relationship with vortex-induced downwash and the resulting effects on the wake behaviour need to be further explored. Here, it may be conjectured that the magnitude of the spanwise velocity component along the wing leading edge, which changes with changing angle of sweep, is responsible for this observation; this is another conjecture put to test in our present work.

We perform linear TriGlobal modal analysis of separated flow over finite aspect ratio three-dimensional wings, followed by data-driven modal analysis \citep{Taira-modal-1} once the leading three-dimensional global mode has led the flow to nonlinear saturation. Since large scale separation already occurs at the conditions of interest, Tollmien-Schlichting and cross-flow mechanisms are neither expected to play a role, nor considered. TriGlobal stability analysis is first performed on steady or stationary unstable flows, in order to identify the mechanisms leading to the formation of unsteady wake. Data-driven analysis is then conducted past the first bifurcation, with the objective of classifying the dominant structures of the separated flow during nonlinear saturation and assessing the effects of wing geometry and angles of attack on the leading flow structures. During both facets of the work performed, we address the above mentioned open questions and construct links of earlier observations with intrinsic flow eigenmodes during the linear or nonlinear regime.

%As discussed above previous efforts aimed at understanding the instabilities of wing flows mostly focused on spanwise-homogeneous configurations or specific areas of the three-dimensional field associated with a certain flow phenomenon. These works have undoubtedly enriched the understanding of instability mechanisms. However, the nature of the three-dimensional finite aspect ratio swept wing flow calls on the fully inhomogeneous three-dimensional global (TriGlobal) linear instability analysis in order to improve the understanding of different phenomena of this flow, unravel the instability mechanisms behind the complex vortical interactions observed on such wings and document the effects of wing geometry and the angles of sweep and attack on these mechanisms.
The paper is organised as follows. The theory behind the modal analysis methods used is discussed in \S\ref{sec:theory} followed by the explanation of computational setup and numerical methods in \S\ref{sec:numeric}. Results are reported in \S\ref{sec:results} starting with the discussion of the base flow and followed by that on linear global modes. Finally, data-driven analysis results of the nonlinear saturated wake are presented with the effects of angle of attack, angle of sweep and wing aspect ratio on the dominant modes are reported.

%% file: sections/theory.tex
\section{Theory}
\label{sec:theory}
\subsection{TriGlobal linear stability theory}
\label{sec:lst}
The flow under consideration is governed by the nondimensional incompressible Navier-Stokes equations and continuity equations:
\begin{equation}
	\partial_t {\bm u} + {\bm u} \cdot \nabla{\bm u} = -\nabla p + Re^{-1} \nabla^2 \bm u,  \quad \quad  \nabla \cdot \bm u = 0, \\
	\label{eq:incNS}
\end{equation}
where the Reynolds number, $Re \equiv U_\infty c/\nu$, is defined by reference to the free-stream velocity, $U_\infty$, the chord, $c$, and the kinematic viscosity, $\nu$. The flow field can be expressed on an orthogonal coordinate system as a function of the unsteady velocity components and pressure
\begin{equation}
	\bm{q}(\bm{x},t) = (\bm{u}, p)^T,
\end{equation}
which are decomposed into a base flow component $\bm{\bar{q}}$ and a superimposed small perturbation $\bm{\tilde{q}}$, such that
\begin{equation}
	\bm{q} = \bm{\bar{q}} + \varepsilon \bm{\tilde{q}}, \quad \quad \varepsilon \ll 1.
	\label{eq:perturbation}
\end{equation}
The approach followed to obtain steady stable, or stationary unstable base flows will be discussed in \S \ref{sec:sfd}. Substituting (\ref{eq:perturbation}) into (\ref{eq:incNS}), subtracting the base flow at \textit{O}(1) and neglecting \textit{O}($\varepsilon^2$) terms leads to the linearised Navier-Stokes equations (LNSE)
\begin{equation}
	\partial_t {\tilde{\bm{u}}} + \bar{\bm{u}} \cdot \nabla{\tilde{\bm{u}}} + \tilde{\bm{u}} \cdot \nabla{\bar{\bm{u}}} = -\nabla p + Re^{-1} \nabla^2 \tilde{\bm{u}},  \quad \quad  \nabla \cdot \tilde{\bm{u}} = 0. \\
	\label{eq:LNSE}
\end{equation}
For the incompressible flow of interest the pressure perturbation can be related to the velocity perturbation through $\tilde{p} = -\nabla^{-2} (\nabla \cdot (\bar{\bm{u}} \cdot \nabla \tilde{\bm{u}} + \tilde{\bm{u}} \cdot \nabla \bar{\bm{ u}}))$. Now the LNSE can be written compactly as the evolution operator $\mathcal{L}$ forming an initial value problem (IVP)
\begin{equation}
	\partial_t {\mathbf{\tilde{u}}} = \mathcal{L} \mathbf{\tilde{u}}.
\label{eq:IVP}
\end{equation}
For steady basic flows, the separability between time and space coordinates in (\ref{eq:IVP}) permits introducing a Fourier decomposition in time of the general form $\tilde{\bm{u}} = \hat{\bm{u}}(\bm{x})e^{-i\omega t}$. Depending on the number of inhomogeneous spatial directions in the base flow analysed and the related number of periodic directions assumed different forms of the ansatz for $\tilde{\bm{u}}$ can be used \citep{theofilis2003advances,Juniper2014}. Since the flow in question is fully three-dimensional, no homogeneity assumption is permissible. This requires the use of TriGlobal linear stability theory, in which both the base flow $\bar{\bm{q}}$ and the perturbation $\tilde{\bm{u}}$ are inhomogeneous functions of all three spatial coordinates giving the following ansatz
\begin{equation}
	\tilde{\mathbf{u}}(x,y,z,t) = \hat{\mathbf{u}}(x,y,z)e^{-i\omega t} + c.c..
\label{eq:ansatz}
\end{equation}
Here, $\hat{\bm{u}}$ is the amplitude function, and $c.c.$ is a complex conjugate to ensure real-valued perturbations. Substituting (\ref{eq:ansatz}) into (\ref{eq:IVP}) leads to the TriGlobal eigenvalue problem (EVP)
\begin{equation}
	\mathsfbi{A} \hat{\bm{u}} =-i\omega \hat{\mathbf{u}}.
\label{eq:evp}
\end{equation}
The matrix $\mathsfbi{A}$ results from spatial discretisation of the operator $\mathcal{L}$ and comprises of the basic state $\bar{\bm{q}}(\bm{x})$ and its spatial derivatives, as well as the Reynolds number as a parameter.

The IVP solution in (\ref{eq:IVP}) can be obtained for finite time horizons $t \to \tau$ over a time interval $\tau$ or in asymptotically large time $t \to \infty$. The modal approach considers the limit $t \to \infty$ and solves the three-dimensional (TriGlobal) eigenvalue problem (\ref{eq:evp}). In the nonmodal approach, a finite-time horizon is introduced to describe the disturbance behaviour at arbitrary times and the solution of the IVP is sought. This allows transient growth analysis where the energy growth of perturbations over a finite time interval is monitored. In view of the absence of modal analysis in the problem at hand, the present work focuses on modal stability analysis.
The TriGlobal EVP (\ref{eq:evp}) is solved numerically, using time-stepping tools implemented in {\em nektar++} \citep{Cantwell2015}, to obtain the complex eigenvalues $\omega$ and the corresponding eigenvectors $\hat{\bf u}$, which are referred to as the global modes. The real and imaginary components of the complex eigenvalue $\omega=\omega_r + i\omega_i$ correspond to the frequency and the growth/decay rate of the global mode.

\subsection{Residuals algorithm}
\label{sec:ra}
 When the unsteady equations of motion are advanced in time and a steady state is reached, the eigenvalue problem pertaining to that state can be solved. Alternatively, linear global modes can be computed by employing the residuals algorithm \citep{Theofilis_ra} during linear decay of perturbations; this algorithm has also been used here, at conditions applicable to steady laminar three-dimensional flow. In its simplest version, when the DNS has reached the state of monotonic exponential decay of residuals, it is straightforward to extract both the steady state to which the simulation marches and the amplitude function of the least-damped global mode pertinent to the flow by simple algebraic expressions.  
Given transient DNS flow field solutions $\bm{q}_{1,2}$ at arbitrary times $t_1$, and $t_2$ during the exponential decay of the signal, the damping rate of a monotonically-decaying signal is obtained from the logarithmic derivative 
\begin{equation}
\label{eq:sigma}
	\omega_i = \frac{{\rm ln} {q}_2 - {\rm ln} {q}_1}{\Delta t},
\end{equation}
where $\Delta t = t_2 - t_1$ and $q_{1,2}$ represent field values of any flow quantity at arbitrary probes in the flow, extracted from the flow filed at times $t_1$ and $t_2$. A system of equations for the flow at the two instances in time can be written
\begin{align}
	\left\{
		\begin{array}{ll}
      		\bm{q}_1 = \bar{\bm{q}} + \varepsilon \hat{\bm{q}}e^{\omega t_1}, \\
      		\bm{q}_2 = \bar{\bm{q}} + \varepsilon \hat{\bm{q}}e^{\omega t_2}, \\
		\end{array} 
	\right.
\end{align}
where $\bar{\bm{q}}$ is the steady base state toward which the simulation is marching, $\hat{\bm{q}}$ is the amplitude function of the least-damped global mode and $\sigma$ is the damping rate. Expressions pertinent to oscillatory signals have been presented by %\cite{theofilis2011global} and 
\cite{Theofilis_Colonius_2003}. Solving this system of equations at arbitrary times $t_1$ and $t_2$ allows predicting the steady state (well ahead of convergence in time) and recover the spatial structure of the amplitude function of the global mode. Several examples of successful application of the algorithm can be found in the literature \citep[e.g.][]{Gomez_2012, Tumuklu_2017,Tumuklu_2018}. The selective frequency damping (SFD) method employed to obtain unstable stationary three-dimensional base states will be discussed in \S 3.3.

\subsection{Proper orthogonal decomposition}
\label{sec:POD}
Data-driven modal analysis is performed in the nonlinear saturation regime, with the aim of understanding the physics of massively separated flow past the primary flow bifurcation. Proper orthogonal decomposition (POD) provides a decomposition of the flow field data into a minimum number of modes that capture the most energy of the flow at any one time. POD, referred to in the literature also as principal component analysis or empirical eigenfunction decomposition, is applicable to signals taken either during linear growth of perturbation, as well as during nonlinear signal saturation. It was first introduced in fluid dynamics by \citet{Lumley_1967} and its reviews can be found in \cite{holmes_lumley_berkooz_1996, Berkooz_1993} and \cite{Taira-modal-1,Taira-modal-2}.

A given flow field $\bm{q}(\bm{x},t)$ is stacked as column vectors, each representing a fluctuating portion of a flow component (for example, $u, v, w$) with the time-averaged value $\bar{\bm{q}}(\bm{x})$ removed. This is done at a number of snapshots chosen such that the important flow features are well resolved in time, leading to:
\begin{equation}
	\bm{\chi}(t) = \bm{q}(\bm{x},t) - \bar{\bm{q}}(\bm{x})   \in \mathbb{R}^{n}, \quad  t=t_1, t_2, ..., t_m.
	\label{eq:fuct-flow}
\end{equation}
These column vectors are combined into an $n$ by $m$ data matrix $\mathsfbi{X}$, where $n$ is the physical size of the problem and $m$ is the number of snapshots. %For example, a POD of velocity components of a 2D $n_x$ by $n_y$ flow field will have $n=2n_xn_y$.
%\begin{equation}
%	\mathsfbi{X}=[\bm{x}(t_1) \quad \bm{x}(t_2) \quad ... \quad \bm{x}(t_m)]  \in \mathbb{R}^{n \times m}.
%	\label{eq:Xmat}
%\end{equation}
POD seeks optimal basis vectors ${\bm{\phi}}_j$ to represent the flow field data $\bm{q}(\bm{x})$ with the least number of modes. This can be determined by solving the eigenvalue problem:
\begin{equation}
	\mathsfbi{X}\mathsfbi{X}^T \bm{\phi}_j = \lambda_j\bm{\phi}_j \quad \bm{\phi}_j \in \mathbb{R}^n \quad \lambda_1 \geq ... \geq \lambda_n \geq 0.
	\label{eq:classicPOD}
\end{equation}
 known as the spatial or classical POD. The issue with this method is the fact that a $n \times n$ eigenvalue problem has to be solved, with $n$ the total degree of freedoms used in the spatial discretization. This makes the computational requirements of classic POD directly dependent on the size of the data field and would be impractical to analyse flow past 3D wings, where a significant portion of the wake has to be included in the data. However the temporal correlation matrix yields the same dominant spatial modes, while giving rise to a much smaller $m \times m$ eigenvalue problem as pointed out by \citet{Sirovich_1987}:
\begin{equation}
	\mathsfbi{X}^T \mathsfbi{X} \bm{\psi}_j = \lambda_j \bm{\psi}_j, \quad \bm{\psi}_j \in \mathbb{R}^m, \quad m \ll n.
	\label{eq:snapPOD}
\end{equation}
The nonzero eigenvalues of (\ref{eq:snapPOD}) are same as those of (\ref{eq:classicPOD}), while the eigenvectors of (\ref{eq:snapPOD}) can be related to POD modes through
\begin{equation}
	\bm{\phi}_j = \mathsfbi{X}\bm{\psi}_j \frac{1}{\sqrt{\lambda_j}} \in \mathbb{R}^n, \quad j=1,2,...,m.
\end{equation}
When POD is used to describe flow development during linear instability, it is straightforward to identify in the decomposition
\begin{equation}
    \tilde{\bm{q}}(\bm{x},t) = \sum_{j}\bm{a}_j(t) \bm{\phi}_j(\bm{x}).
\end{equation}
Following the terminology used by \citet{Aubry_1991b} the coefficients $\bm{a}_j$ are termed \emph{chronos}, as they represent the time behaviour of the mode, as opposed to the spatial structures $\bm{\phi}_j$, which are referred to as \emph{topos}. When the signal analysed is in the phase of linear growth/decay, chronos and topos can be respectively identified with the exponential time factor and the amplitude functions shown in (\ref{eq:ansatz}), and POD has been utilised to cross-verify damped global modes computed by solution of the TriGlobal eigenvalue problem.   

Numerically POD coefficients are obtained by projecting the snapshots onto the POD modes
\begin{equation}
	\bm{a}_j = \bm{\Phi}^T \bm{\chi}_j, \quad j=1,2,...,r,
\end{equation}
where $\bm{\Phi} = [\phi_1, \phi_2, ..., \phi_r]$ and $r$ can be a subset of the total number of snapshots $m$.

%% file: sections/numerical.tex
\section{Numerical work}
\label{sec:numeric}
\subsection{Geometry and mesh}

\begin{figure}		% Overal geometry
 \begin{center}
    \includegraphics[width=0.8\textwidth]{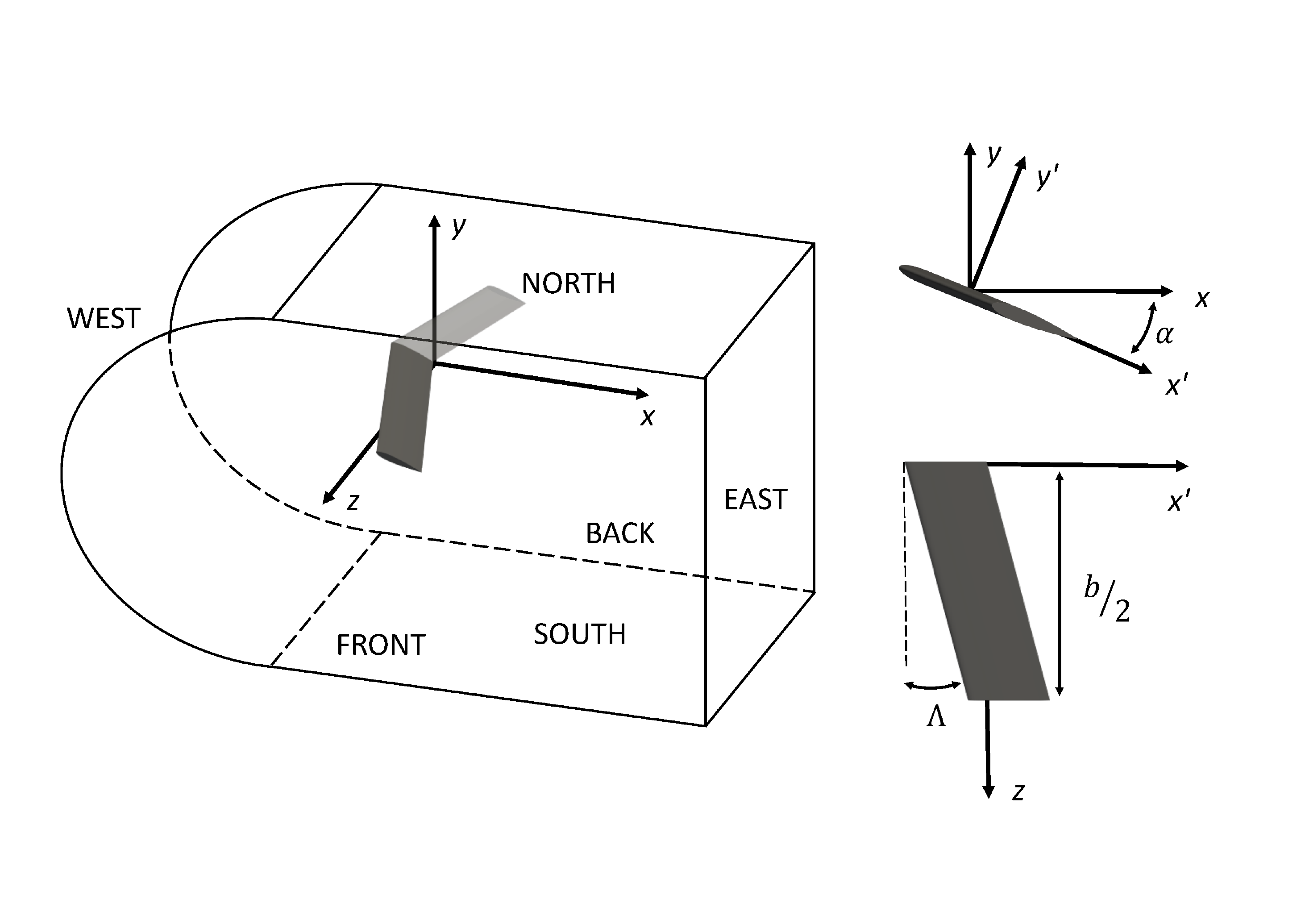}
    \caption{Problem setup showing wing and the computational domain. The symmetry condition is applied at the BACK plane. The half wing model is shown in grey and is not to scale. Light grey indicates the opposite side of the wing when mirrored in the symmetry plane and is shown for visualisation purposes only.}
    \label{fig:geometry}
\end{center}
\end{figure}

The geometry under consideration is a untapered wing based on the symmetric NACA 0015 airfoil with a sharp trailing edge and a straight cut wing tip. Taking advantage of the symmetry of the problem, half of the wing is considered as shown in figure \ref{fig:geometry}. The chord-based Reynolds number $Re = 400$ is held constant, while the wing sweep $(\Lambda)$, aspect ratio $(sAR)$ and angle of attack $(\alpha)$ are varied. Here, we use the symmetry aspect ratio defined as $sAR = b/2c$, where $b$ is the wingspan defined from wing tip to wing tip and $c$ is the wing chord.

It is important to take into account the order of the operations performed to construct a swept wing at an angle of attack. Translating the wing tip along $x$ first and then rotating the wing by $\alpha$ or first rotating and then translating the wing tip along $x$ will give two different geometries. The latter order of operations would also result in a wing that has a dihedral angle. To obtain the wing geometry used here the first a two-dimensional mesh was generated and the airfoil was rotated by $\alpha$. It was then extruded along a vector $\{x,y,z\}=\{b/2 \tan{\Lambda}\cos{\alpha}, -b/2 \tan{\Lambda}\sin{\alpha}, b/2\}$. This is equivalent to rotating the wing about an axis parallel to the leading edge and achieves a swept back wing without dihedral angle.

\begin{figure}		% Mesh figure
\begin{center}		
\begin{minipage}[b]{0.55\textwidth}
    \begin{subfigure}[b]{\textwidth}
        \begin{overpic}[width=1\textwidth]{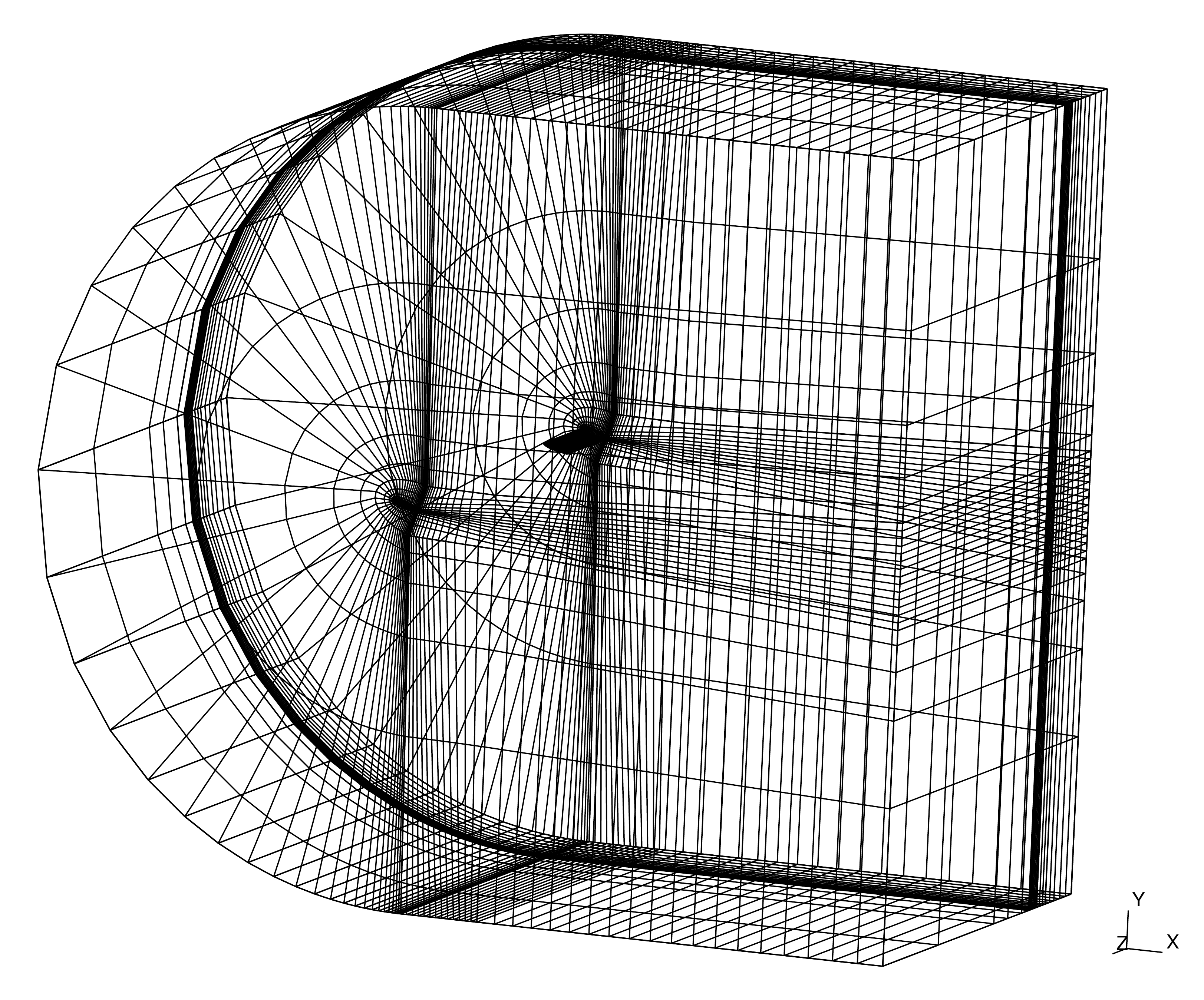}
		\put(0, 80) {$(a)$}
		\end{overpic}
        \label{fig:nek_mesh}
    \end{subfigure}

\end{minipage}
\begin{minipage}[b]{0.35\textwidth}
    \begin{subfigure}[b]{\textwidth}
    		\begin{overpic}[width=1\textwidth]{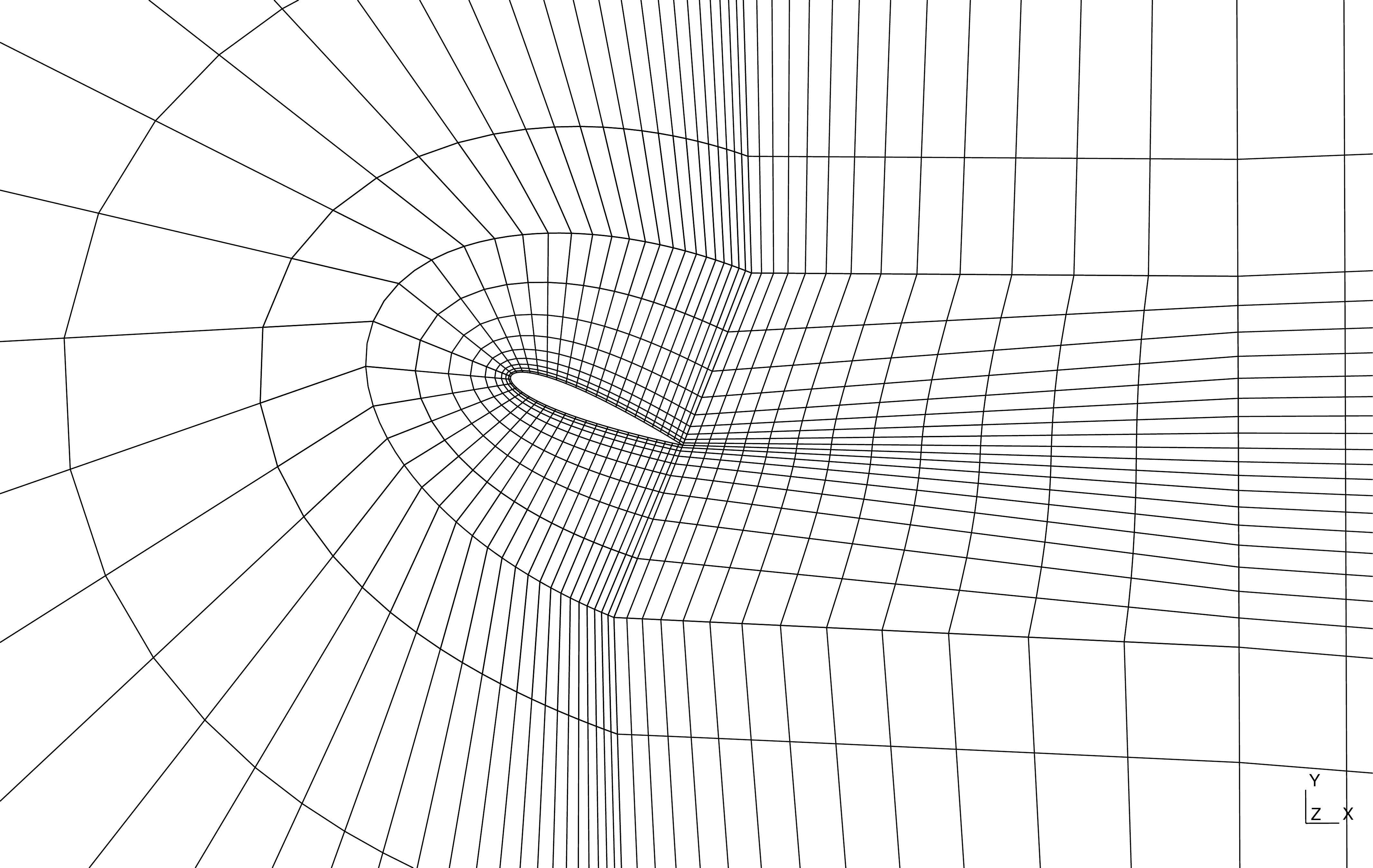}
		\put(-8, 60) {$(b)$}
		\end{overpic}
        \label{fig:mesh_close}
    \end{subfigure}
        \begin{subfigure}[b]{\textwidth}
        \begin{overpic}[width=1\textwidth]{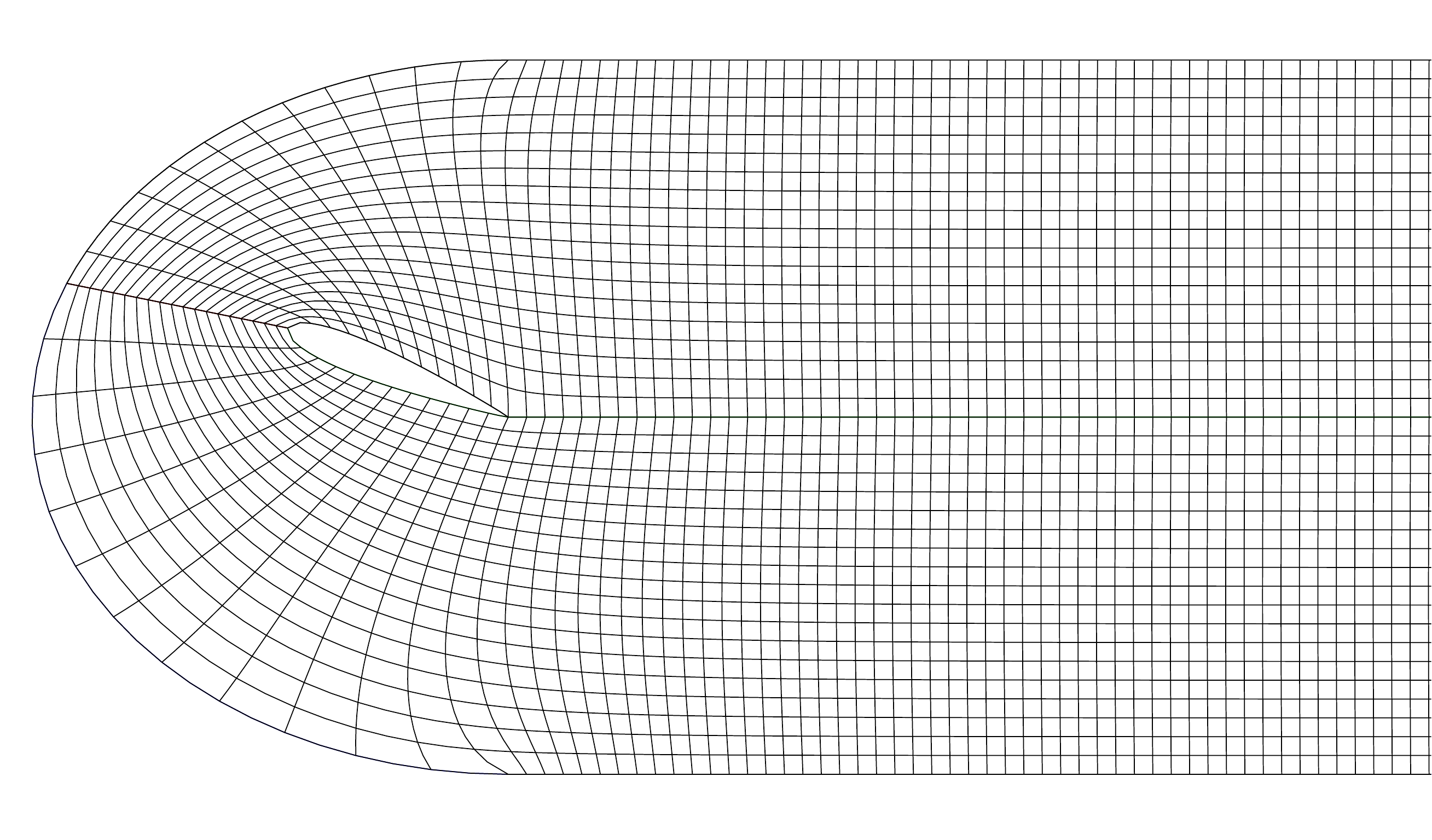}
		\put(-8, 50) {$(c)$}
		\end{overpic}
        \label{fig:pod_mesh}
    \end{subfigure}
\end{minipage}
\end{center}	
\caption{Computational mesh. ($a$) DNS mesh showing full domain ($b$) Close up of the DNS mesh near the airfoil ($c$) section of the mesh used for modal analysis. For clarity only the macroscopic elements are shown in ($a$) and ($b$), while the internal field and the mesh resulting from a high-order polynomial fitting are not shown. Similarly, only every second grid line is shown in ($c$).}
\label{fig:meshes}
\end{figure}

The computational extent is $(x,y,z)={[-15:20]\times[-15:15]\times[0:15]}$. The half wing was meshed using {\em Gmsh} \citep{gmsh}, with a structured C type mesh around the wing. Macroscopic elements for a typical $sAR = 4$ straight wing mesh are shown in figure \ref{fig:meshes}($a$), the closeup in \ref{fig:meshes}($b$) shows refinement near the wing. Within each element both spectral code (discussed in \S \ref{sec:codes}) resolve flow quantities by use of high-order polynomials, the degree of which is adjusted until convergence is achieved. Figure \ref{fig:meshes}($c$) shows the mesh used for data-driven analysis that will be explained below. Several computational meshes having different number of macroscopic elements were tested with different polynomial order $p$ to ensure spacial and temporal convergence. A combination of 46735 hexahedra and prisms as macroscopic elements for an $sAR = 4$ wing and polynomial order of $5$ was selected.

For analysing the effect angle of attack, the aspect ratio and sweep angle are kept constant at $sAR=4$ and $\Lambda=0^\circ$. While the effects of sweep are analysed at a constant angle of attack of $22^\circ$ at which the flow is separated with sweep angle varied between $0^\circ$ and $30^\circ$ for wings of $sAR=4$ and 2. Length and velocity is nondimensionalized by wing chord $c$ and $U_\infty$ respectively. Time refers to nondimensional convective time normalised by $c/U_\infty$ and the Strouhal number is defined as $St=fc \sin(\alpha)/U_\infty$. For modal stability results shown in further section, each perturbation component is normalised by its own absolute maximum, for example $u=u/\max(|u|)$.

% Due to the high computational costs associated with computing the three-dimensional base flow, mesh convergence was evaluated on a two-dimensional slice with mesh identical to that of the full three-dimensional field. Table \ref{tab:domain} shows the change of vertical velocity component and 
% wing lift with polynomial order $p$ and number of macroscopic elements $h$ for {\em \textit{Nektar++}}. A $p=5$ and $h=1888$ was chosen based on this. The number of mesh layers in the $z$ direction was set to maintain uniform discretization in all three directions in the near wake, with extra refinement near the wing tip, resulting in a total number of 46735 hexahedra and prisms as macroscopic elements for an $sAR = 4$ wing.

% \begin{table}{}
%  \begin{center}
% \begin{tabular}{ c c c c c c c }
% 	 & \multicolumn{2}{c}{$v$ at P1(2, 0)} & \multicolumn{2}{c}{$v$ at P2(4, 0)} & \multicolumn{2}{c}{$C_l$}  \\
% 	$p$ & $h$ = 1888 & $h$ = 3689 & $h$ = 1888 & $h$ = 3689 & $h$ = 1888 & $h$ = 3689 \\
% 	3  & 0.371516 & 0.382928 & 0.191197 & 0.221460 & 0.436170 & 0.438744 \\
% 	5  & 0.382060 & 0.383806 & 0.198087 & 0.213150 & 0.439123 & 0.439378 \\
% 	7  & 0.383675 & 0.383178 & 0.210402 & 0.206991 & 0.439319 & 0.439179 \\
% 	9 & 0.383664 &     -    & 0.200428 &    -     & 0.439341 &    - \\
%   \end{tabular}
%   \caption{Two dimensional convergence of vertical velocity measured at two probe locations and airfoil lift for different $h$ and $p$ values.}{\label{tab:domain}}
%  \end{center}
% \end{table}

\subsection{Solvers and boundary conditions}
\label{sec:codes}

\begin{table}			% Boundary conditions
\centering
 \renewcommand{\arraystretch}{1.2}
\begin{tabular}{lccccccc}
             & WING & NORTH & SOUTH & WEST & EAST & FRONT  & BACK\\

$\bar{u}$    & D    & U & U & U & O  & N & S\\
$\bar{v}$    & D    & U & U & U & O  & N & S\\
$\bar{w}$    & D    & U & U & U & O  & N & S\\	[2ex]

$\hat{u}$    & D    & D & D & D & N  & N & S\\
$\hat{v}$    & D    & D & D & D & N  & N & S\\
$\hat{w}$    & D    & D & D & D & N  & N & S\\
\end{tabular}
\caption{\label{tab:bcs} Boundary conditions for the base flow $\bar{\bm q}$ and perturbation $\hat{\bm q}$ components.}
\end{table}

Direct numerical simulation is used to solve equations of motion using either of the \textit{nek5000} \citep{nek5000} or \textit{nektar++} \citep{Cantwell2015} spectral element codes. \textit{nektar++} was used for computing artificially stationery base flows and triglobal stability analysis via time stepping, while \textit{nek5000} was chosen to generate flow fields for data-driven modal analysis due to its high speed. 

\begin{figure}		% Nektar Nek5000 probe comparisson
 \begin{center}
    \includegraphics[width=0.9\textwidth]{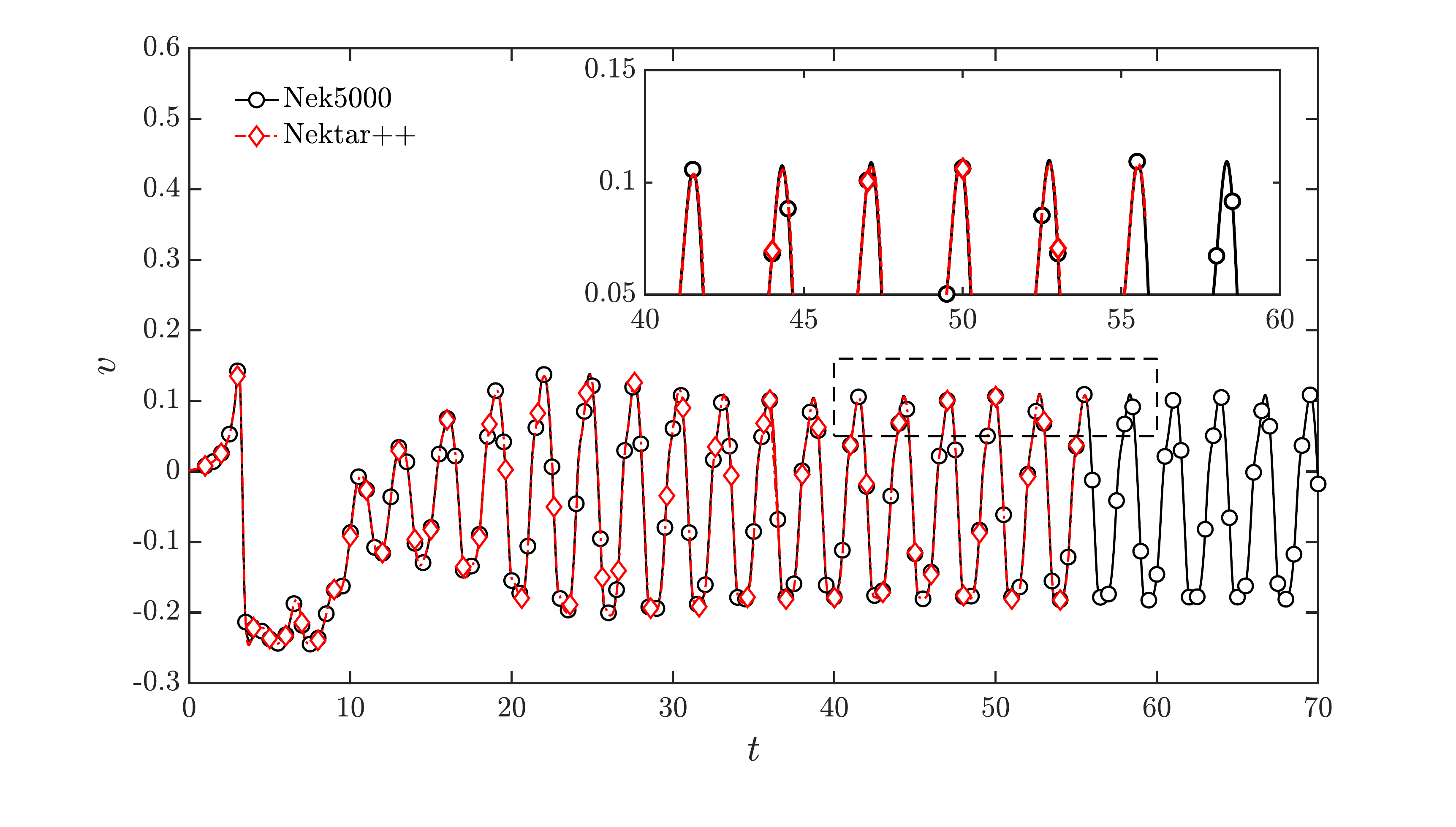}
    \caption{Comparison of $v$ velocity signal between \textit{nek5000} and \textit{nektar++} for $(sAR,\Lambda,\alpha)=(2,0^\circ,22^\circ)$ at $(x,y,z) = (4,0,1)$.}
    \label{fig:nek-nektar}
 \end{center}
\end{figure}

In order to close the systems of equations solved, the boundary conditions defined in table \ref{tab:bcs} are used. In this table, {\em D} denotes homogeneous Dirichlet, {\em N} denotes homogeneous Neumann, {\em S} denotes symmetry boundary and {\em O} denotes \emph{robust outflow} \citep{dong2014jcp} boundary conditions that were used in \textit{nektar++}, while at the \emph{inlet} uniform  flow ${\bm{U}}=(U_\infty,0,0)^T$ is imposed. The base flow solutions obtained by both codes were compared to ensure that identical results are achieved. Figure \ref{fig:nek-nektar} shows good agreement in variation of vertical velocity with time for a given wing geometry between the two codes. In general, for the configurations considered good agreement between the two codes is achieved when using time steps $\Delta t \leqslant 5\times 10^{-4}$ and polynomial orders $p \geqslant 5$.

\begin{table}{}	% Cl and St comparisson with Kai
 \begin{center}
\begin{tabular}{ c c @{\hskip 8mm} c c c @{\hskip 8mm} c c }
%\hline
 & & \multicolumn{3}{c}{Present results} & \multicolumn{2}{c}{Zhang \etal (2020a)}\\
 Case & $\alpha$ & $C_L$ & POD & DMD & $C_L$ & $St$\\[1ex]
\multirow{2}{*}{$sAR=4$}	& 12$^\circ$ & 0.365 & -- & -- & 0.362 & --\\
						    & 22$^\circ$ & 0.571 & 0.1398 & 0.1395 & 0.578 & 0.1387\\
\multirow{2}{*}{$sAR=2$}	& 12$^\circ$ & 0.330 & -- & -- & 0.327 & --\\
						    & 22$^\circ$ & 0.500 & 0.1338 & 0.1338 & 0.504 & 0.1336\\
					  2D	& 22$^\circ$ & 0.765 & --  & -- & 0.771 & --\\	
  \end{tabular}
  \caption{Comparison of mean lift coefficient over unswept NACA 0015 wings at $Re=400$ with literature.}{\label{tab:validation}}
 \end{center}
\end{table}

The values of the average lift coefficient ($C_L$), presented in table \ref{tab:validation} are compared to results of \citet{zhang_2020} showing good agreement. Further comparisons between the \textit{CharLES} and \textit{nektar++} solvers have been presented in \citet{He_2019} and \citet{zhang_2020}. The frequencies of the dominant modes as will be discussed in detail in \S \ref{sec:wake_modes} of straight wings of different aspect ratios are compared in table \ref{tab:validation} with the frequencies extracted from the lift coefficient time signal reported by \citet{zhang_2020}. Very good agreement between the dominant frequencies of the $C_L$ signal and dominant modes can be seen.

\subsection{Base flows and signal processing}
\label{sec:sfd}
%%% Residual algorithm
As discussed in \S \ref{sec:lst}, at conditions at which a steady state exists, it is used as the base flow for the analysis, either by converging the DNS solution in time, or by applying the residuals algorithm discussed in \S \ref{sec:ra}. 

%%%% SFD
Past the first bifurcation, unsteady flow ensues and obtaining a steady base flow is not as straight forward. A number of numerical techniques have been developed for the recovery of basic flows for the conditions where global linear instability is expected. These include approaches based on continuation \citep{Keller1977}, selective frequency damping (SFD) \citep{AkervikEtAl2006}, and more recently residual recombination procedure \citep{Citro2017} and minimal gain marching \citep{Teixeira2017}. 

Here the SFD method, as implemented in \textit{nektar++}, has been used to compute an artificially stationary, unstable base state that is used for subsequent modal analysis. This methodology was shown to recover {\em amplified} global modes of a sphere \citep{He2019}. SFD uses filtering and control of unstable temporal frequencies in the flow, the time continuous formulation in \textit{nektar++} can be expressed as
\begin{align}
	\left\{
		\begin{array}{ll}
      		\dot{\bm{q}} &= NS(\bm{q})- \gamma(\bm{q} - \bar{\bm{q}}), \\
      		\dot{\bar{\bm{q}}} &= (\bm{q} - \bar{\bm{q}})/\Delta \\
		\end{array} 
	\right.
\end{align}
where $\bm{q}$ represents the problem unknown(s), the dot represents the time derivative, $NS$ represents the Navier-Stokes equations, $\gamma \in \mathbb{R}_+$ is the control coefficient, $\bar{\bm{q}}$ is a filtered version of $\bm{q}$, and $\Delta \in {\mathbb{R}_+}^*$ is the filter width of a first-order low-pass time filter \citep{Jordi_sfd}. Choice of the parameters $\gamma$ and $\Delta$ affects the convergence to the steady-state solution when $\bm{q} = \bar{\bm{q}}$. If the dominant mode is known, for example through DMD, and specified as input one can adjust the filter parameters to accelerate convergence.

\subsection{Modal analyses}
TriGlobal instability analysis was performed using time-stepper algorithm implemented in \textit{nektar++} \citep{Cantwell2015} with proper boundary conditions shown in table \ref{tab:bcs}, using the implicitly restarted Arnoldi method and a Krylov subspace dimension of 32. Two most amplified eigenmodes are solved with a tolerance of $10^{-4}$ using the fully three-dimensional base flow. %The typical computational requirement is 8 Intel Xeon nodes each with 385GB memory 160k CPUhours.
POD and DMD were carried out using an in house Fortran code that uses LAPACK \citep{lapack} implementations of singular value decomposition and eigenvalue solvers. The flow fields were computed with \textit{nek5000}. Snapshots were recorded every 0.1 unit of characteristic time once the flow has reached periodic shedding, with 300 snapshots collected in the interval from 40 to 70 time-units. Each snapshot was interpolated using built in spectral interpolation of \textit{nek5000} onto the equidistant mesh shown in figure \ref{fig:meshes}($c$) such that scaling due to changing cell volume for each date point does not have to be taken into account. This simplifies the construction of the covariance matrix. The new mesh was generated by extruding an elliptic two-dimensional mesh around the airfoil, which resulted in areas of changing cell volume and some mesh distortion directly above and below the wing, especially below the pressure side. Hyperbolic mesh could have offered better control over the cell shape at the boundary but would not allow to define the far field boundary itself. It should be emphasised that data-driven methods do not require computation of gradients, so the effects of mesh non-orthogonality at the boundary is minimal, especially since there are minimal changes in the base flow at these locations. Finally, the mode structures associated with massively separated flows such as the high-$\alpha$ wings considered here are expected to lie in the wake where cell volume is constant. This mesh extended $7c$ downstream of the training edge, $1.5c$ above and below the wing, and $1c$ beyond the wing tip in z direction. With 26 points per characteristic length this resulted in $412 \times 39 \times 130$ mesh for $sAR = 4$ and $412 \times 39 \times 78$ mesh for $sAR = 2$ wings. Three velocity components were stacked to generate the data vector for DMD with the addition of pressure for POD the resulting data matrix size is shown in table \ref{tab:matsize}.

\begin{table}{}			% POD, DMD mesh
 \begin{center}
 \renewcommand{\arraystretch}{1.2}
\begin{tabular}{ c c c c c c }
	 \multicolumn{2}{c}{Case} & Variables & $n_x \times n_y \times n_z$ & $n$ & Memory (GB) \\
	 \hline
	 	\multirow{2}{*}{$sAR=2$} & POD & $u, v, w, p$ & \multirow{2}{*}{$412 \times 39 \times 78$} & 5 013 216 & 11.2 \\
	 						    & DMD & $u, v, w$   & 											& 3 759 912 & 8.4 \\
\\%\vspace{2mm}
	 	\multirow{2}{*}{$sAR=4$} & POD & $u, v, w, p$ & \multirow{2}{*}{$412 \times 39 \times 130$} & 8 355 360 & 18.7 \\
	 						    & DMD & $u, v, w$    & 											 & 6 266 520 & 14.0 \\
  \end{tabular}
  \caption{Mesh and data matrix sizes for modal analysis.}{\label{tab:matsize}}
 \end{center}
\end{table}

%% file: sections/results.tex
\section{Results}
\label{sec:results}

%=======================================================================
\subsection{Wake dynamics}
\label{sec:bf}
%=======================================================================
\begin{figure}		% Baseflow AoA
	\begin{center}
        \begin{overpic}[width=0.85\textwidth]{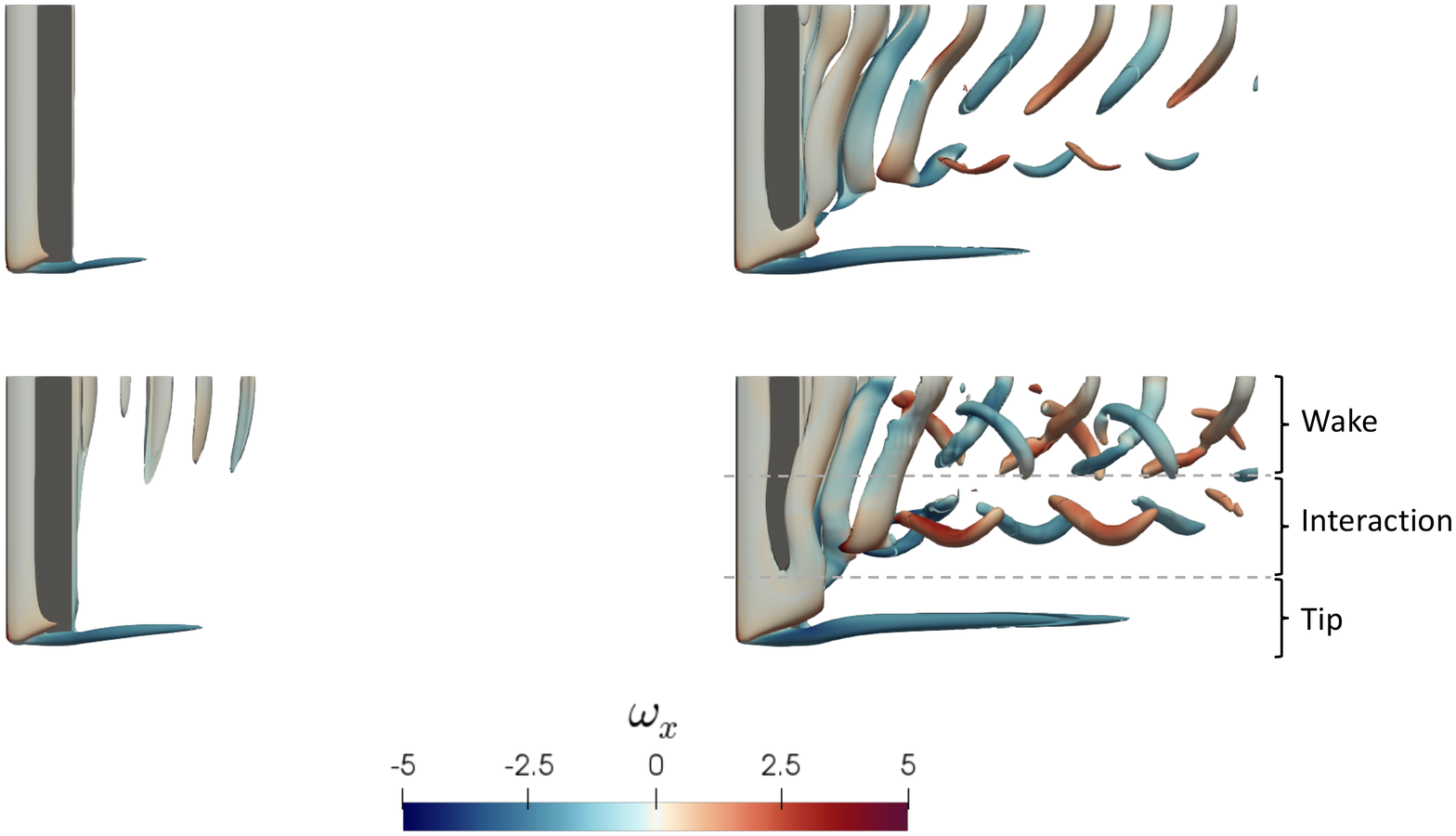}
 		\put(-2, 50) {$\alpha=10^\circ$}
 		\put(-2, 28) {$\alpha=14^\circ$}
 		\put(43, 50) {$\alpha=18^\circ$}
 		\put(43, 28) {$\alpha=22^\circ$}
		\end{overpic}	
	\end{center}
	\vskip -3em 
    \caption{Isocontours of $Q$-criterion $Q = 1$ coloured by streamwise vorticity showing the effect of angle of attack on instantaneous DNS solution at $sAR=4$, $\Lambda=0^\circ$.}
    \label{fig:bfAoA}
\end{figure}

\begin{figure}
	\begin{center}
        \begin{overpic}[width=1\textwidth]{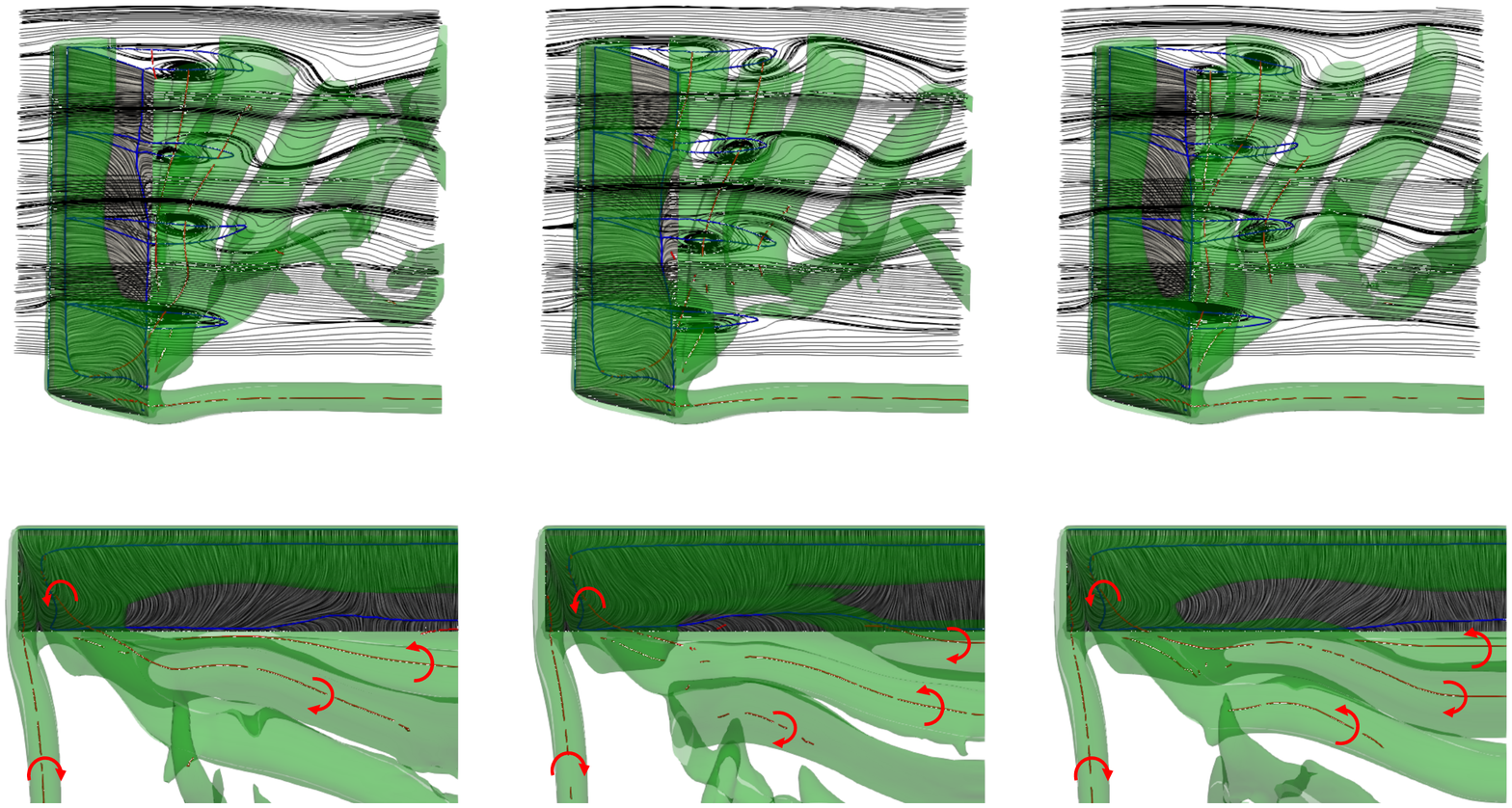}
		\put(3, 58) {$(a)$}
		\put(34, 58) {$(b)$}
		\put(63.5, 58) {$(c)$}
		%\begin{overpic}[width=0.9\textwidth]{figures/vortexcores.pdf}
	    %\put(-3, 50) {$(a)$}
		%\put(32, 50) {$(b)$}
		%\put(66.5, 50) {$(c)$}
		\end{overpic}
	\end{center}
	\vskip -5em 
    \caption{Snapshots of DNS for $(AR,~\Lambda,~\alpha) = (4,~0^\circ,~22^\circ)$ at three different times $(a)$ $t=50$, $(b)$ $t=60$, $(c)$ $t=71$.  Grey/white contour shows the surface-friction lines on the wing.  Black lines show the streamlines in slices corresponding to spanwise location at $z=0$, 1, 2 and 3. The separation zone is highighted by blue lines. Vortices are visualised by isocontour of $Q=1$ and its center location and rotational direction are highlighted by red lines and arrows, respectively.}
    \label{fig:vortexcores}
\end{figure}

\begin{figure}		% Baseflow sweep
	\begin{center}
        \begin{overpic}[width=0.85\textwidth]{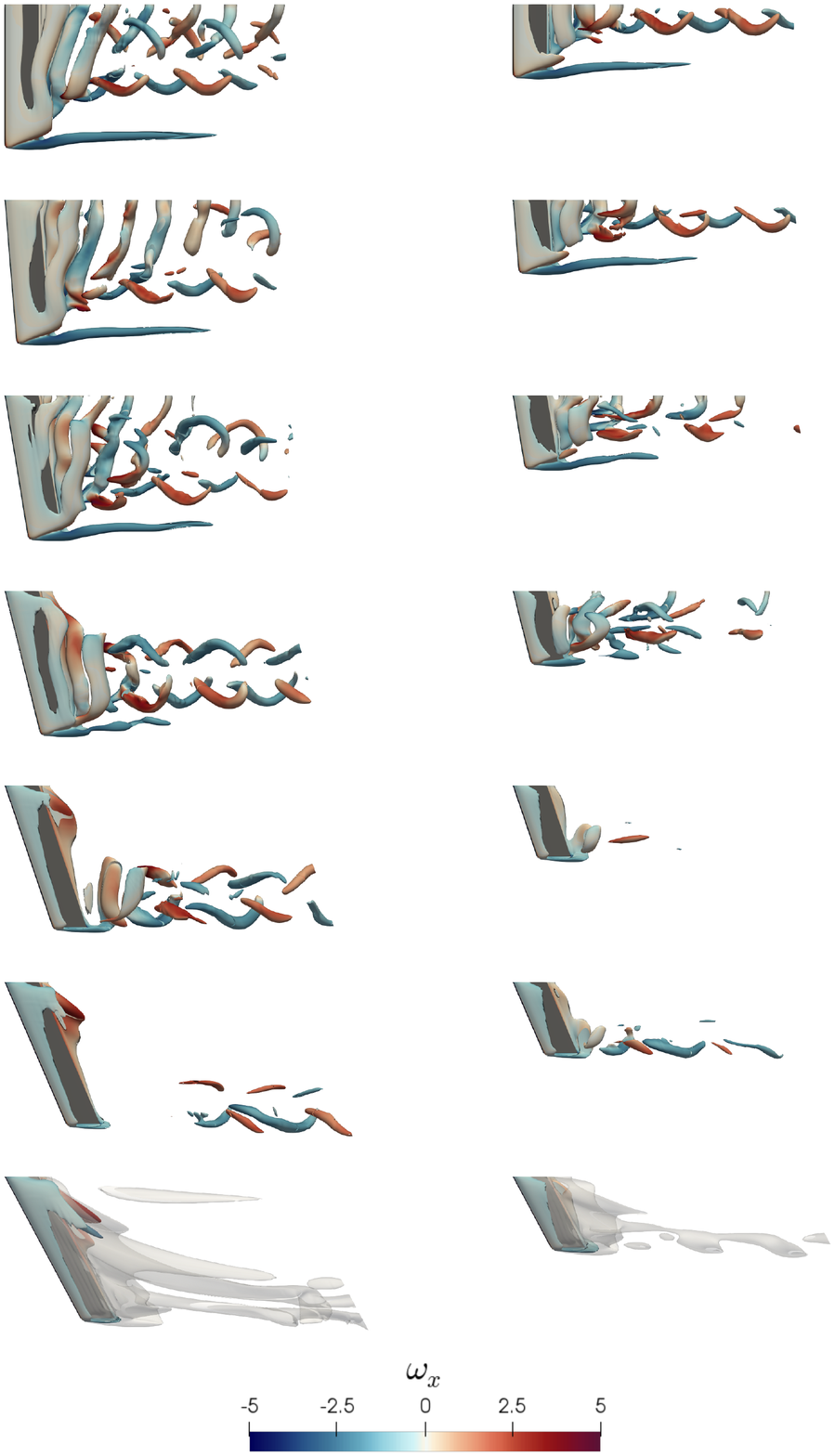}
        \put(13,96) {$sAR=4$}
        \put(48,96) {$sAR=2$}
		\put(0, 89) {$\Lambda=0^\circ$}
		\put(0, 76) {$\Lambda=5^\circ$}
		\put(0, 64) {$\Lambda=10^\circ$}
		\put(0, 51) {$\Lambda=15^\circ$}
		\put(0, 39) {$\Lambda=20^\circ$}
		\put(0, 26) {$\Lambda=25^\circ$}
		\put(0, 14) {$\Lambda=30^\circ$}
		\end{overpic}	
	\end{center}
	\vskip -1em
    \caption{Isocontours of $Q$-criterion $Q = 1$ coloured by streamwise vorticity showing the effect of sweep on instantaneous DNS solution at $Re = 400$ and two wing aspect ratios. Top to bottom are  $\Lambda=0^\circ - 30^\circ$ with increment of $5^\circ$.
    For clarity $\Lambda=30^\circ$ is shown with $Q = 0.1$ in transparent grey.}
    \label{fig:bfSweep}
\end{figure}

The evolution of the flow over the unswept $sAR=4$ wing with angle of attack is shown in figure \ref{fig:bfAoA}. The vortical structures of the three-dimensional wake over unswept wings is in agreement with the DNS results of \citet{zhang_2020}. For the separated flows at high angles of attack, three regions can be identified behind the wing. These are the wake region close to the symmetry plane containing mostly spanwise vortices, the tip region containing the tip vortex and the interaction region in between where a pair of counter-rotating wake vortices are connected into a closed loop by braid-like structures \citep{zhang_2020}.

At $\alpha=10^\circ$, shown in figure \ref{fig:bfAoA}, the flow is steady with separation occurring at approximately two-thirds of the chord, independently of the angle of sweep. At $\alpha=14^\circ$, an unsteady wake is formed, the shed vortices being practically parallel to the trailing edge of the wing. The separation location moves approximately at half-chord and the spanwise region of the flow affected by the tip vortex is reduced, with the separation bubble extending closer to the tip. At the higher angles of attack of $18^\circ$ and $22^\circ$, also shown in figure \ref{fig:bfAoA}, the three distinct regions first identified by \citet{zhang_2020} develop: these regions are the wake, consisting of spanwise vortices near the symmetry plane, the essentially steady tip vortex, and the interaction region between the wake and tip characterised by the  braid-like vortices. 

With increasing $\alpha$ the separation location moves closer to the leading edge and the tip vortex becomes stronger.
At $\alpha=18^\circ$ and $22^\circ$ the reattachment line close to the trailing edge of the airfoil starts to oscillate periodically, as can be seen in the surface streamlines movies available as online supplementary material. The origin of this motion is presented in detail at a single set of parameters $(AR,~\Lambda,~\alpha) = (4,~0^\circ,~22^\circ)$, in figure~\ref{fig:vortexcores}. Vortex centers and surface streamlines are plotted at three instances in time. When a counter-clockwise (on the $xy$ plane, as viewed in the top row of figure \ref{fig:vortexcores}) trailing edge vortex starts to separate from the wing near the symmetry plane the reattachment line (shown in blue) is pushed further away from the trailing edge of the wing as in figure~\ref{fig:vortexcores}($a$) . The shedding of a clockwise leading edge vortex causes the reattachment location to return closer to the trailing edge as in figure~\ref{fig:vortexcores}($b$). Alternating shedding of clockwise and counter-clockwise vortices that is non-uniform in the spanwise direction leads to the wavy-like motion of the reattachment line.  Further downstream, each pair of counter-rotating spanwise vortices connect through the braid-like structures that are shed between $2 \le z \le 2.5$, forming the vortex loops discussed by \citet{zhang_2020}.

The frequency of these motions was measured by extracting a line parallel to the trailing edge of the wing, located $0.1c$ above the trailing edge, and performing a fast Fourier transform of the signal on this line. The dominant frequencies based on the time signal of vertical velocity component were found to be $St=0.139$, $0.139$ and $0.143$ for the $sAR=4$ wing at sweep angles of $\Lambda=0^\circ$, $5^\circ$ and $10^\circ$ respectively. The spanwise locations of these peaks of the power spectral density correspond well with the extremities of the reattachment line motion shown in figure \ref{fig:vortexcores}.

The effect of sweep on the flow over the longer $sAR=4$ and the shorter $sAR=2$ wing are shown in figure \ref{fig:bfSweep}. As the longer wing is swept back, the braid-like region is moved closer to the wing tip by the increased spanwise cross flow, which results in the tip vortex becoming noticeably less steady. Interestingly, stronger periodic vortices are observed in the near wake behind the wing at $(sAR,\Lambda)=(4,5^\circ)$, shown in figure \ref{fig:bfSweep}, that meet at an oblique angle on the symmetry plane and form chevron-like patterns reminiscent of those reported in the wake of large aspect ratio cylinders by \citet{Williamson_1989}. These vortices extend about four chord lengths downstream of the wing with four cores clearly visible before the wake changes to the two distinct regions seen behind the straight wing. 

\begin{figure}		% spanwise vs induced
  \begin{center}	 
    \begin{subfigure}[b]{0.47\textwidth}
        \begin{overpic}[width=1\textwidth]{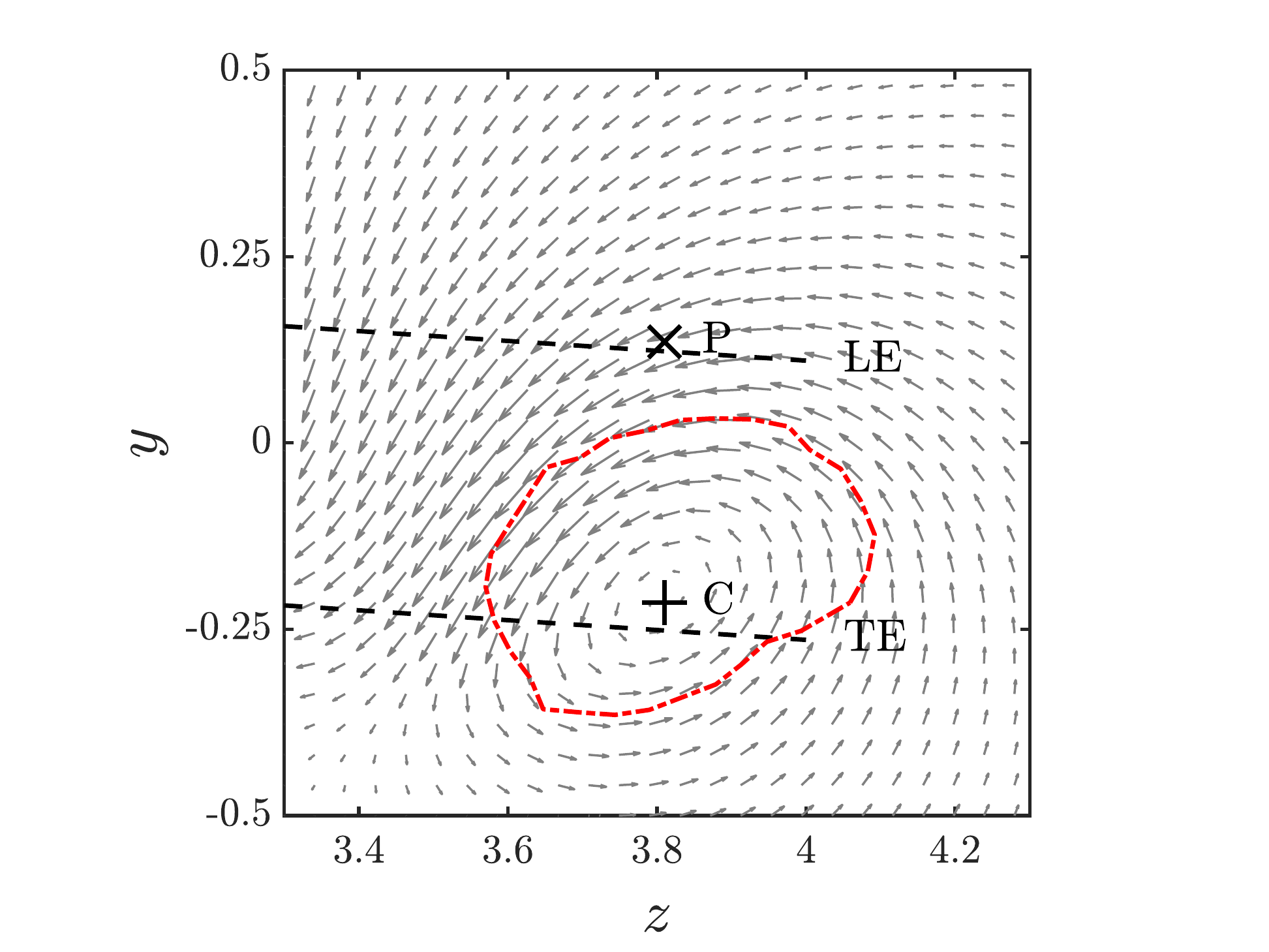}
		\put(2, 68) {$($a$)$}
		\end{overpic}
    \end{subfigure}
    %~ 
    \begin{subfigure}[b]{0.47\textwidth}
        \begin{overpic}[width=1\textwidth]{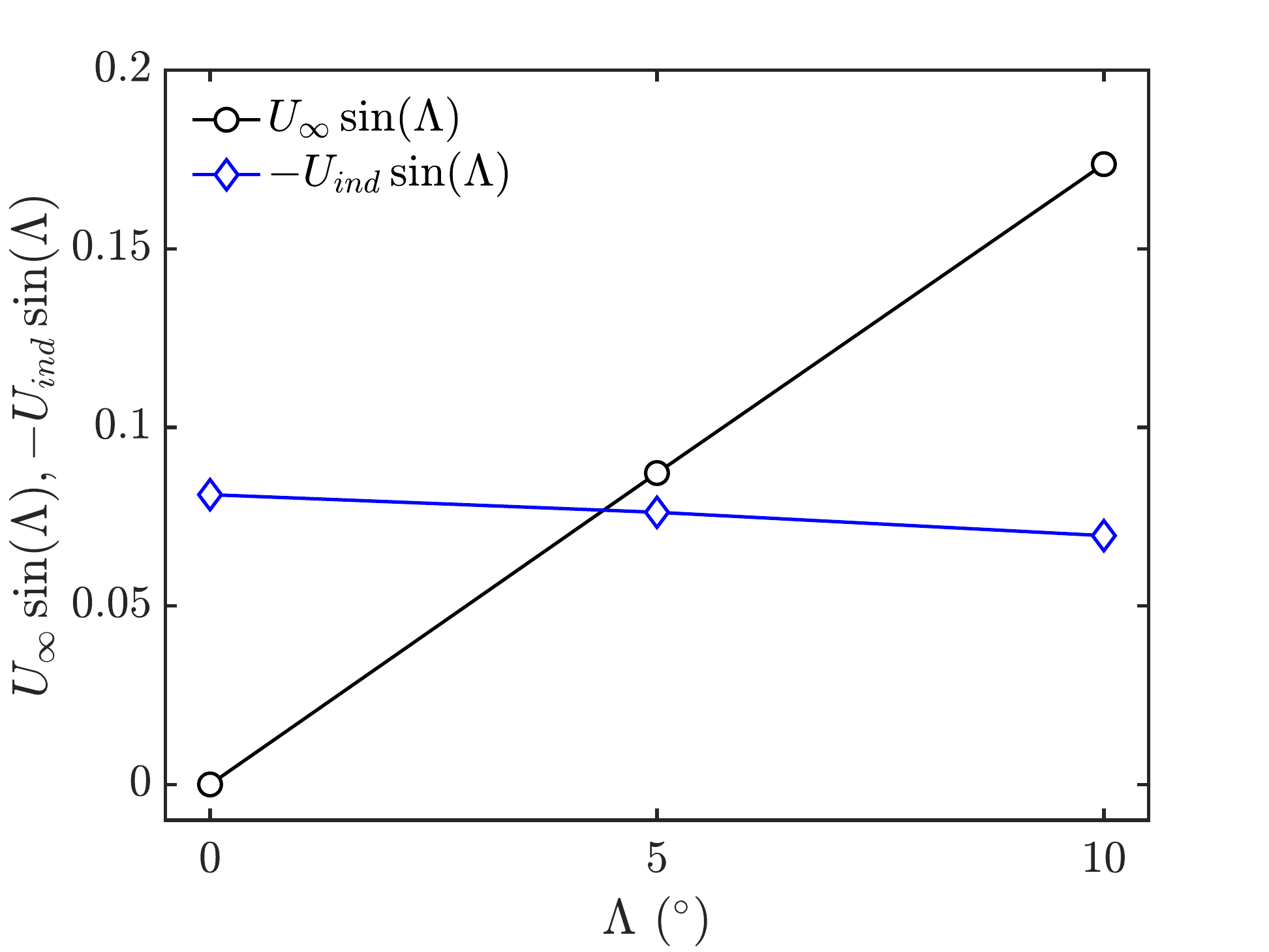}
		\put(-1, 68) {$(b)$}
		\end{overpic}
    \end{subfigure}
    \caption{Effects of spanwise and vortex induced flow ($a$) tip vortex region $(sAR,\Lambda,\alpha)=(4,10^\circ,22^\circ)$ one $c$ downstream of wing trailing edge showing vortex center and core, ($b$) comparison of components parallel to the wing leading edge of free-stream velocity $U_\infty$ and vortex induced velocity $U_{ind}$ for low angles of sweep.}
    \label{fig:induced}
  \end{center}
\end{figure}

In order to analyse the cause of this inherently two-dimensional shedding at $\Lambda=5^\circ$, the spanwise component of velocity (parallel to wing leading edge) caused by the sweep angle and the velocity induced by the tip vortex along the opposite direction were compared. The circulation $\Gamma$ of the tip vortex was calculated from time-averaged flow at one chord downstream of the wing. Figure \ref{fig:induced}(a) shows the location of the tip vortex relative to the projections of wing LE and trailing edge (TE) as well as the vortex core size defined here as the contour of maximum azimuthal velocity for the sweep angle of $10^\circ$. The vortex core is located in the wake behind the wing with its spanwise position moving inboard further downstream of the wing.  For a point $P$ outside the vortex core the induced velocity can be approximated as $U_{ind} = \Gamma/4\pi r$ where $r$ is the distance between vortex center $C$ and $P$. The induced velocity by the tip vortex is compared to the spanwise component of the free-stream flow upstream of the wing $(U_\infty \sin(\Lambda))$ in figure \ref{fig:induced}($b$) as a function of the sweep angle $\Lambda$. With increasing sweep the strength of the tip vortex and hence $U_{ind}$ slightly decreases while the spanwise component of the free-stream that is opposite in direction increases significantly. Near $\Lambda=5^\circ$ the magnitudes of the spanwise velocity of both components are similar, which supports the previous finding that at this sweep angle the wake evolves in a quasi-two-dimensional manner.

The oscillations of the reattachment line seen on the unswept case are also observed with sweep. However, at $\Lambda=10^\circ$ a steady region of attached flow forms at the inboard trailing edge of the wing. There is a qualitative change in the structure $sAR=4$ wing wake as sweep angle reaches $\Lambda=15^\circ$. The periodic vortices passing though the symmetry plane are no longer visible, and the wake now consists of two series of braid-like vortices forming behind the half-wing, that do not pass through the symmetry plane. The tip vortex is now less pronounced and clearly unsteady.  As the sweep angle increases further to $20^\circ$, the tip vortex is no longer visible and the wake is dominated by the two braid-like regions. The size of the attached flow region at the inboard root of the wing increases with sweep and at $\Lambda=20^\circ$ the separation bubble splits into two halves on either side of the symmetric wing, with attached flow forming in the middle. Interestingly, the presence of such region of attached flow at the root of a swept wing was also reported by \citet{Visbal_2019} for turbulent flow at much higher Reynolds numbers.  At $\Lambda=25^\circ$ the braid-like regions become narrower and vortices extending from the inboard section of the wing into the wake behind the tip are starting to form; these structures are sometimes referred to as "ram’s horn" vortices \citep{Black_1956}. The growth of these structures results in another fundamental change in the wake at the maximum sweep of $\Lambda=30^\circ$ examined in our work. A "ram’s horn" vortex is generated on the suction side of the wing close to the leading edge at the symmetry plane and a stronger counter-rotating vortex emanates from the trailing edge. These two vortices form a closed structure further downstream that starts to shed with the shed vortices reminiscent of the hairpin-vortex.
Overall higher angle of sweep has a stabilising effect on the flow. It was shown by \citet{zhang_2020b} that as the sweep is further increased the flow wake will become steady at $\Lambda \approx 45^\circ$.

The effects of sweep are qualitatively analogous on the shorter wing. The breakdown of vortices at the symmetry plane to two braid-like regions happens at $\Lambda=20^\circ$ rather then $15^\circ$. Horn-like vortices similar to the larger wing form at $\Lambda=30^\circ$. Unlike the $sAR=4$ case, these structures do not form hairpin vortices further in the wake, but rather merge into a spiral vortex, an effect attributed to the stronger tip effects on the shorter wing.

%=======================================================================
\subsection{Linear global modes}
%=======================================================================

\begin{figure}		% Global Spectrum
\begin{center}
	\begin{overpic}[width=1\textwidth]{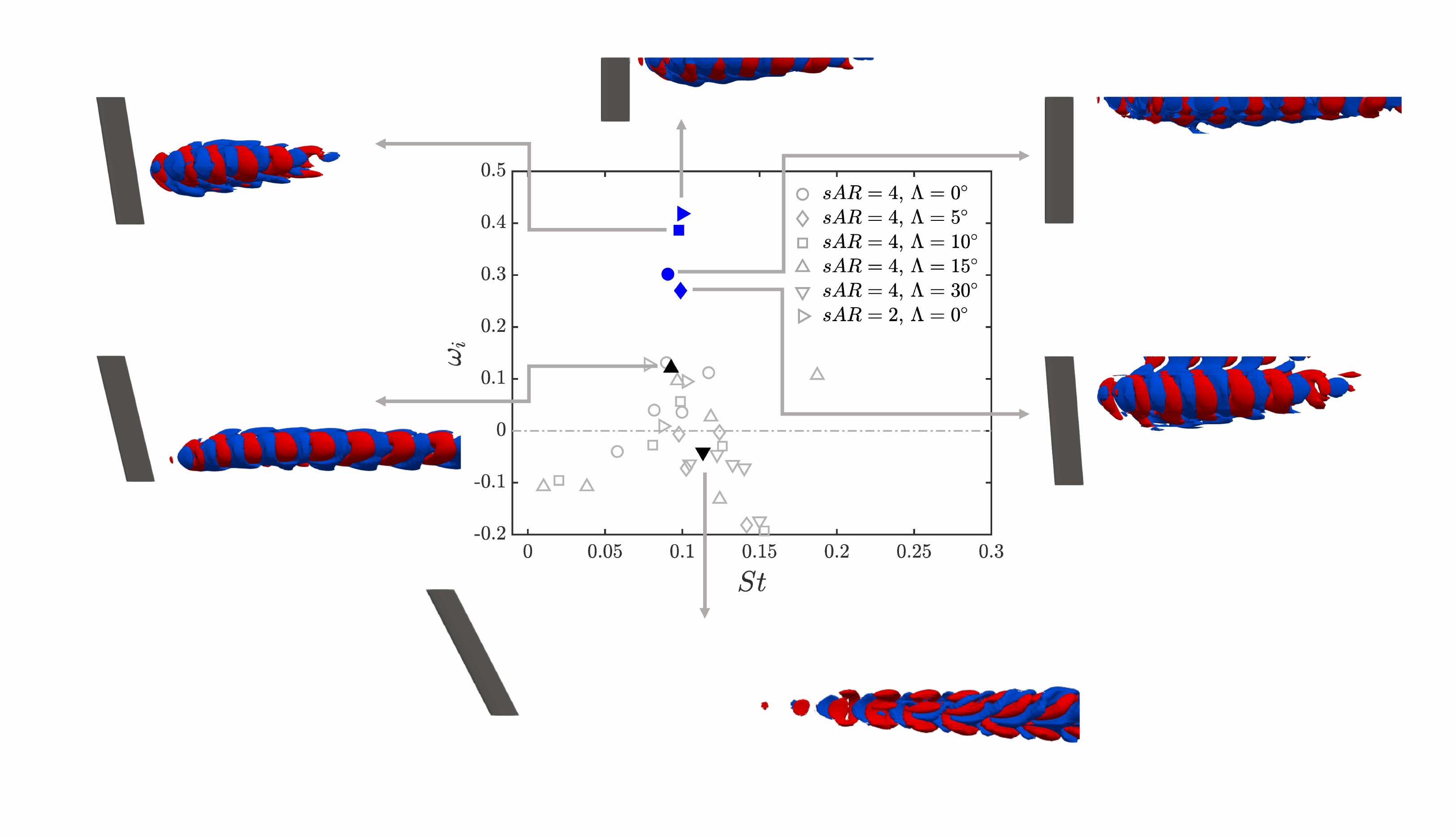}
	\end{overpic}
\end{center}
\vskip -3em 
    \caption{Leading global modes of different wing geometries for a constant $\alpha=22^\circ$. Most unstable or least stable modes are indicated with filled symbols with corresponding spatial structures visualised with contours of $Q=\pm1$ shown from the top. Here $St$ is defined as $St=\omega_r c \sin \alpha / 2 \pi U_{\infty}$.}
    \label{fig:global_spectrum}
\end{figure}

TriGlobal modal linear stability analysis was performed at conditions at which steady flow naturally existed or could be computed using the selective frequency damping method discussed in \S \ref{sec:sfd} and the effect of sweep angle and aspect ratio on the least stable global modes was documented. Due to the high computational cost of the SFD method, analysis was performed at a limited (but representative) number of configurations at which an unstable stationary flow could be computed with the focus on the least stable dominant modes. Figure \ref{fig:global_spectrum} summarises the findings of TriGlobal analyses carried out over different wing geometries and flow parameters. The amplification rate and frequency of the leading global mode at each set of conditions is plotted on the same graph for comparison, alongside the spatial structure of the most amplified mode at each set of conditions, visualised with contours of the same, arbitrarily chosen value of the $Q$-critetion. 

\begin{figure}		% TriGlobal AR=4 0,5,10 and AR=2, 0
\vskip 2em 
\begin{center}
        \begin{overpic}[width=1\textwidth]{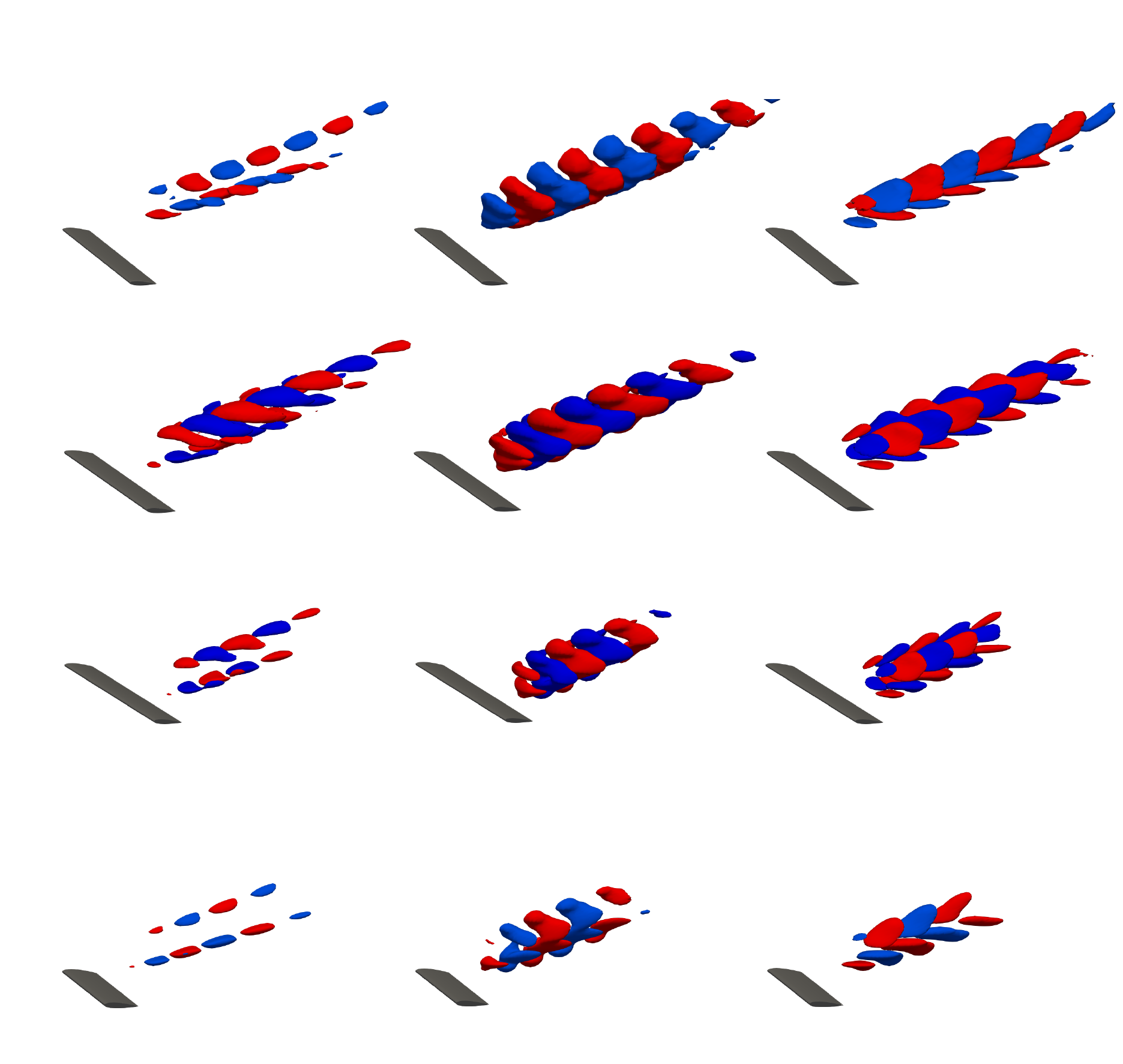}
        \put(45, 92) {$sAR=4$}
        \put(1, 88) {$\Lambda$:}
 		\put(20, 88) {$\hat{u}$}
 		\put(55, 88) {$\hat{v}$}
 		\put(85, 88) {$\hat{w}$}
 		\put(1, 75) {$0^\circ$}
 		\put(1, 55) {$5^\circ$}
 		\put(1, 35) {$10^\circ$}
 		\put(45, 20) {$sAR=2$}
 		\put(1, 9) {$0^\circ$}
		\end{overpic}
\end{center}
\vskip -1em
    \caption{Most unstable TriGlobal modes for $sAR=4$ and $2$ wings at different angles of sweep shown with contours of $\hat{u}$, $\hat{v}$, and $\hat{w}$ at $\pm 0.2$.}
    \label{fig:TG-low-sweep}
\end{figure}

The most unstable mode for the unswept $sAR=4$ wing takes the form of vortical structures localised at the symmetry plane. The components of the normalised perturbation velocity plotted in figure \ref{fig:TG-low-sweep} show the presence of two branches in the $\hat{u}$ and $\hat{w}$ perturbation velocity components, originating from the leading and trailing edges. The leading mode of the shorter $sAR=2$ wing is more amplified but is qualitatively identical to the $sAR=4$ result. 
With increasing sweep angle, the peak of the mode structure moves along the spanwise direction towards the tip. Qualitatively, the structures observed at $\Lambda=0^\circ$, $5^\circ$, and $10^\circ$ represent the same wake mode as it is evolving with sweep angle and are highlighted in blue in figure \ref{fig:global_spectrum}. The difference in the non-dimensional frequency of these modes remains below 10\% in this range of sweep angles. 

\begin{figure}		% Sweep 15 and 30 Vorticity slices
\begin{center}
    \begin{subfigure}[b]{0.49\textwidth}
        \begin{overpic}[width=1\textwidth]{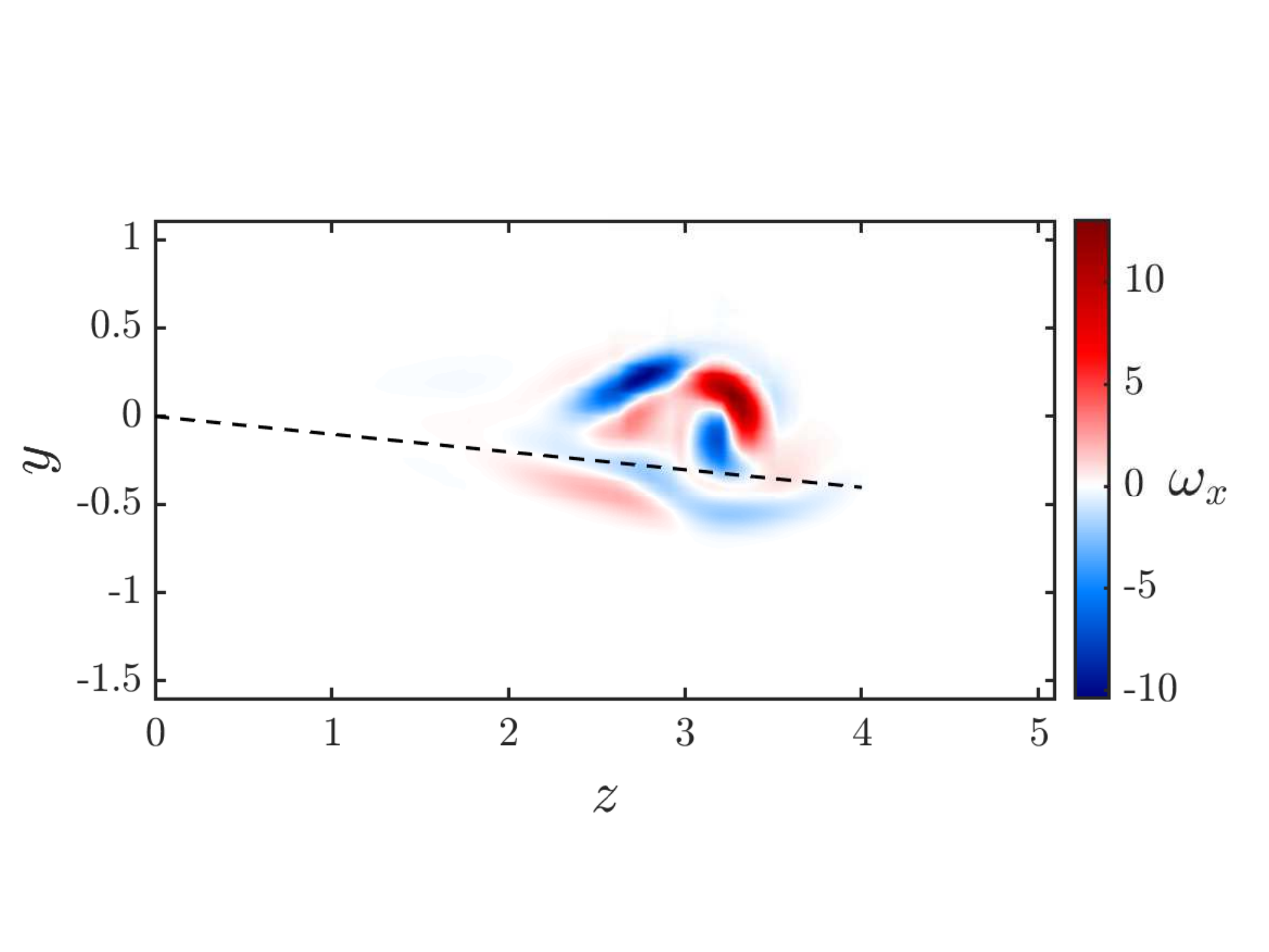}
		\put(0, 55) {$(a)$}
		\end{overpic}
    \end{subfigure}
    %~ 
    \begin{subfigure}[b]{0.49\textwidth}
        \begin{overpic}[width=1\textwidth]{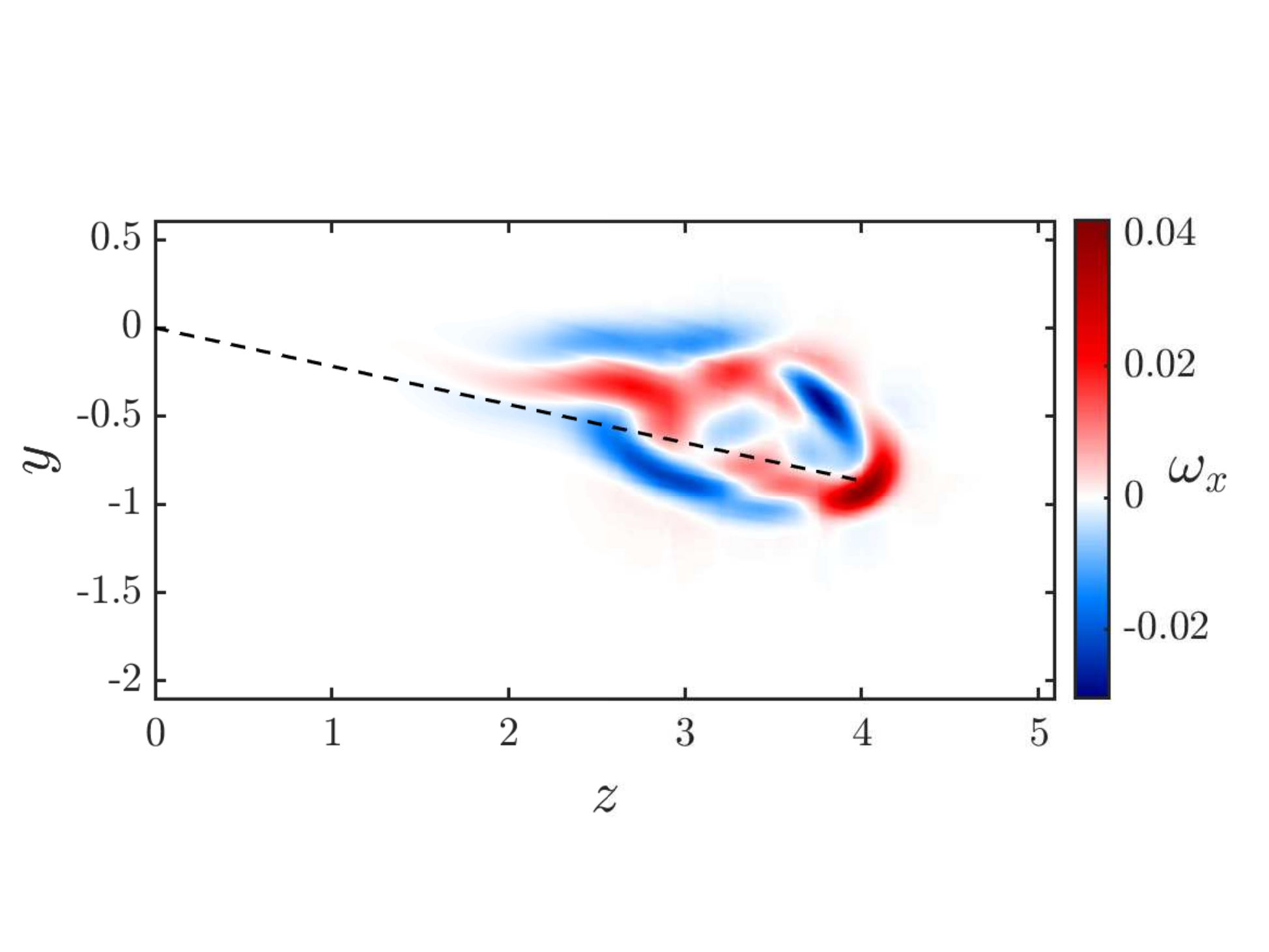}
		\put(0, 55) {$(b)$}
		\end{overpic}
    \end{subfigure}
\vskip -2em   
    \begin{subfigure}[b]{0.49\textwidth}
        \begin{overpic}[width=1\textwidth]{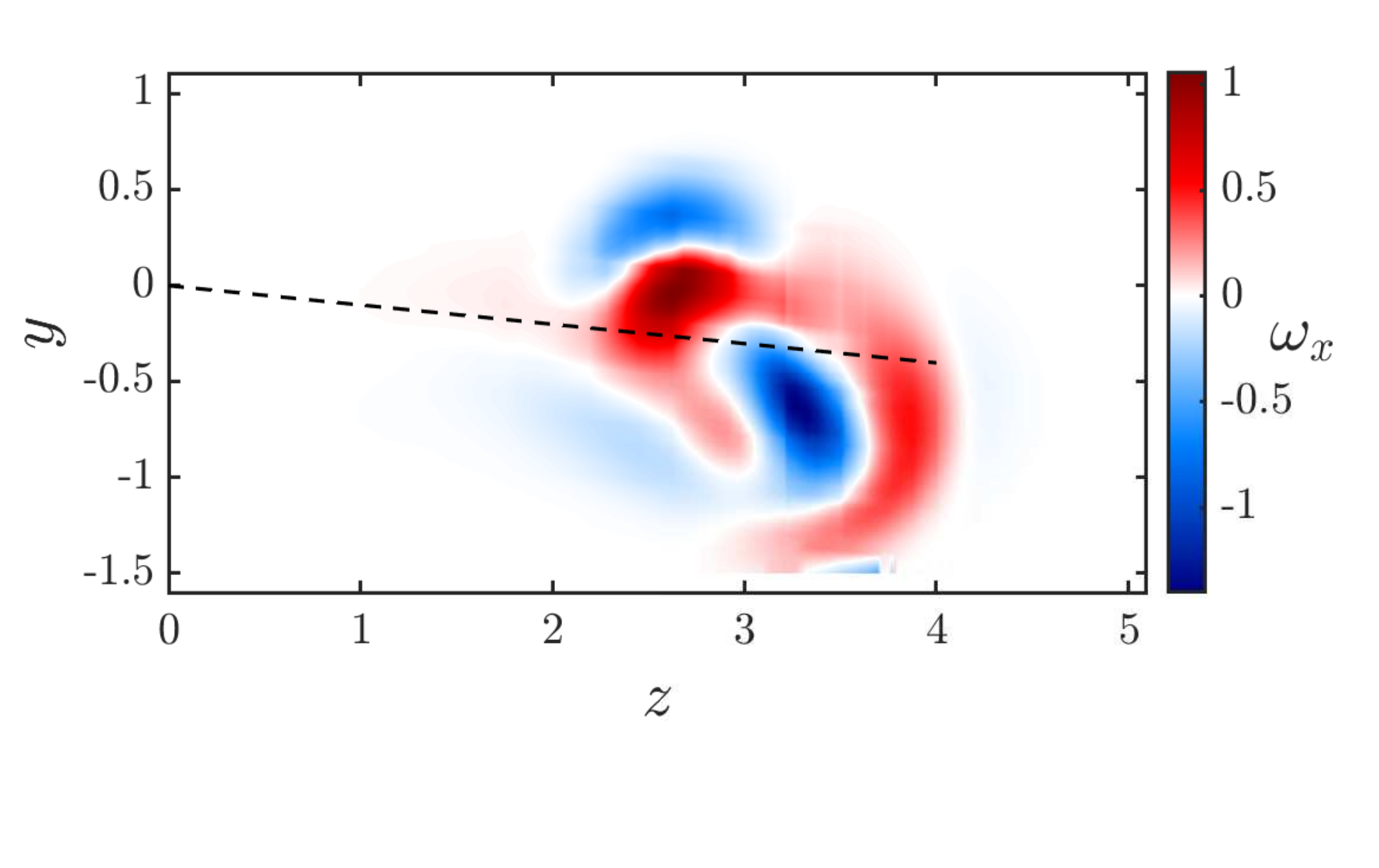}
		\put(0, 55) {$(c)$}
		\end{overpic}
    \end{subfigure}
    %~ 
    \begin{subfigure}[b]{0.49\textwidth}
        \begin{overpic}[width=1\textwidth]{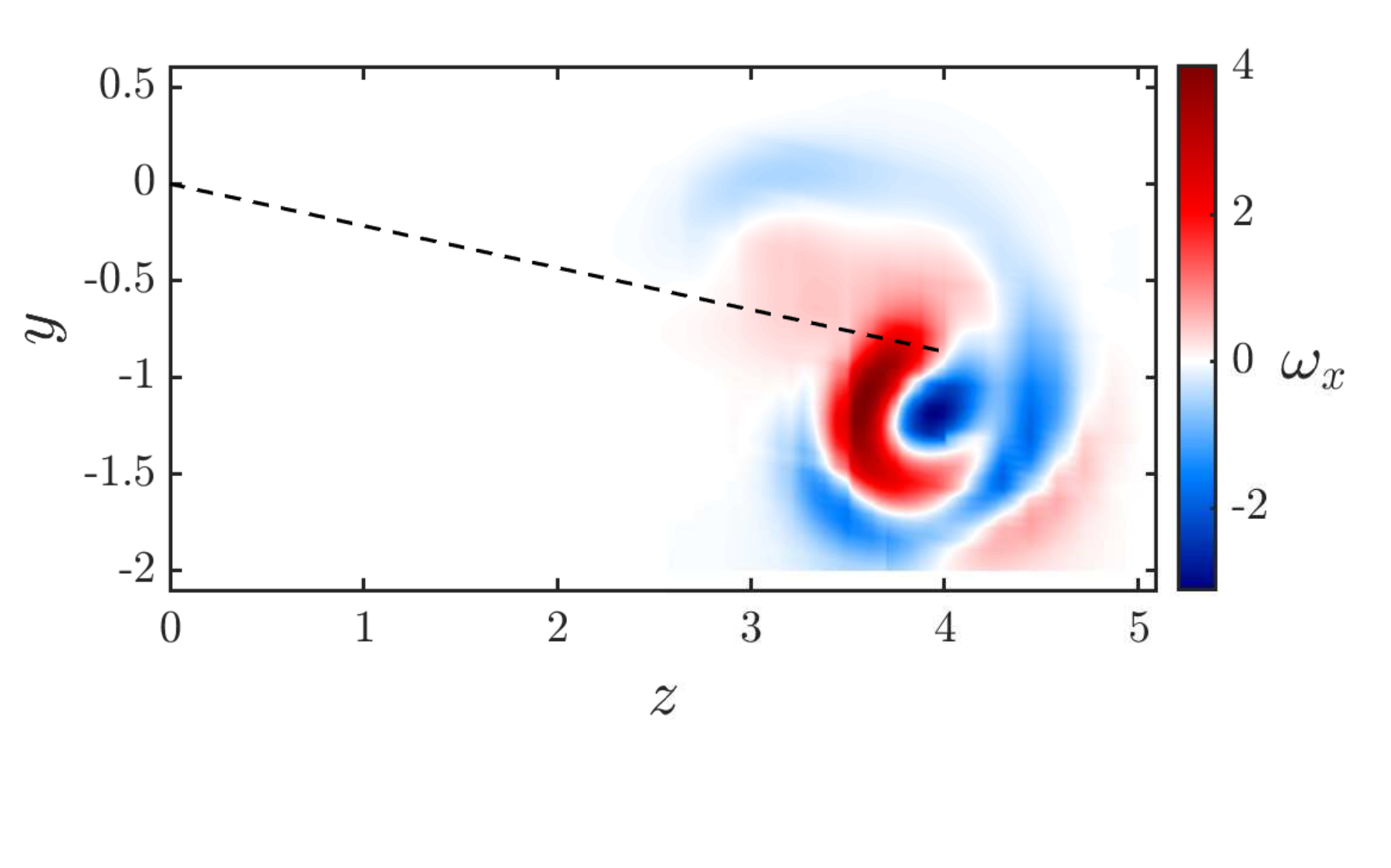}
		\put(0, 55) {$(d)$}
		\end{overpic}
    \end{subfigure}
\end{center}
\vskip -2em
    \caption{Streamwise vorticity of leading global modes for ($a,c$) $(sAR,\Lambda,\alpha)=(4,15^\circ,22^\circ)$ and ($b,d$) $(4,30^\circ,22^\circ)$ plotted at $x=5$ in ($a,b$) and $x=20$ in ($c,d$). The dashed line indicates the projection of the wing trailing edge onto the the slice.}
    \label{fig:vorticity}
\end{figure}

There is a change in the structure of the leading mode at $\Lambda=15^\circ$. Although the structures of vortices close to the wing still resemble the lower sweep results, as visible in figure \ref{fig:global_spectrum}, closer examination of the velocity components reveals significant changes. While mode structures at the lower sweep angles were mostly streamwise periodic, with little change in the $x$ direction, the leading mode at $\Lambda=15^\circ$ transitions to a vortical instability further downstream of the wing as evident from the vorticity plots in figure \ref{fig:vorticity}$(a,c)$ where positive $\omega_x$ indicates clockwise rotation (counter-rotating with respect to the tip vortex). At the maximum considered sweep of $\Lambda=30^\circ$ the leading mode is stable and takes the form of a vortical instability that grows with $x$ downstream of the wing. Unlike the $\Lambda=15^\circ$ mode that exhibits vortical structures centred around counter rotating vortices of the SFD base flow, the $\Lambda=30^\circ$ global mode clearly shows the influence of the tip vortex, as can be seen in $yz$-plane at $x=20$ shown in figure~\ref{fig:vorticity}$(d)$. The change of amplification rate with sweep is significant, showing an overall trend of lower amplification and fewer unstable modes with increased sweep, until no amplified modes at all are observed at $\Lambda=30^\circ$ with one notable exception of $\Lambda=10^\circ$.

\begin{figure}		% 4-0-22 perturbation growth
  \begin{center}	 
    \begin{subfigure}[b]{0.49\textwidth}
        \begin{overpic}[width=1\textwidth]{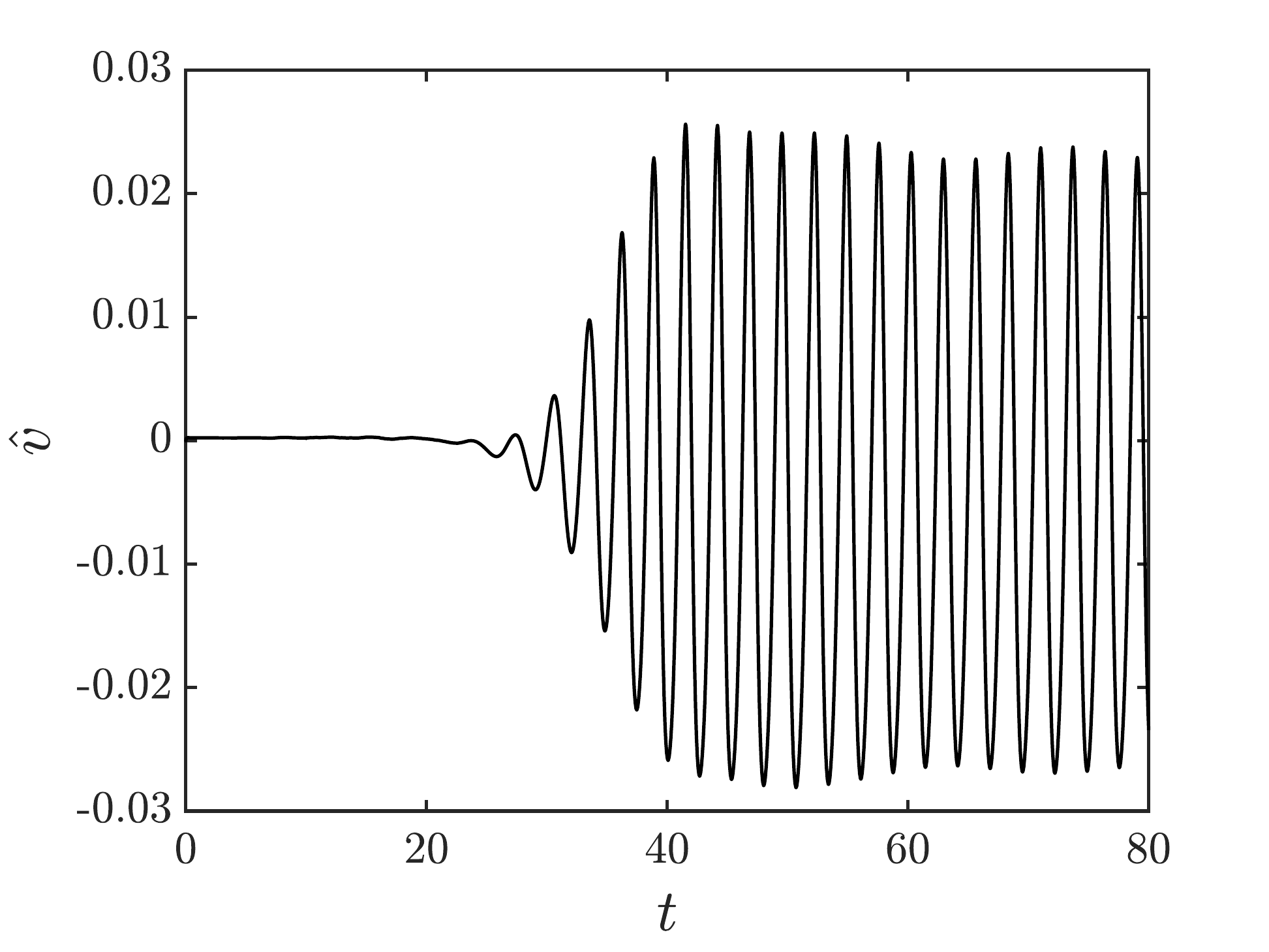}
		\put(-1, 68) {$(a)$}
		\end{overpic}
    \end{subfigure}
    %~ 
    \begin{subfigure}[b]{0.49\textwidth}
        \begin{overpic}[width=1\textwidth]{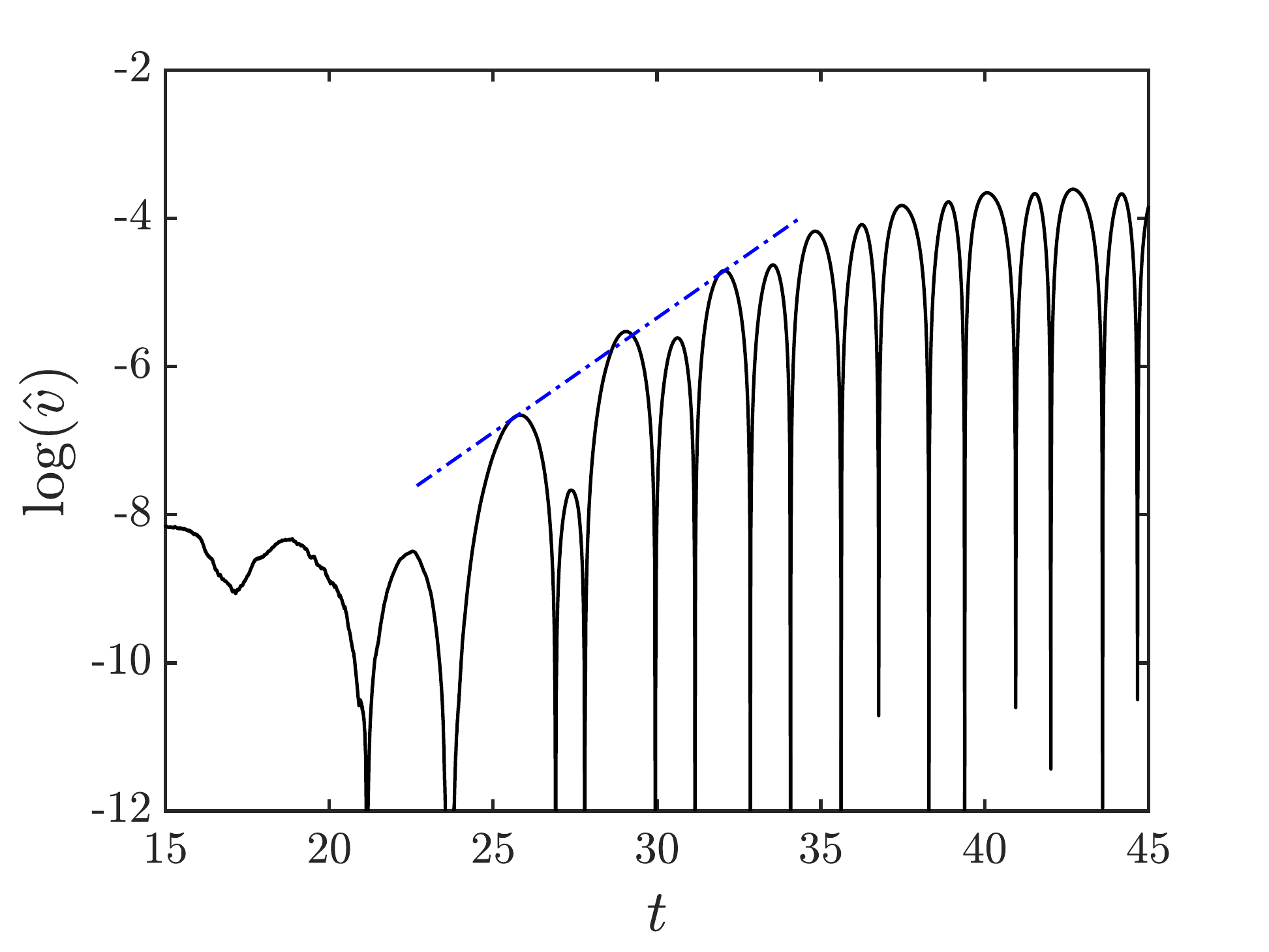}
		\put(-1, 68) {$(b)$}
		\end{overpic}
    \end{subfigure}
    \caption{Growth of the perturbation for $(sAR,\Lambda,\alpha)=(4,0^\circ,22^\circ)$ ($a$) perturbation with time showing linear growth and nonlinear saturation ($b$) close up plotted with $\log$ scale showing exponential growth.}
    \label{fig:4-0-22-pert}
  \end{center}
\end{figure}

In order to validate the above findings, full direct numerical simulations have been initialised using selected global modes as initial condition. As an example, the TriGlobal mode at $(sAR,\Lambda,\alpha)=(4,0^\circ,22^\circ)$ is superimposed at an amplitude $O(10^{-8})$ onto the SFD-obtained base flow. The time history of the $\hat{v}$ component of the perturbation at a probe located at the symmetry plane is shown in figure \ref{fig:4-0-22-pert}($a$), where the linear growth region leading to nonlinear saturation is clearly visible. The slope, measured from the logarithmic plot in figure~\ref{fig:4-0-22-pert}($b$) is equal to $0.3106$ which is $3\%$ larger than the growth rate of the global mode. The small discrepancy between the growth rate is attributed to the presence of additional amplified modes in the stability analysis. %Therefore the unstable global mode is responsible for the formation of the unstable wake.
%
%It should be pointed out that if snapshots for the modal analysis were recorded in this region, POD and DMD would pick up the global mode just like in the cylinder example shown in appendix \ref{sec:appB-2}. However, this is not the intention here, rather data-driven analysis is used to study the wake in the nonlinear regime (see \S 4.3).

\begin{figure}		% TriGlobal Sweep 15 and 30
\begin{center}
        \begin{overpic}[width=1\textwidth]{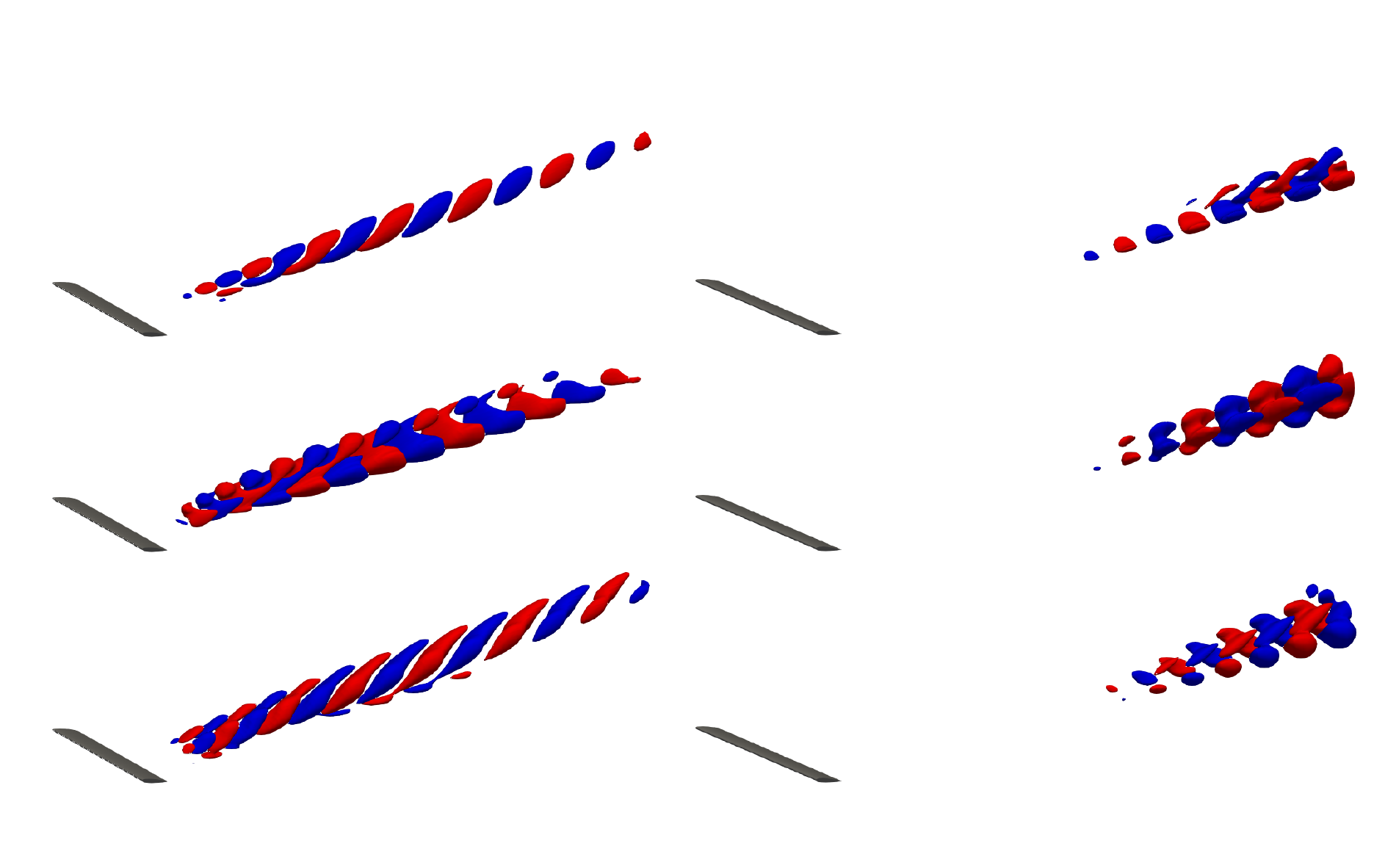}
 		\put(20, 55) {$\Lambda=15^\circ$}
 		\put(70, 55) {$\Lambda=30^\circ$}
 		\put(1, 45) {$\hat{u}$}
 		\put(1, 30) {$\hat{v}$}
 		\put(1, 12) {$\hat{v}$}
		\end{overpic}
\end{center}
\vskip -1em
    \caption{Most unstable TriGlobal mode for $(sAR,\Lambda,\alpha)=(4,15^\circ,22^\circ)$ and least stable mode for $(4,30^\circ,22^\circ)$ $\hat{u}$, $\hat{v}$ and $\hat{w}$ at $\pm 0.2$.}
    \label{fig:TG15-30}
\end{figure}

%=======================================================================
\subsection{Comparison with local theory}

\begin{figure}		% Local comparison
  \begin{center}	 
        \begin{overpic}[width=0.6\textwidth]{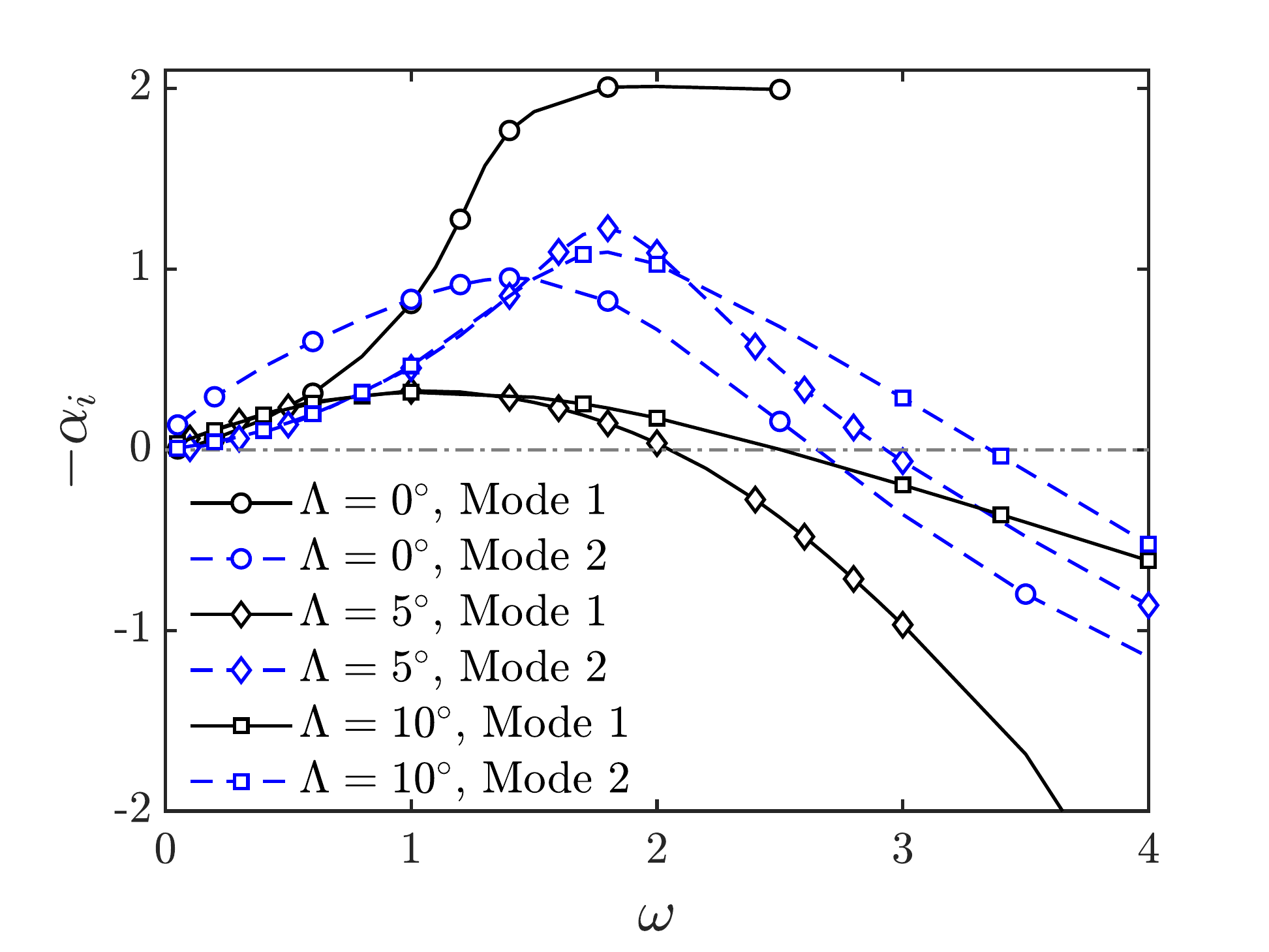}
		\put(-1, 68) {$(b)$}
		\end{overpic}
    \caption{Local spatial stability analysis results for small sweep $\Lambda=0^\circ-10^\circ$ showing variation of the spatial amplification rate $\alpha_i$ with frequency $\omega$.}
    \label{fig:local}
  \end{center}
\end{figure}

The relative simplicity and low computational cost of local stability analysis make it an attractive tool to compare its results with those of the computationally challenging TriGlobal analysis. The obvious shortcoming of local theory is the assumption of flow homogeneity in two spatial directions; however it might be argued that locally this assumption is fulfilled in the wake of at least the high aspect ratio wings. Moreover, previous works have demonstrated that it can deliver reasonably accurate results when applied at such locations in an overall three-dimensional flow that mostly satisfy the parallel flow assumption \citep{HeTCFD,Paredes_2016}. %It is reasonable to pose a question whether local stability theory can adequately predict the physics of the instability for finite swept wing geometry.

To address this question, solutions of the spatial Orr-Sommerfeld equation were obtained using steady base flow profiles of streamwise velocity $u$ computed by the SFD method. The larger aspect ratio wing was chosen, and profiles were extracted at the midspan for the unswept wing, so as to stay away of the tip vortex and the three-dimensional vortical structures at quarter-span. For the low sweep angles of $\Lambda=5^\circ$ and $10^\circ$, spanwise locations of $z/c=1$ and $2$ were chosen based on the maximum velocity deficit in the streamwise velocity profile. Due to high three-dimensionality of both the base flow and the global mode at larger sweep angles, local analysis was not applied at those conditions. The perturbation was assumed to be homogeneous in the streamwise direction and the analysis was carried out at spawnwise wavenumber, $\beta$ of zero.

\begin{figure}		% Local comparison
  \begin{center}	 
    \begin{subfigure}[b]{0.325\textwidth}
        \begin{overpic}[width=1\textwidth]{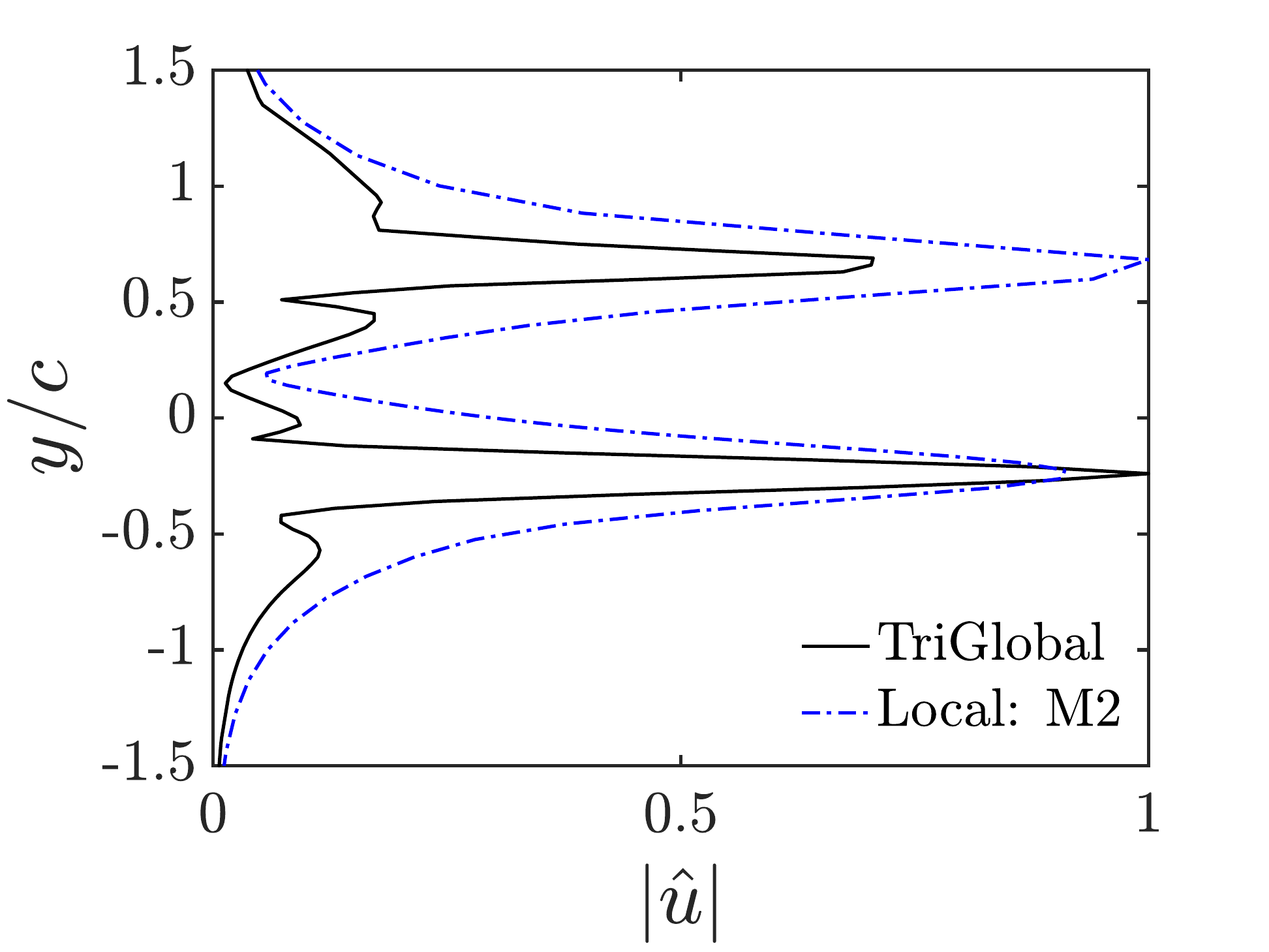}
		\put(-1, 68) {$(a)$}
		\end{overpic}
    \end{subfigure}
    %~ 
    \begin{subfigure}[b]{0.325\textwidth}
        \begin{overpic}[width=1\textwidth]{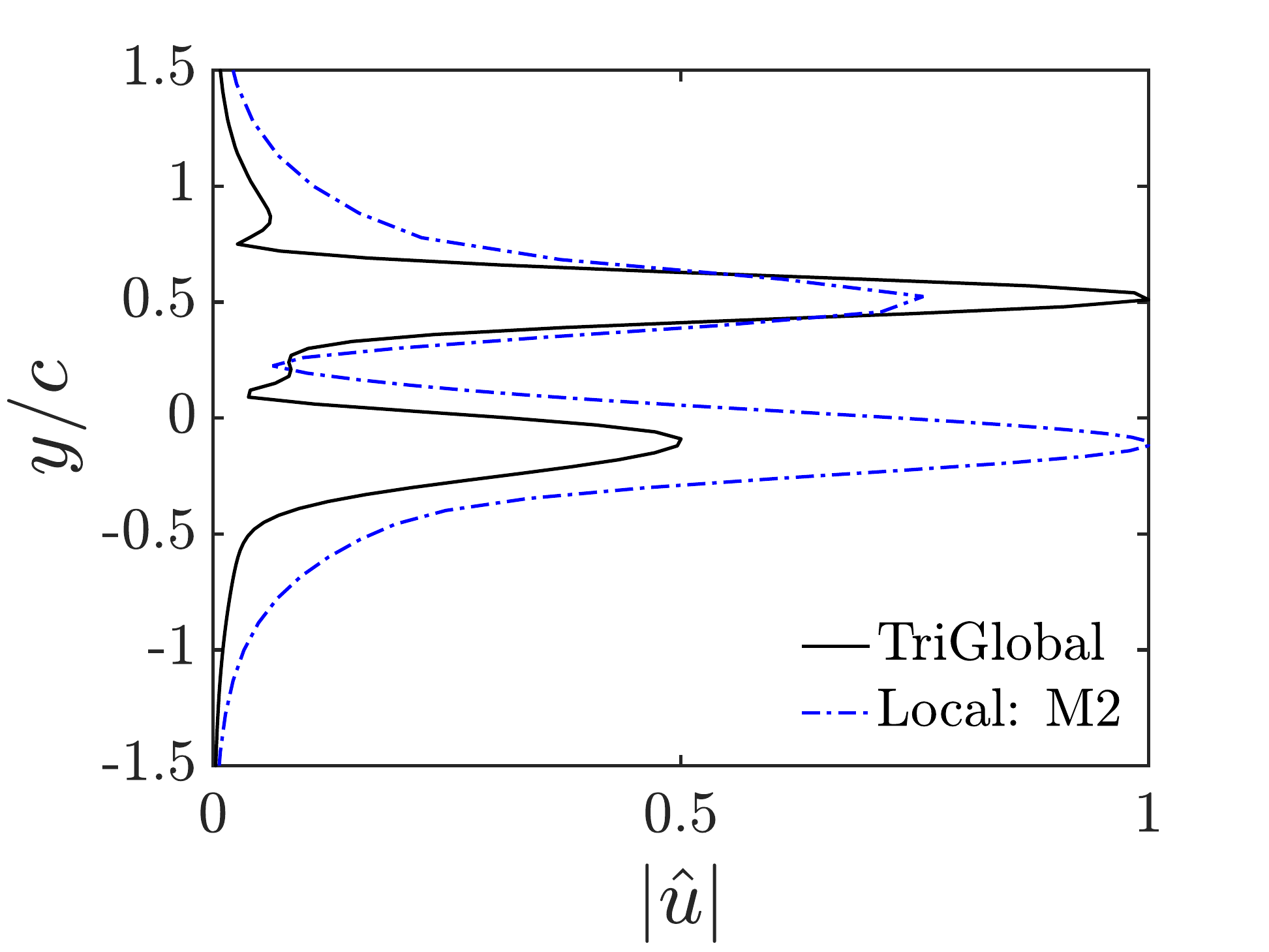}
		\put(-1, 68) {$(b)$}
		\end{overpic}
    \end{subfigure}
        %~ 
    \begin{subfigure}[b]{0.325\textwidth}
        \begin{overpic}[width=1\textwidth]{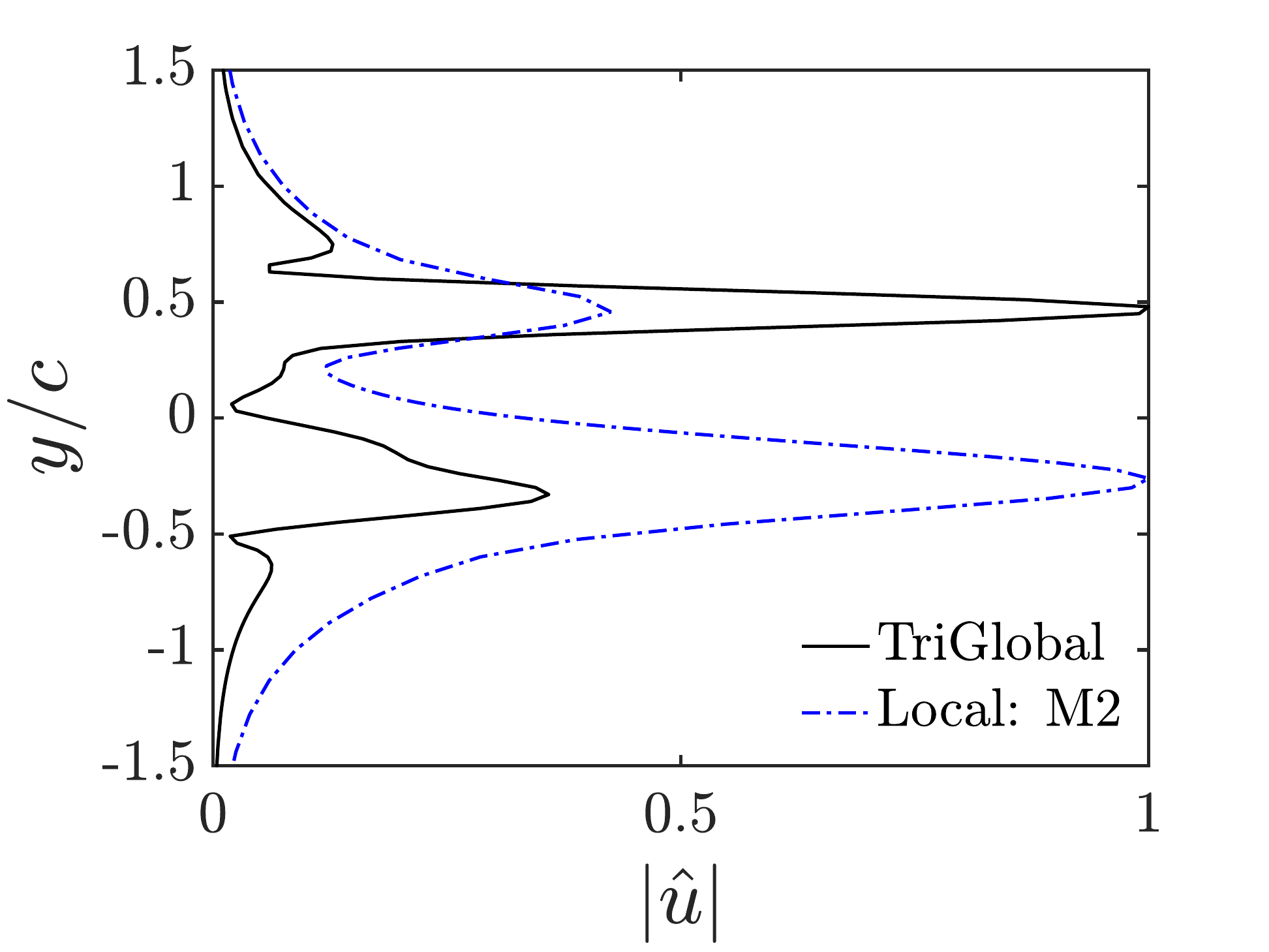}
		\put(-1, 68) {$(c)$}
		\end{overpic}
    \end{subfigure}
    \caption{Comparison of the spatial local analysis eigenfunctions with corresponding slices through the global modes at $\omega$ equal to the frequency of the global mode ($a$) $\Lambda=0^\circ$ $\omega=1.520$, ($b$) $\Lambda=5^\circ$, $\omega=1.658$ ($c$) $\Lambda=10^\circ$ $\omega=1.639$.}
    \label{fig:local-ef}
  \end{center}
\end{figure}

Spatial analysis results in figure \ref{fig:local}($b$) show two unstable modes. For the unswept case mode 1 reaches maximum amplification at $\omega \approx 2$ as its $\alpha_r$ approaches zero while mode 2 peaks at $\omega \approx 1.5$. For $\Lambda=5^\circ$ and $10^\circ$ mode 1 reaches maximum amplification at $\omega \approx 1$ and is overtaken at larger frequencies by mode 2 that peaks around $\omega=1.8$. The eigenfunctions of local spatial analysis are compared to corresponding slices of the global mode in figure \ref{fig:local-ef}. It can be seen that the normalised eigenfunctions of unstable local spatial mode 2 correlate reasonably well with the location of the dominant peaks of the global mode, while the agreement of other local modes with the result of TriGlobal analysis is poorer. Quantitatively, the difference between the maximum amplification rates of local mode 1 at $\Lambda=5^\circ$ and $10^\circ$ and amplification rate of the global mode is within 20\% and 24\% respectively. On the other hand, mode 2 has an order-of-magnitude larger amplification than that of the global mode. The frequency of the global modes lies approximately in between the frequencies corresponding to maximum amplification of the local modes 1 and 2.  

In summary, and somewhat unsurprisingly due to the highly inhomogeneous nature of the base flows, this comparison shows that, despite some qualitative agreement between global and local eigenfunction results, large differences in quantitative predictions of the amplification rates and frequencies exist. These differences underline the inability of local analysis to adequately capture the physics of linear instability in this configuration even for locations close to the symmetry plane and away from the main sources of three-dimensionality.

%=======================================================================
\subsection{Wake modes in the nonlinear saturation regime}
\label{sec:wake_modes}
%=======================================================================

%=======================================================================
\subsubsection{Effect of angle of attack}
\begin{figure}			% 4-0-10 POD and RA			
  \begin{center}  
    \begin{subfigure}[b]{0.32\textwidth}
        \begin{overpic}[width=1\textwidth]{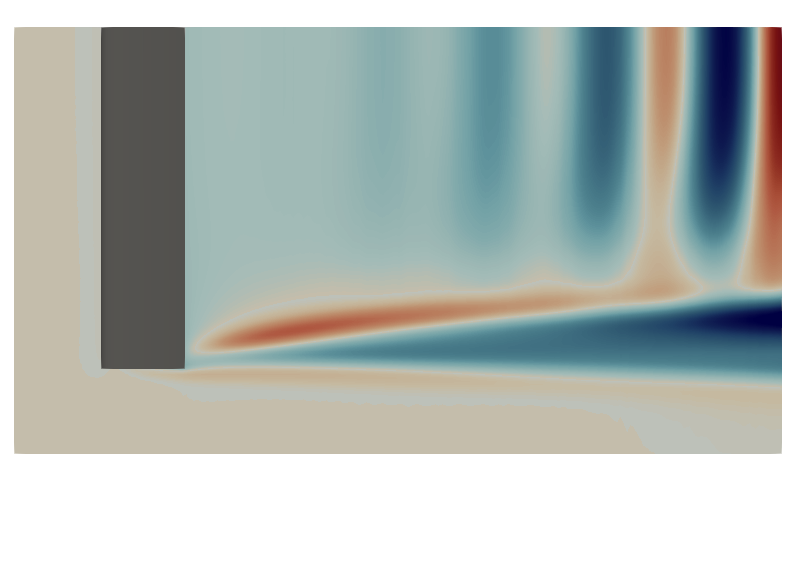}
		\put(1, 61) {$(a)$}
		\end{overpic}
    \end{subfigure}
    \begin{subfigure}[b]{0.32\textwidth}
        \begin{overpic}[width=1\textwidth]{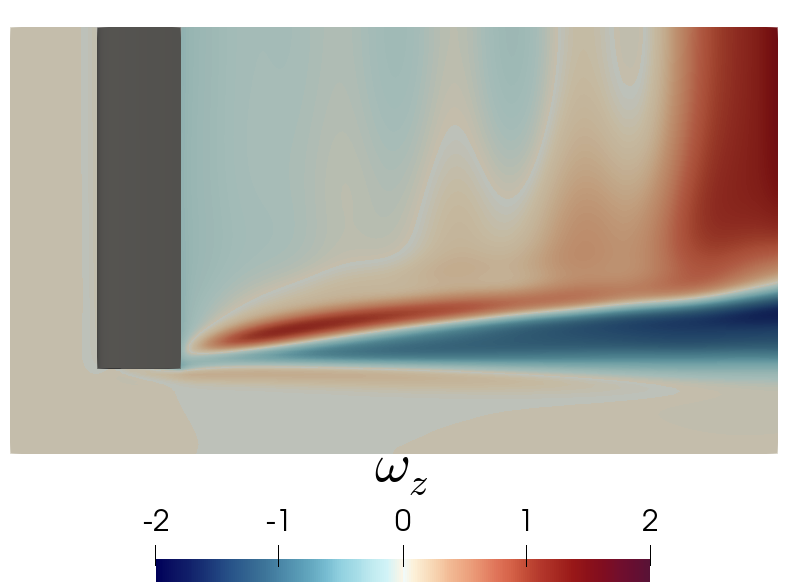}
		\put(1, 61) {$(b)$}
		\end{overpic}
    \end{subfigure}
    \begin{subfigure}[b]{0.32\textwidth}
        \begin{overpic}[width=1\textwidth]{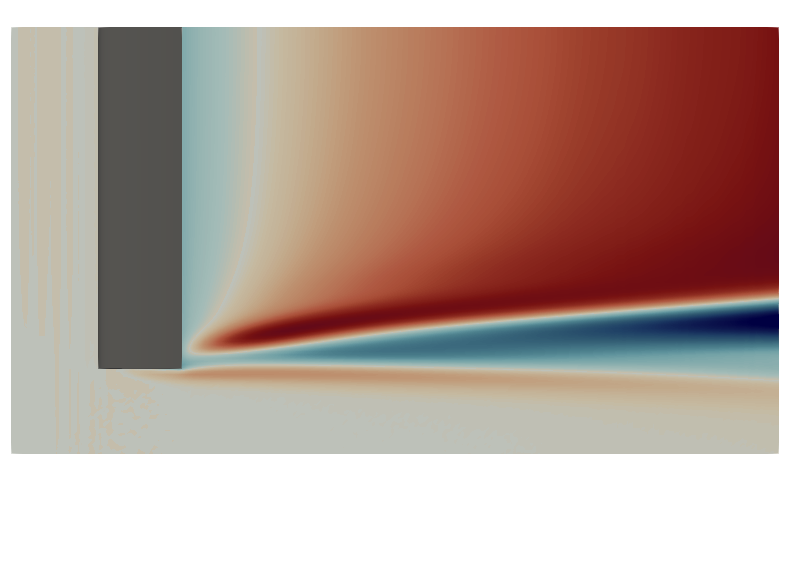}
		\put(1, 61) {$(c)$}
		\end{overpic}
    \end{subfigure}
    \end{center}
    \caption{Modes of $(sAR,\Lambda,\alpha)=(4,0^\circ,10^\circ)$ wing ($a$) global mode obtained with residual algorithm ($b$) most energetic POD mode based on sample $10 \leqslant t \leqslant 40$ ($c$) damped wake instability POD mode based on sample $40 \leqslant t \leqslant 70$. Contours of spanwise vorticity plotted at a $xz$ slice located at the wing trailing edge.}
    \label{fig:4-0-10}
\end{figure}

%=======================================================================
Data-driven analysis was performed in the nonlinear saturation regime, in order to analyse the effects of wing geometry on the wake. Both POD as described in \S \ref{sec:POD} and DMD as described in Appendix \ref{appA}  was carried out for all cases. Before moving to unsteady flow results the stable case of $(sAR,~\Lambda,~\alpha)=(4,~0^\circ,~10^\circ)$ is first considered. Two sets of snapshots were taken: one in the region of a exponential decay, at $10 \leqslant t \leqslant 40$, and one when steady flow has been reached, $40 \leqslant t \leqslant 70$. At this set of parameters the residual algorithm is used and its results, shown in figure~\ref{fig:4-0-10}($a$), are compared to dominant mode delivered by POD, the latter applied in the two aforementioned time-windows. It can be seen that the leading POD mode during exponential decay corresponds to the decaying wake instability/tip-vortex structure shown by the global mode. The damped wake instability recovered in the $\alpha=10^\circ$ case on the $sAR=4$ finite aspect ratio wing is a Kelvin–Helmholtz mode akin to the two-dimensional mode reported by \citet{HeJFM} in their analysis of spanwise homogeneous base flow.

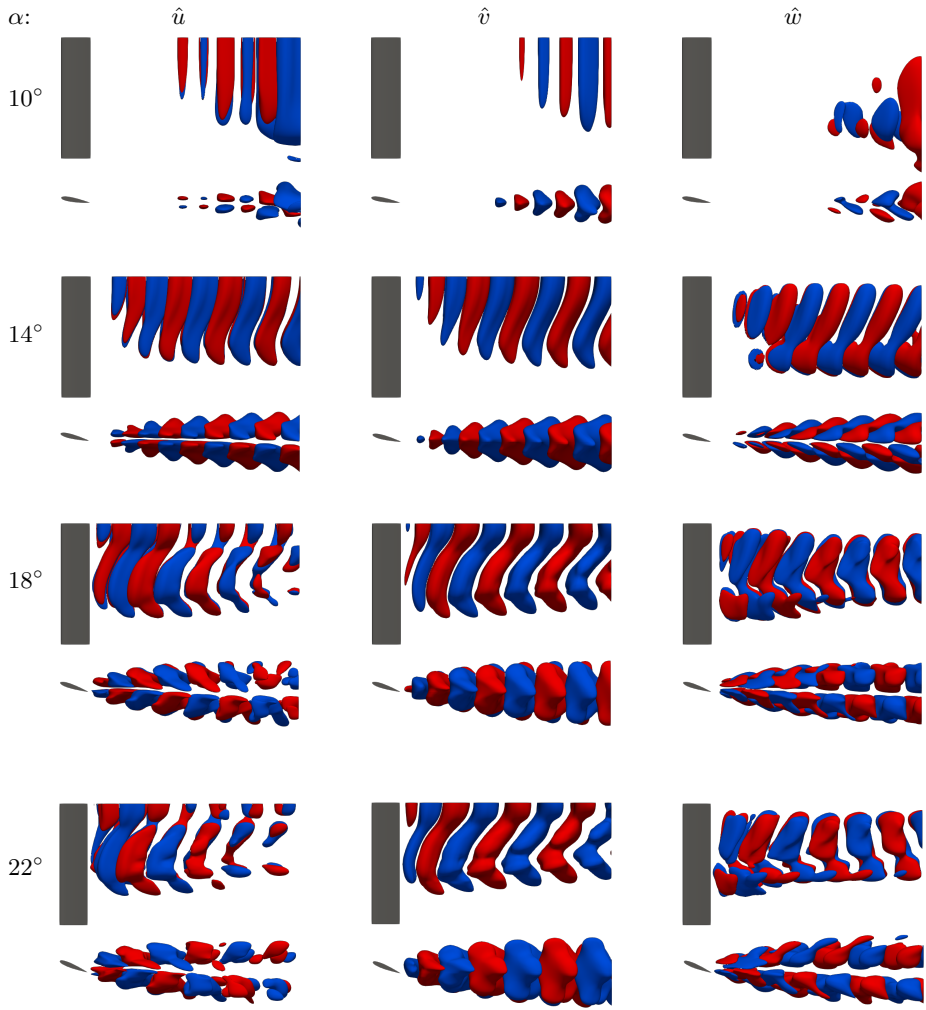
\begin{figure}		% AoA: top/side of dominant mode fig:ar4-aoa-ev1
\vskip 2em 
\begin{center}
        \begin{overpic}[width=0.95\textwidth]{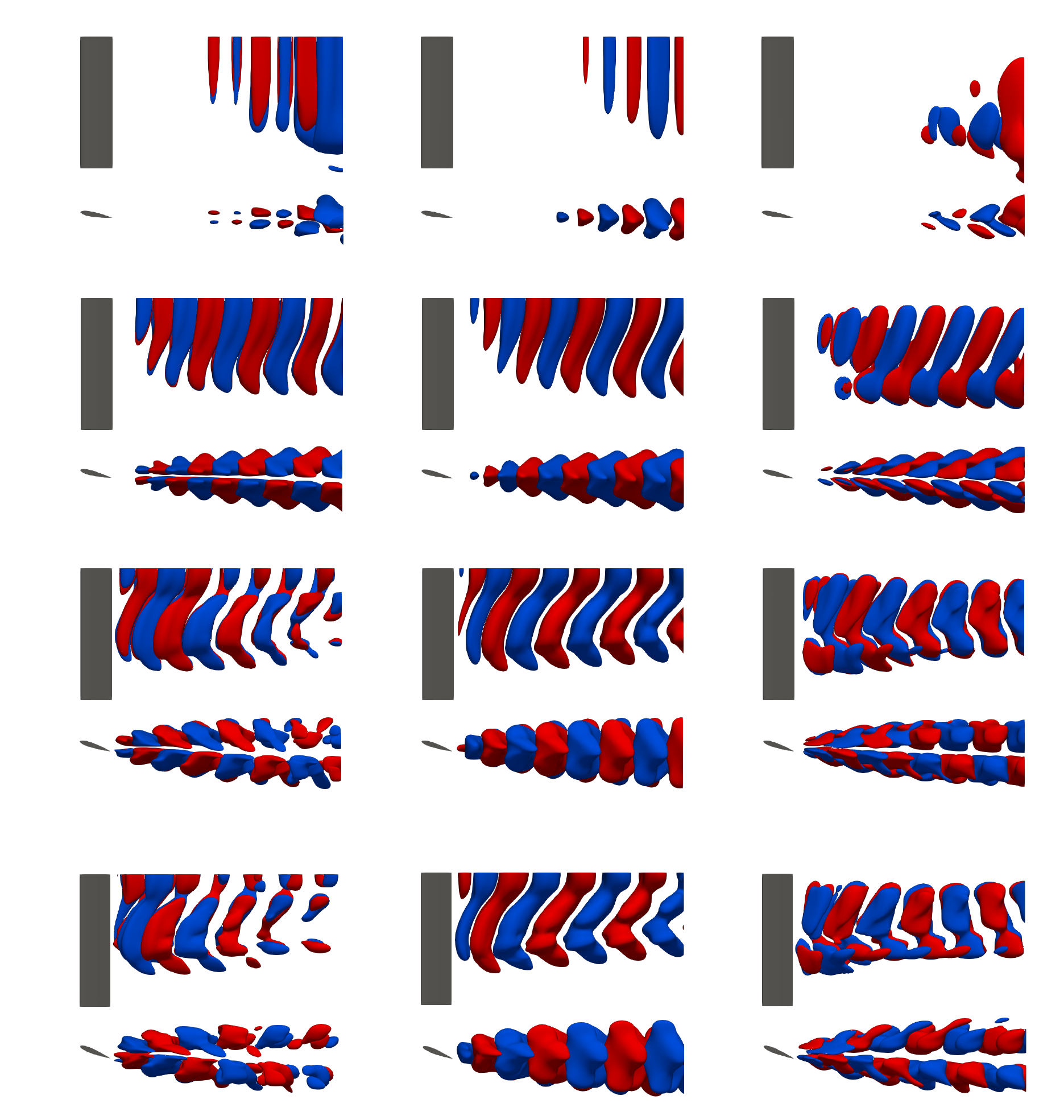}
        \put(2, 98) {$\alpha$:}
 		\put(18, 98) {$\hat{u}$}
 		\put(48, 98) {$\hat{v}$}
 		\put(78, 98) {$\hat{w}$}
 		\put(2, 90) {$10^\circ$}
 		\put(2, 67) {$14^\circ$}
 		\put(2, 43) {$18^\circ$}
 		\put(2, 15) {$22^\circ$}
		\end{overpic}
\end{center}
\vskip -1em
    \caption{Evolution of the leading mode for $sAR=4$ $\Lambda=0^\circ$ wing with angle of attack for $\alpha=10^\circ$,  $14^\circ$, $18^\circ$ and $22^\circ$ visualised with isosurfaces of $\hat{u}$ in $\hat{v}$ and $\hat{w}$ at $\pm0.2$ shown in top and side view.}
    \label{fig:ar4-aoa-ev1}
\end{figure}

The spatial structure of the dominant mode as the angle of attack is increased, is shown in figure \ref{fig:ar4-aoa-ev1}. The results correspond to modes obtained by POD analysis, since DMD gives practically identical, in both frequencies, shown in table \ref{tab:ar4-aoa}, and spatial structure of the modes. Due to this POD modes will be shown throughout the figures, unless otherwise stated. At low angles of attack the leading mode at these conditions represents a Kelvin-Helmholtz instability and takes the form of periodic spanwise structures in the wake. At $\alpha=10^\circ$ these structures are parallel to the trailing edge of the wing, while as the angle of attack increases they are deformed, following the curvature of the vortex cores in the DNS results now meeting at an oblique angle to the symmetry plane. It can be seen that at $\alpha=14^\circ$ the mode still resembles Kelvin–Helmholtz instability, while at higher angles of attack the structure is increasingly more complex, as seen in the side views of $\hat{u},\hat{v},\hat{w}$, shown in figure \ref{fig:ar4-aoa-ev1}. This increased three-dimensionality and the peaks in $\hat{u},\hat{v}$ developing at about $1/4b$ indicate the significant effect of the interaction region vortices on the mode. We refer to this structure as {\em the interaction mode}. This mode contains up to 47\% of the flow kinetic energy as shown in table \ref{tab:ar4-aoa}. It should be noted that POD modes are grouped in pairs, with nearly identical frequencies and similar energies, the table shows only the first mode of the pair, similarly when classifying modes only the first member of the pair is shown. When combined, the two modes of the POD pair would actually contain twice the $E_k$ quoted in this table. 

\begin{figure}		% Effect of AoA: 3D Q of 3 modes fig:ar4-aoa
%\vskip -1em 
\begin{center}
        \begin{overpic}[width=0.9\textwidth]{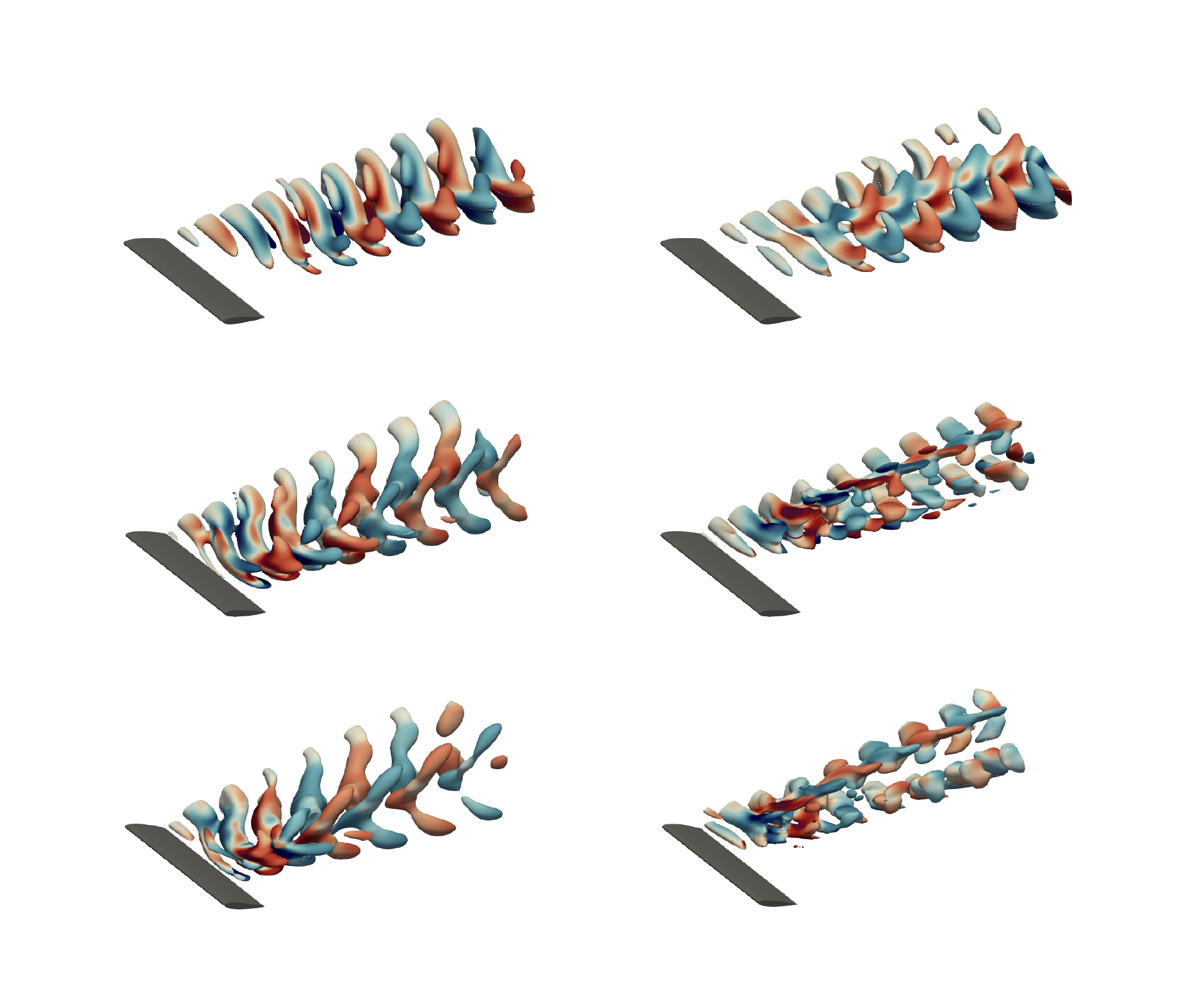}
         \put(25, 78) {IM}
         \put(70, 78) {WM}
         \put(10, 72) {$St=0.122$}
         \put(58, 72) {$St=0.116$}
 		\put(0, 61) {$\alpha=14^\circ$}
 		\put(10, 48) {$St=0.136$}
        \put(58, 48) {$St=0.181$}
 		\put(0, 37) {$\alpha=18^\circ$}
 		\put(10, 23) {$St=0.140$}
        \put(58, 23) {$St=0.160$}
 		\put(0, 12) {$\alpha=22^\circ$}
		\end{overpic}
\end{center}
\vskip -1em
    \caption{Modes of $\Lambda=0^\circ$, $sAR = 4$ wings visualised with contours of $Q=1$ and coloured by streamwise vorticity for $\alpha = 14^\circ$, $18^\circ$, and $22^\circ$, showing the interaction mode (IM) and the wake mode  (WM).}
    \label{fig:ar4-aoa}
\end{figure}

The dependence of the three most dominant modes at $sAR=4$ on the angle of attack is visualised in figure~\ref{fig:ar4-aoa}, using contours of 
$Q$-criterion computed from the normalised perturbation components. The interaction mode (IM) shown at $\alpha=14^\circ, 18^\circ$ and $22^\circ$ in figure \ref{fig:ar4-aoa}, clearly shows spanwise vortical structures in the wake. The pattern of these IM structures near the wing from $\alpha \geqslant 18^\circ$ shows a clear connection to the curvature of vortex core lines shown in figure \ref{fig:vortexcores}. It is also worth mentioning that the frequency of the IM is in close agreement with the dominant frequencies of the reattachment line motion discussed in \S \ref{sec:bf}. 
The structures visible closer to the tip at $2 \leqslant z \leqslant 2.5$ reflect the braid-like vortices of the interaction region of the flow. The increasing strength of the mode structures in this region mimics the growth of the braid-like vortices with increasing $\alpha$ in the DNS results. Higher harmonics of the dominant IM also appear in the spectrum with the most energetic one labelled IM2 shown in table \ref{tab:ar4-aoa}.
The wake mode (WM) appears to be concentrated in the wake close to the symmetry plane, as also seen in \ref{fig:ar4-aoa}. It has a lower frequency compared to that of the IM and shows the least contribution from the braid-like structures of the interaction region. Also, unlike the IM and its harmonic IM2, the structure of the WM mode does not follow the vortex line curvature. The frequencies of IM and WM at $\alpha=22^\circ$ as shown in table \ref{tab:ar4-aoa} match very well with the two distinct frequency peaks of the lift coefficient signal observed by \citet{zhang_2020} at same conditions on $sAR=4$ wing.

\begin{figure}		% POD DMD summary with AoA
    \begin{subfigure}[b]{0.49\textwidth}
        \begin{overpic}[width=1\textwidth]{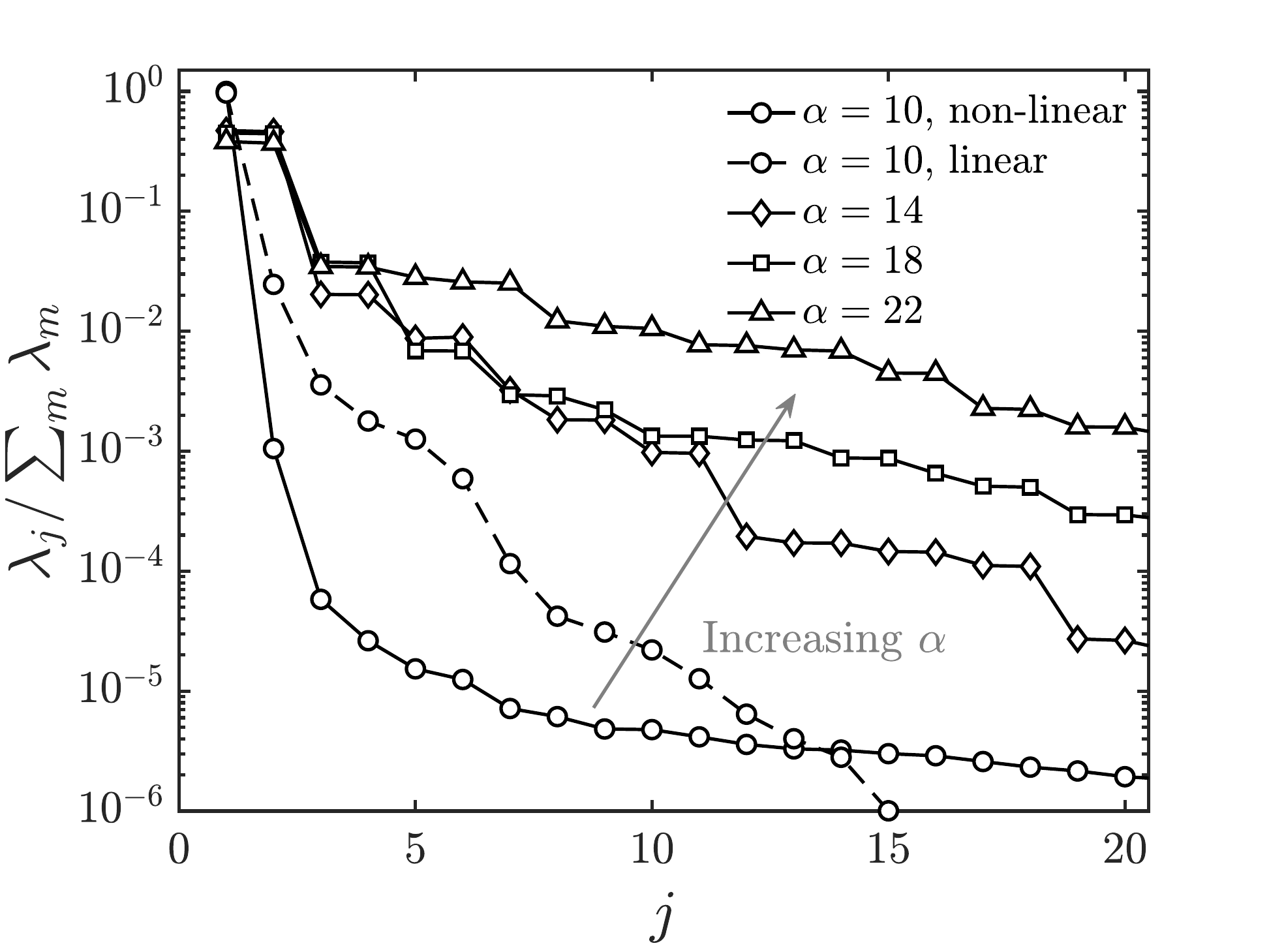}
		\put(-1, 68) {$(a)$}
		\end{overpic}
    \end{subfigure}
    ~ %
    \begin{subfigure}[b]{0.49\textwidth}
    		\begin{overpic}[width=1\textwidth]{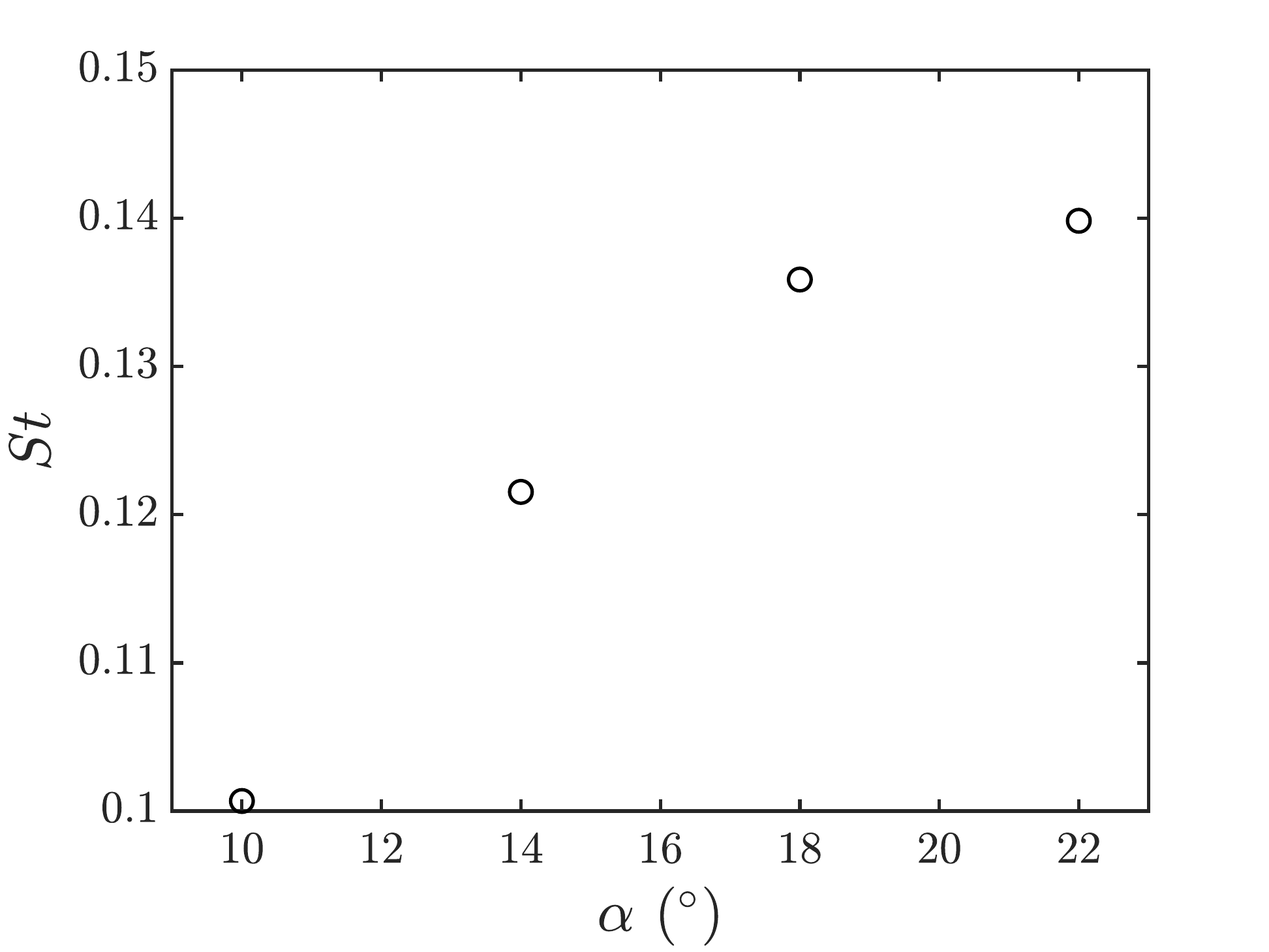}
		\put(-1, 68) {$(b)$}
		\end{overpic}
    \end{subfigure}
    \caption{Effects of angle of attack: $(a)$ spectrum of POD and $(b)$ variation of the interaction mode frequency for unswept $sAR=4$ wing.
    }
    \label{fig:aoa-summary}
\end{figure}

Finally, the dependence of POD eigenvalues on angle of attack is plotted in figure \ref{fig:aoa-summary}($a$). At $\alpha=10^\circ$ the results for analysis in both linear and nonlinear regimes show that the first few modes contain the majority of flow energy. The ratio $\lambda_j/\sum_m \lambda_m$ between the mode eigenvalue and the sum of eigenvalues decreases for a given mode with increasing $\alpha$ meaning that higher order modes become more important as their relative energy contribution increases. The Strouhal number of the dominant mode increases with angle of attack as shown in figure \ref{fig:aoa-summary}($b$) indicating an increase of dominant mode frequency.

\begin{table}{}		% AR=4 AoA
 \begin{center}
\begin{tabular}{ c c @{\hskip 8mm} c c c c}
%\hline
 & $\alpha$ & $10^\circ$ & $14^\circ$ & $18^\circ$ & $22^\circ$\\[1ex]
\multirow{4}{*}{K-H (IM)} & $St_{DMD}$ & 0.1011 & 0.1217 & 0.1358 & 0.1395\\
	& $g_j$ & -0.3962 & 0.0001 & 0.0000 & -0.0003\\
	& $St_{POD}$ & 0.1007 & 0.1215 & 0.1359 & 0.1398\\
	&  $E_k$ & 0.1785 & 47.0781 & 44.9946 & 38.1102\\[1ex]
\multirow{4}{*}{IM2} & $St_{DMD}$ & - & 0.2434  & 0.2716 & 0.2789\\
	& $g_j$ & - & 0.0099  & 0.0001 & -0.0011\\
	& $St_{POD}$ & - & 0.2433 & 0.2716 & 0.2794\\
	&  $E_k$ & - & 2.0291 & 3.7739 & 3.4648\\[1ex]
\multirow{4}{*}{WM} & $St_{DMD}$ & - & 0.1151 & 0.1489 & 0.1600\\
	& $g_j$ & - & -0.0194 & -0.0043 & -0.0093\\
	& $St_{POD}$ & - & 0.1159 & 0.1484 & 0.1608\\
	&  $E_k$ & - & 0.8722 & 0.2957 & 2.5794\\[1ex]	
%\multirow{4}{*}{4-V} & $St_{DMD}$ & \\
%	& $g_j$ &\\
%	& $St_{POD}$ & \\
%	&  $E_k$ & \\[1ex]	
  \end{tabular}
  \caption{Change of POD and DMD modes with angle of attack for $sAR = 4$ straight wing. Strouhal numbers according to POD and DMD, growth rate ($g_j$), and percentage of kinetic energy ($E_k$).}{\label{tab:ar4-aoa}}
 \end{center}
\end{table}

%=======================================================================
\subsubsection{Effect of sweep}
%=======================================================================

\begin{table}{}		% AR=4 sweep
 \begin{center}
\begin{tabular}{ c c @{\hskip 8mm} c c c c c c c}
%\hline
 & $\Lambda$ & $0^\circ$ & $5^\circ$ & $10^\circ$ & $15^\circ$ & $20^\circ$ & $25^\circ$ & $30^\circ$\\[1ex]
\multirow{2}{*}{IM} & $St_{POD}$ & 0.1398 & 0.1397 & 0.1428 & 0.1390 & 0.1225 & 0.1096 & 0.1048\\
	&  $E_k$ & 38.1102 & 26.6632 & 40.4665 & 43.6342 & 40.5721 & 43.6234 & 22.1517\\[1ex]
\multirow{2}{*}{IM2} & $St_{POD}$ & 0.2794 & 0.2785 & 0.2855 & 0.2779 & 0.2453 & 0.2190 & 0.2091\\
	&  $E_k$ & 3.4648 & 1.4835 & 3.7203 & 5.0864 & 4.6365 & 3.0208 & 2.1202\\[1ex]
\multirow{2}{*}{WM} & $St_{POD}$ & 0.1608 & 0.1603 & 0.1872 & - & - & - & -\\
	&  $E_k$ & 2.5794 & 12.8017 & 1.4995 & - & - & - & -\\[1ex]	
\multirow{2}{*}{SM} & $St_{POD}$ & - & - & - & - & 0.00884 & 0.0084 & 0.0085\\
	&  $E_k$ & - & - & - & - & 2.8337 & 3.2341 & 39.8751\\[1ex]	
  \end{tabular}
  \caption{Change of POD modes with sweep angle for $sAR = 4$ wing showing Strouhal numbers according to POD coefficients and percentage of kinetic energy.}{\label{tab:ar4-sweep}}
 \end{center}
\end{table}

\begin{figure}		% AR=4, low sweep figure fig:ar4-pod-low-sweep
%\vskip -1em 
\begin{center}
        \begin{overpic}[width=0.9\textwidth]{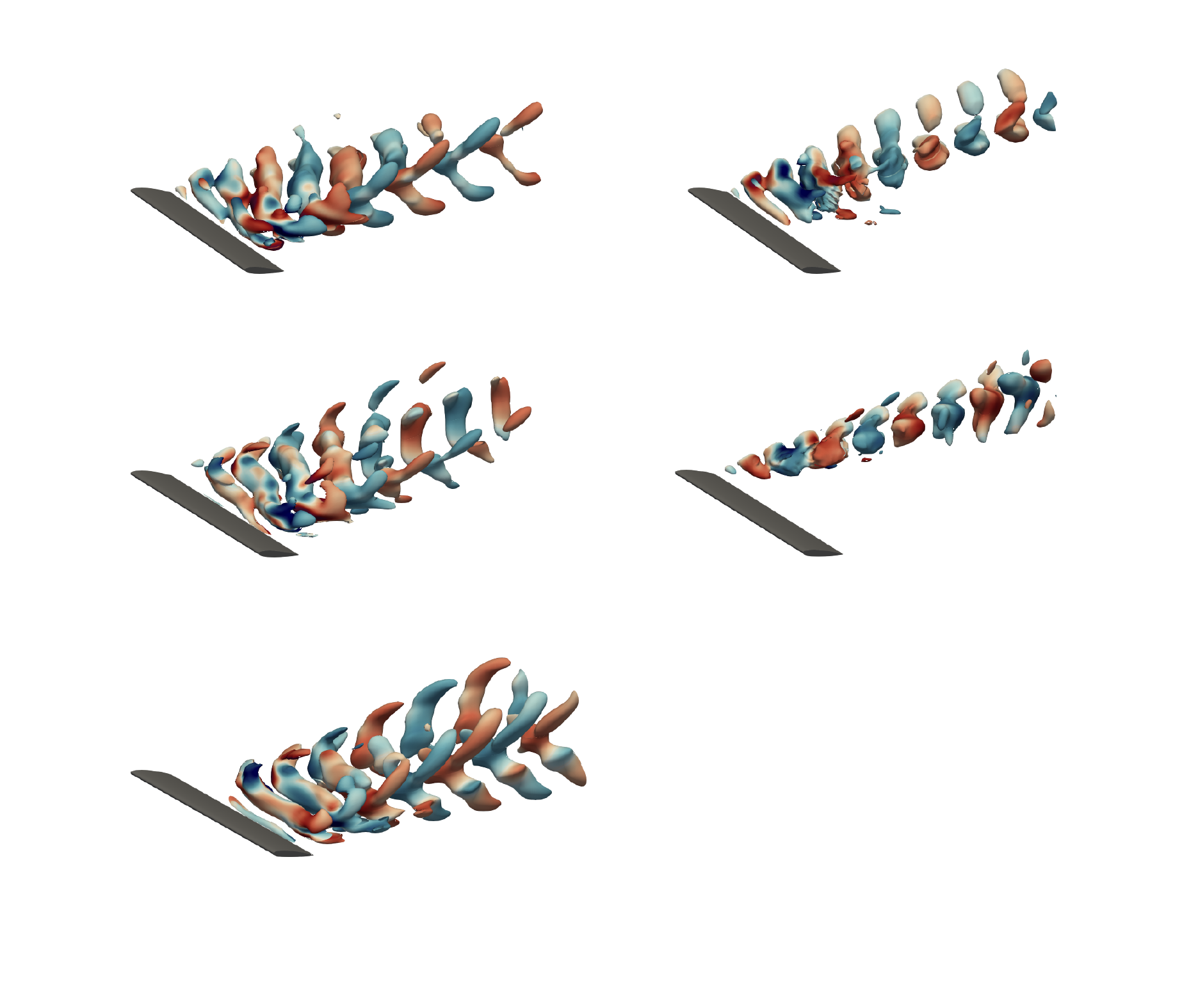}
         \put(25, 78) {IM}
         \put(70, 78) {WM}
         \put(10, 75) {$St=0.140$}
         \put(57, 75) {$St=0.160$}
 		\put(0, 65) {$\Lambda=5^\circ$}
 		\put(10, 50) {$St=0.143$}
        \put(57, 50) {$St=0.186$}
 		\put(0, 42) {$\Lambda=10^\circ$}
 		\put(10, 24) {$St=0.139$}
 		\put(0, 16) {$\Lambda=15^\circ$}
		\end{overpic}
\end{center}
\vskip -2em
    \caption{Modes of $\Lambda=0^\circ$, $sAR = 4$ wings visualised with contours of $Q=1$ and coloured by streamwise vorticity for $\alpha = 14^\circ$, $18^\circ$, and $22^\circ$, showing the interaction mode (IM) and the wake mode (WM).}
    \label{fig:ar4-pod-low-sweep}
\end{figure}
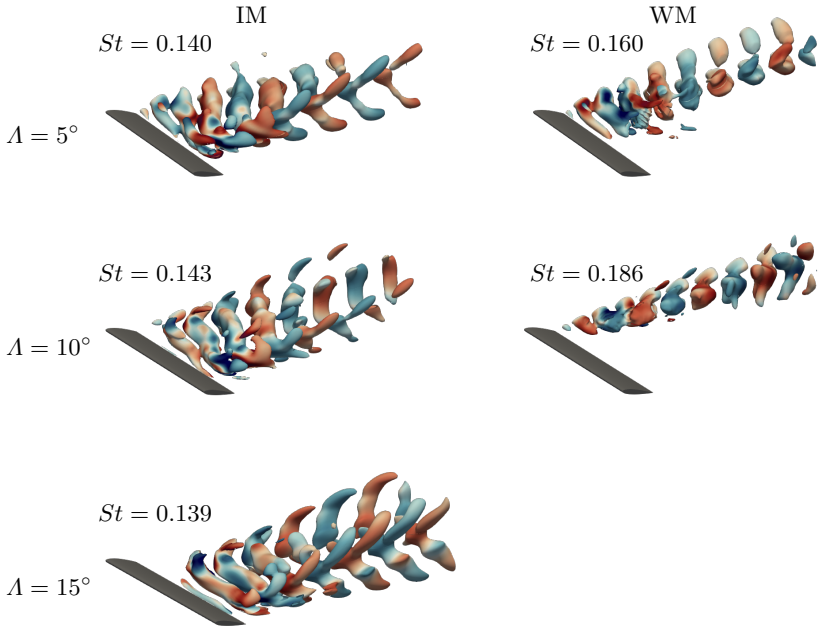

Turning our attention to the effect of angle of sweep, the effect of this parameter on the three dominant modes, the interaction mode (IM), its second harmonic (IM2) and the wake mode (WM) identified in the previous subsection, is monitored at a fixed angle of attack $\alpha=22^\circ$. Table \ref{tab:ar4-sweep} shows the Strouhal numbers obtained from POD coefficients and the percentage of kinetic energy. The dominant interaction mode, having $St=0.122$ at $\alpha=14^\circ$, can be seen in figure \ref{fig:ar4-pod-low-sweep}, its structure reflecting the changes in the corresponding flow fields as the angle of sweep increases. At $\Lambda=10^\circ$ the vortices of the IM no longer pass through the symmetry plane forming two structures on either side of the symmetric wing. 

The wake mode (WM), having $St \approx 0.18$, is shown in figure \ref{fig:ar4-pod-low-sweep}. In a manner analogous to the unswept wing, the WM mode exhibits periodic structures that are strongest in the wake. At $\Lambda=5^\circ$ it has the highest percentage of kinetic energy and overtakes IM2 as the second most energetic mode. At $\Lambda=10^\circ$, WM structures seem to shrink and appear only very close to the symmetry plane. This is the highest sweep angle at which this mode has been detected.

At the sweep angle of $\Lambda=15^\circ$ there is a change in the wake from shedding centred on the symmetry plane to shedding of the braid-like vortices closer to the tip vortex. This is reflected in the interaction mode shown in figure \ref{fig:ar4-pod-low-sweep}, as the mode structures are no longer passing through the symmetry plane. The periodic structures originating from the outboard side of the wing are terminated on either end by spiral vortical structures that correspond to the location of interaction region vortices in the DNS results. Further downstream the periodic structures break down, leaving only the spiral vortices. The frequency spectrum at this angle of sweep is dominated by harmonics of the leading mode, similar to the frequencies of $sAR=2$ modes as will be seen in \S\ref{sec:AReffect}.

\begin{figure}		% AR=4, high sweep figure fig:ar4-pod-high-sweep
% \vskip -1em 
\begin{center}
        \begin{overpic}[width=0.9\textwidth]{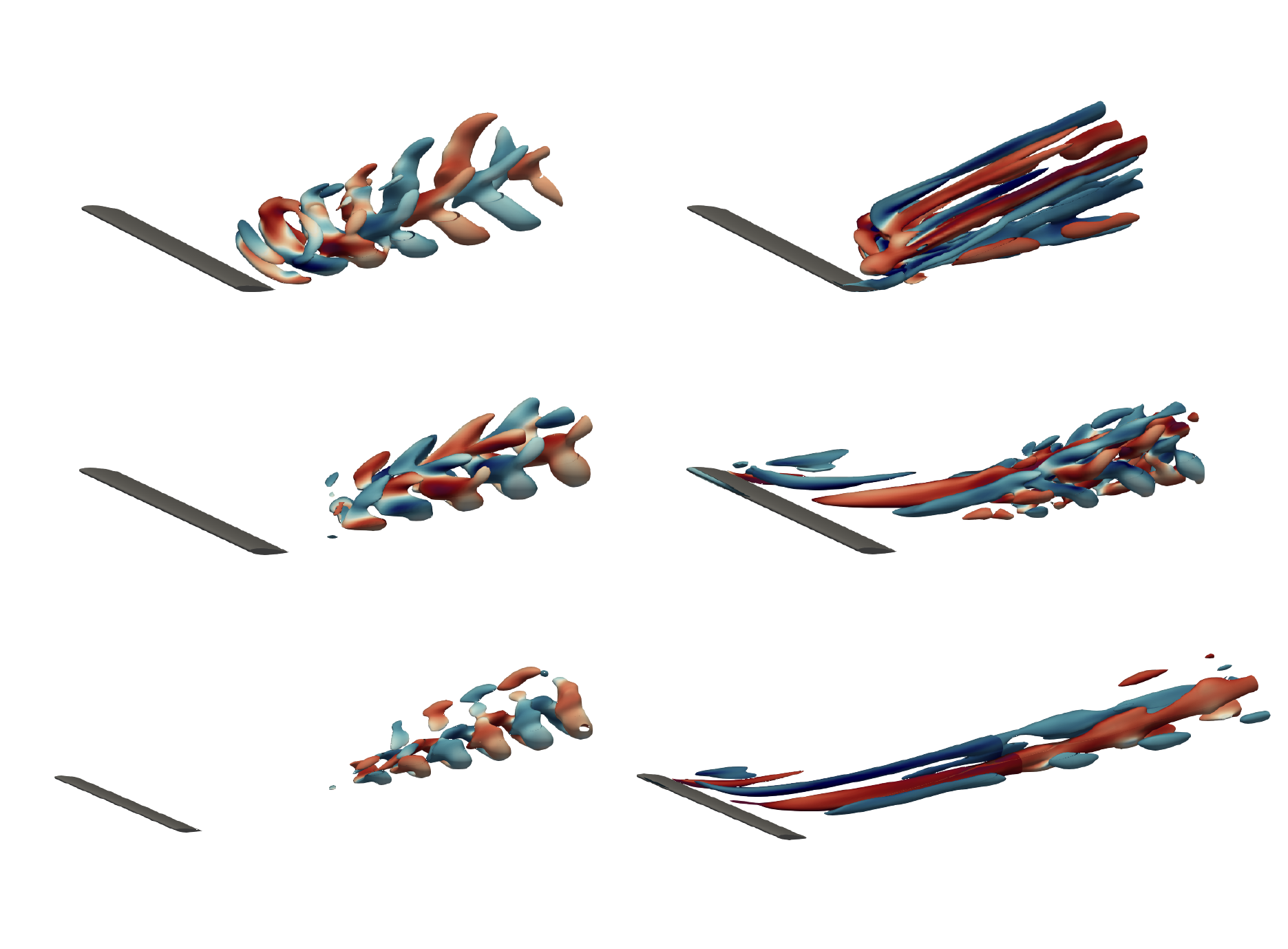}
         \put(20, 68) {IM}
         \put(70, 68) {SM}
         \put(8, 62) {$St=0.123$}
         \put(55, 62) {$St=0.0084$}
  		\put(-3, 53) {$\Lambda=20^\circ$}
 		\put(8, 42) {$St=0.110$}
        \put(55, 42) {$St=0.0084$}
 		\put(-3, 32) {$\Lambda=25^\circ$}
 		\put(8, 19) {$St=0.105$}
 		\put(55, 19) {$St=0.0085$}
 		\put(-3, 8) {$\Lambda=30^\circ$}
		\end{overpic}
\end{center}
\vskip -2em
    \caption{Modes of high $\Lambda$, $sAR=4, \alpha=22^\circ$ wings visualised with contours of $Q = 1$ for $\Lambda = 20^\circ$, $25^\circ$ and $30^\circ$, showing interaction mode (IM) and wake mode WM. $\Lambda=30^\circ$ results are not to scale with the rest of the figure.}
    \label{fig:ar4-pod-high-sweep}
\end{figure}

Dominant modes at higher sweep angles are shown in figure \ref{fig:ar4-pod-high-sweep}. At $\Lambda=20^\circ$ the spatial structure of the dominant interaction mode (IM) is shifted closer to the wing tip, but is still qualitatively similar to the $\Lambda=15^\circ$ case. In contrast, the mode at the highest sweep of $\Lambda=30^\circ$ is fundamentally different and is dominated by tip effects. Note that results for the highest sweep angle are not to scale with the rest. This is because the streamwise extent of the domain used for modal analysis for this case was doubled in order to capture the mode structure resulting from the elongated vortices in the wake; these structures start to shed much further from the wing. A new low-frequency mode with $St \approx 0.008$ appears in the POD spectrum starting at $\Lambda=20^\circ$ and rapidly gains energy from $2.83\%$ at this angle of sweep to $39.86\%$ at $\Lambda=30^\circ$, where it overtakes IM as the mode containing the highest energy. Since the structure of this mode consists of streamwise vortices emanating from near the wing tip, as shown in \ref{fig:ar4-pod-high-sweep}, it is referred to as {\em the streamwise mode} (SM). A pair of elongated structures propagating from the wing downstream visible at $\Lambda=30^\circ$ are associated with the counter rotating streamwise vortices observed in both the instantaneous and the time averaged DNS results. It should be mentioned that it has not been possible to recover these low-frequency modes by DMD at $\Lambda=20^\circ, 30^\circ$; DMD predicts the dominant mode at the highest sweep to be the interaction mode. 

\begin{figure}		% POD frequency waterfall
    \begin{subfigure}[b]{0.49\textwidth}
        \begin{overpic}[width=1\textwidth]{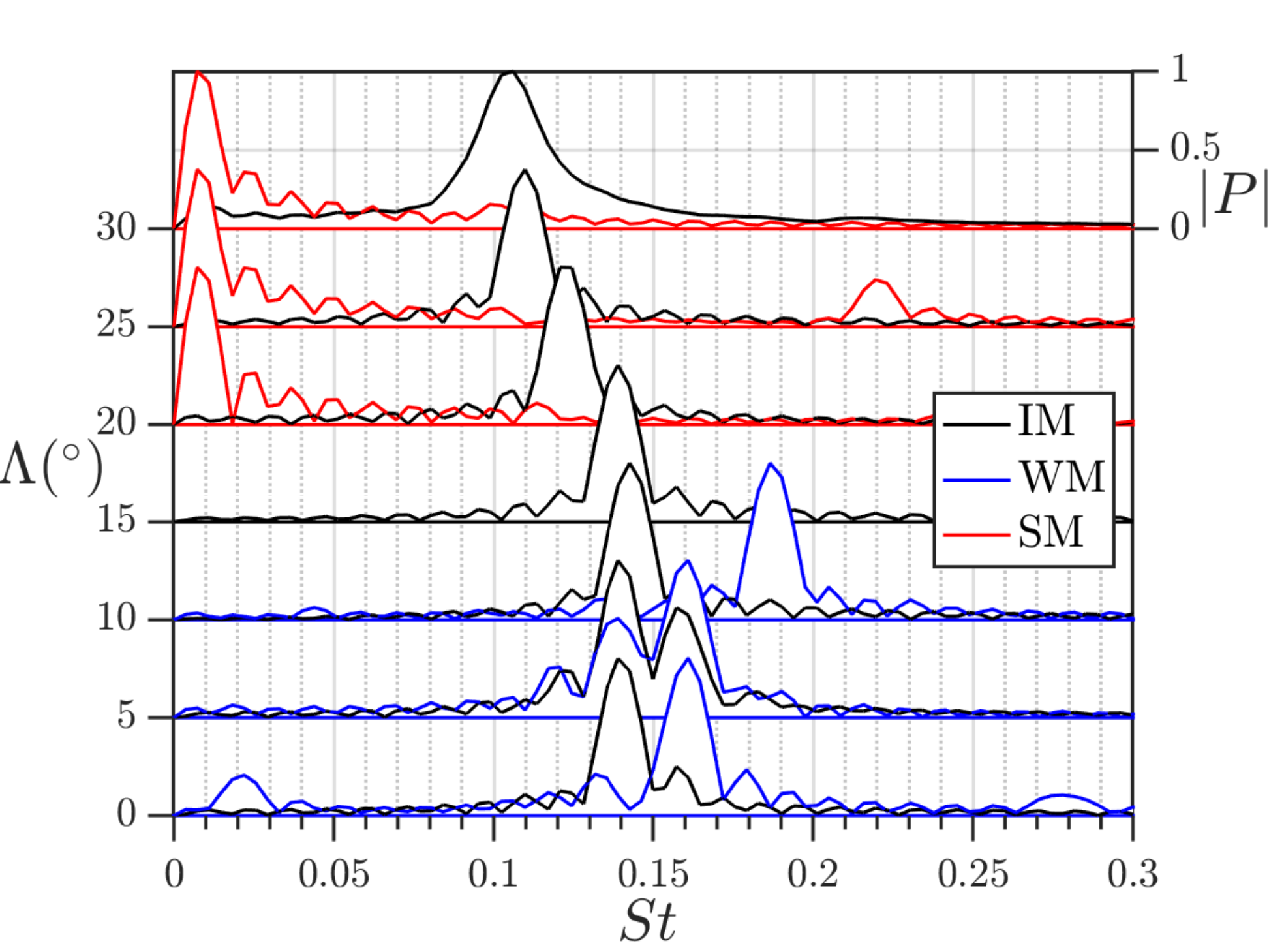}
		\put(-1, 68) {$(a)$}
		\end{overpic}
    \end{subfigure}
    ~ %
    \begin{subfigure}[b]{0.49\textwidth}
    		\begin{overpic}[width=1\textwidth]{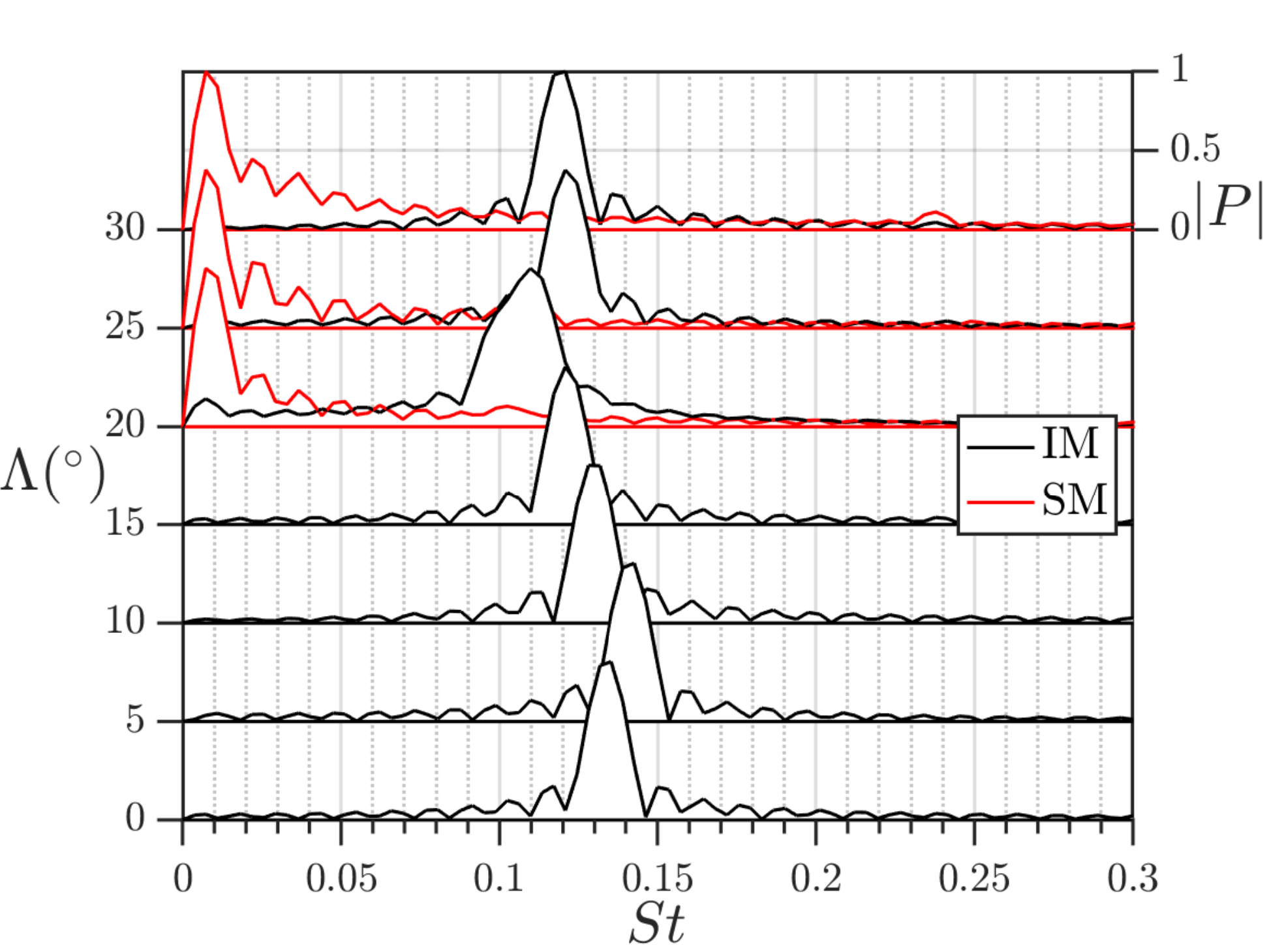}
		\put(-1, 68) {$(b)$}
		\end{overpic}
    \end{subfigure}
    \caption{The frequency spectra of POD coefficients corresponding to dominant modes with sweep ($a$) for $sAR=4$ and ($b$) for $sAR=2$. The spectrum for each sweep angle is normalised by its peak and harmonics of the dominant IM are not plotted.}
    \label{fig:pod_frequency}
\end{figure}

The dependence of the power spectral density of the time coefficients of the dominant POD modes with wing sweep is plotted in figure \ref{fig:pod_frequency}. The interaction mode (IM), wake mode (WM) and the streamwise mode (SM) are shown, but the harmonics of IM are not included for clarity. Although POD modes can in principle contain multiple frequencies, in most of the cases examined modes have been found to correspond to a distinct single frequency, which explains the good overall agreement with DMD results. However, the wake mode on the $sAR=4$ wing shows strong secondary peak in power spectral density at the dominant frequency of the IM at sweep angle of $5^\circ$. Similarly, the IM shows a secondary peak at the dominant frequency of WM at these conditions as can be seen in figure \ref{fig:pod_frequency}$(a)$. The spatial structures of IM recovered by DMD are virtually identical to that of POD while the WM structures are slightly different, showing less influence from the interaction region. This is to be expected as by definition DMD modes contain only one frequency and show the proper wake mode, whereas POD results were contaminated by the interaction mode. Hence the structures WM shown in figures \ref{fig:ar4-aoa} and \ref{fig:ar4-pod-low-sweep} are that produced by DMD. The power spectral density of SM based on the POD time coefficients for different sweeps is shown in figure \ref{fig:pod_frequency}($a$). This mode has predominantly low frequency content ($St \approx 0.01$) with the exception of the $sAR=4$ wing at sweep of $25^\circ$ where a peak at $St \approx 0.22$ is visible.

%=======================================================================
\subsubsection{Effect of aspect ratio}
\label{sec:AReffect}
\begin{figure}		% 2-10-22 POD and Lissajous
    \begin{center}
        \begin{overpic}[width=0.6\textwidth]{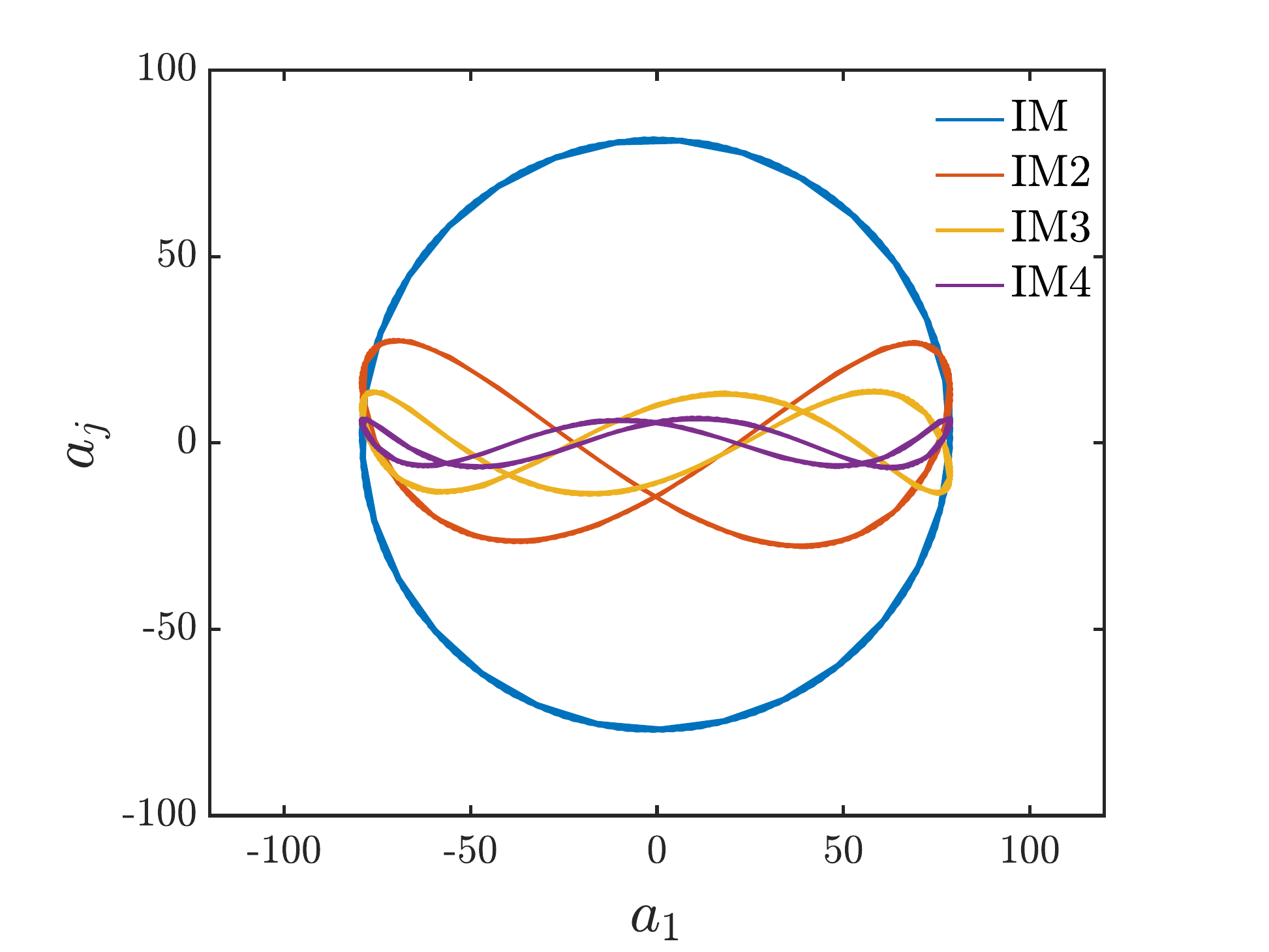}
		\put(0, 70) {$(b)$}
		\end{overpic}
    \end{center}
    \caption{Lissajous curves of higer harmonics of the dominant POD mode for $(sAR,\Lambda,\alpha) = (2,10^\circ,22^\circ)$.}
    \label{fig:pod_lissajous}
\end{figure}

\begin{figure}		% AR=2, low sweep figure fig:ar2-low-sweep
% \vskip -1em 
\begin{center}
        \begin{overpic}[width=0.9\textwidth]{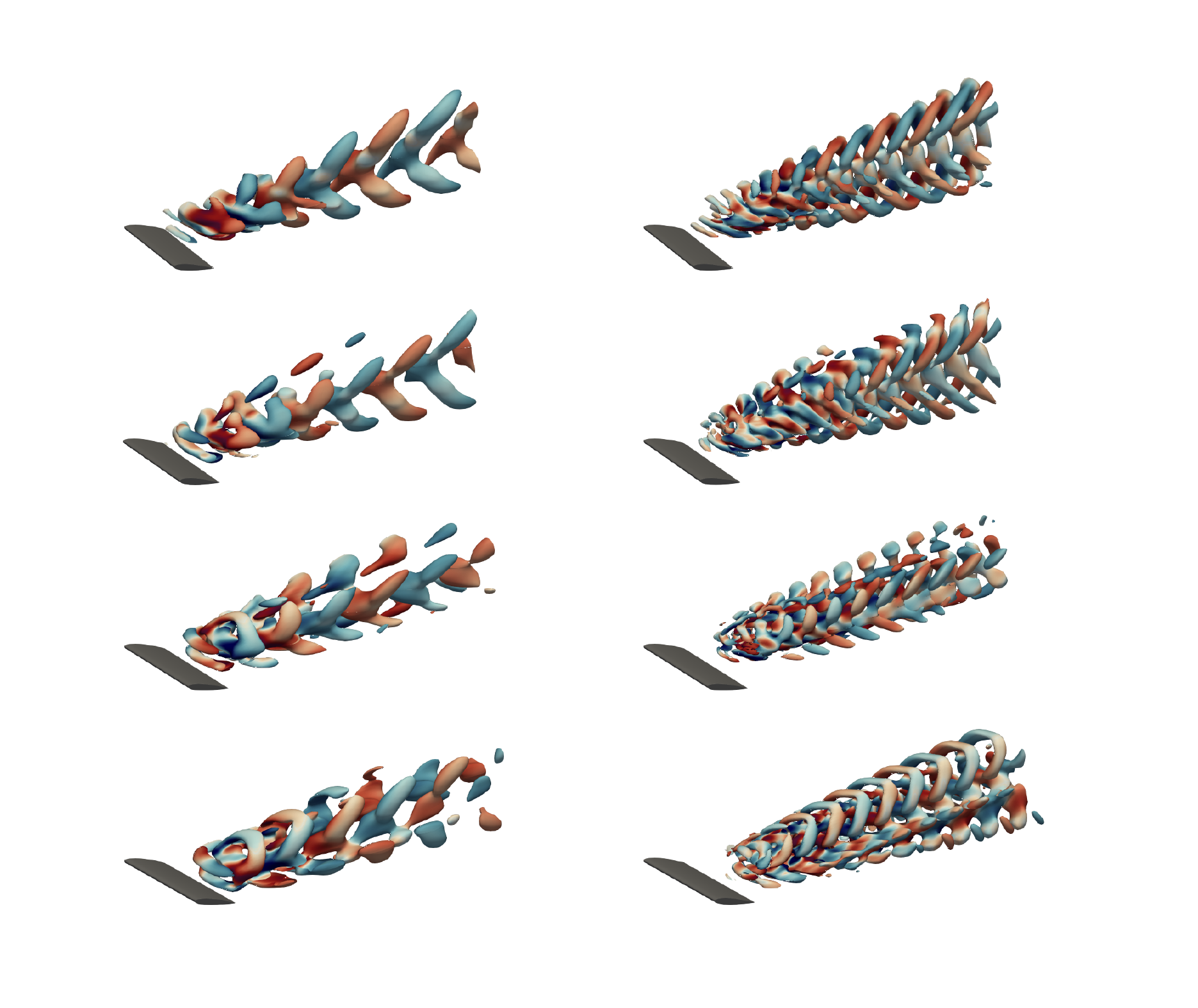}
         \put(23, 78) {IM}
         \put(68, 78) {IM2}
         \put(10, 73) {$St=0.134$}
         \put(52, 73) {$St=0.401$}
   		 \put(0, 65) {$\Lambda=0^\circ$}
         \put(10, 54) {$St=0.141$}
         \put(52, 54) {$St=0.424$}
   		 \put(0, 46) {$\Lambda=5^\circ$}
   		 \put(10, 36) {$St=0.130$}
         \put(52, 36) {$St=0.389$}
   		 \put(0, 29) {$\Lambda=10^\circ$}
   		 \put(10, 17) {$St=0.121$}
         \put(52, 17) {$St=0.363$}
   		 \put(0, 11) {$\Lambda=15^\circ$}
		\end{overpic}
\end{center}
\vskip -2em
    \caption{Modes of low $\Lambda$ $sAR = 2$ wings visualised with contours of $Q = 1$ for $\Lambda=0^\circ$, $5^\circ$, $10^\circ$, and $15^\circ$ showing interaction mode (IM) and its harmonic (IM2).}
    \label{fig:ar2-low-sweep}
\end{figure}

%=======================================================================
Finally, the effect of wing sweep on the leading modes is discussed at the $sAR=2$. Frequencies of the three most dominant modes at low angles of attack are presented in table \ref{tab:ar2-sweep} and the corresponding spatial structures are shown in figure \ref{fig:ar2-low-sweep}. At $0^\circ \leqslant \Lambda \leqslant 15^\circ$ the overall structure of modes IM and its harmonic IM2 is analogous with that found at $sAR=4$. However, unlike the larger wing, no true wake modes have been found, that would be independent of the interaction region vortices discussed in section \ref{sec:bf}. This can be attributed to the stronger downwash effects of the tip vortex on the short wing. As a result, the spectrum of physically relevant modes (contributing to $E_k \geqslant 0.01$) consists entirely of higher harmonics of IM. The relationship of amplitude, frequency and phase between the harmonics of the IM can be concisely reflected in the Lissajous image shown in figure \ref{fig:pod_lissajous}, where the coefficients of the first four harmonics of the IM mode are plotted against each other at $(sAR,\Lambda,\alpha) = (2,10^\circ,22^\circ)$. The first line, labelled IM, corresponds to the coefficient of the second member of the first POD pair plotted against that of the IM. Since POD mode pairs have the same amplitude and a phase difference of $90^\circ$ the resulting pattern is a circle. The ratio of frequencies determines the number of "lobes" in the curve with rational ratios resulting in closed curves. It can be seen that lines corresponding to the second, third and fourth harmonics of IM have two, three and four lobes respectively. The smaller relative size of the patterns corresponding to higher harmonics reflects the ratio of amplitudes, which are smaller for higher order modes.

\begin{figure}		% AR=2, high sweep figure fig:ar2-pod-high-sweep
% \vskip -1em 
\begin{center}
        \begin{overpic}[width=0.9\textwidth]{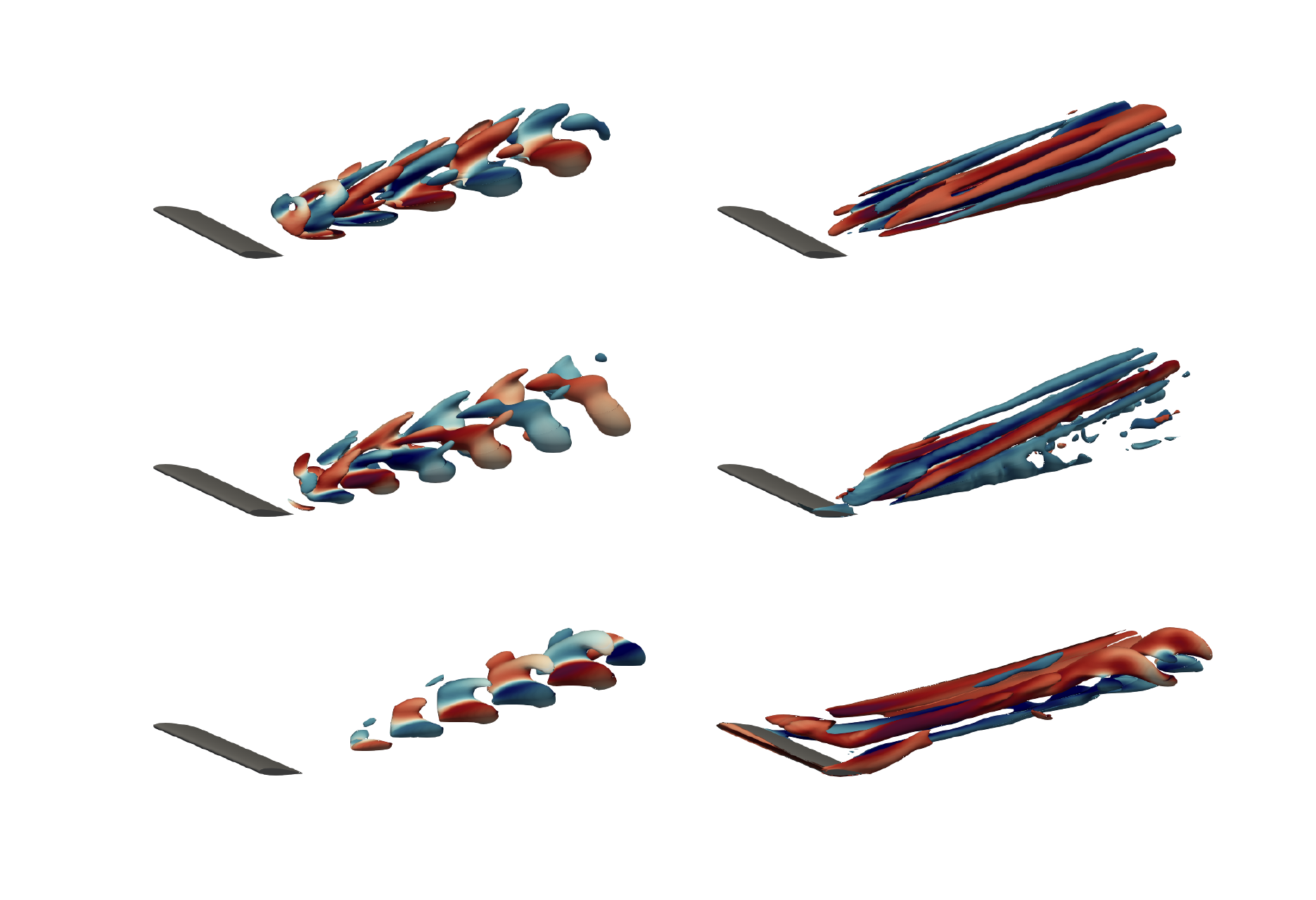}
         \put(27, 62) {IM}
         \put(70, 62) {SM}
         \put(12, 58) {$St=0.110$}
         \put(55, 58) {$St=0.0086$}
   		 \put(0, 52) {$\Lambda=20^\circ$}
         \put(12, 38) {$St=0.122$}
         \put(55, 38) {$St=0.0082$}
   		 \put(0, 33) {$\Lambda=25^\circ$}
   		 \put(12, 17) {$St=0.119$}
         \put(55, 17) {$St=0.0083$}
   		 \put(0, 12) {$\Lambda=30^\circ$}
		\end{overpic}
\end{center}
\vskip -2em
    \caption{Modes of high $\Lambda$, $sAR=2$ wings visualised with contours of $Q = 1$ for $\Lambda=20^\circ$, $25^\circ$ and $30^\circ$ showing interaction mode (IM) and streamwise mode (SM).}
    \label{fig:ar2-pod-high-sweep}
\end{figure}
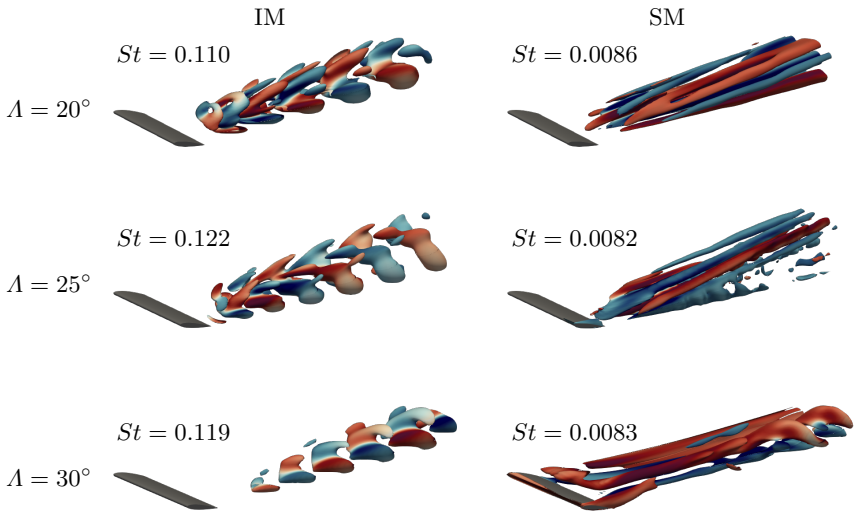

The spatial structures of $sAR=2$ wing modes at high angles of sweep are visualised in figure \ref{fig:ar2-pod-high-sweep}, where higher harmonics of IM are not shown for clarity. In a manner analogous to the larger wing, the structures of the dominant mode show spiral vortices at spatial locations where vortices are present in the flow; these vortices move towards the tip with increased sweep angles. Also analogously to the $sAR=4$ case, a low frequency SM mode with $St \approx 0.008$ appears at $\Lambda \geqslant 20^\circ$ and is shown in figure \ref{fig:ar4-pod-high-sweep}. At $\Lambda=30^\circ$ the structure of SM reflects the presence of the "ram's horn" vortices in the flow. However, unlike the $sAR=4$ wing this mode does not become the dominant at $\Lambda=30^\circ$. Instead, on the low aspect ratio wing, the IM clearly shows strong tip effects and retains the largest portion of kinetic energy. This supports the point made in \S\ref{sec:bf} that laminar separation over the shorter wing is dominated by wing tip effects. The velocity induced by the tip counters the spanwise flow due to high angle of sweep resulting in the dominant structure being a tip instability.

\begin{table}{}	% AR=2
 \begin{center}
\begin{tabular}{ c c @{\hskip 8mm} c c c c c c c}
%\hline
 & $\Lambda$ & $0^\circ$ & $5^\circ$ & $10^\circ$ & $15^\circ$ & $20^\circ$ & $25^\circ$ & $30^\circ$\\[1ex]
\multirow{2}{*}{IM} & $St_{POD}$ & 0.1338 & 0.1412 & 0.1299 & 0.1212 & 0.1103 & 0.1214 & 0.1192\\
	&  $E_k$ & 42.7072 & 45.3875 & 42.6281 & 40.8454 & 35.3662 & 44.9421 & 48.0198\\[1ex]
\multirow{2}{*}{IM2} & $St_{POD}$ & 0.2676 & 0.2826 & 0.2594 & 0.2422 & 0.2159 & 0.2428 & 0.2381\\
	&  $E_k$ & 6.2301 & 3.9885 & 5.2020 & 7.1408 & 2.1053 & 3.7166 & 0.4682\\[1ex]
\multirow{2}{*}{IM3} & $St_{POD}$ & 0.4014 & 0.4239 & 0.3891 & 0.3634 & 0.3244 & 0.3643 & 0.3571\\
	&  $E_k$ & 1.3333 & 0.9576 & 1.3242 & 1.7727 & 0.2182 & 0.6555 & 0.0097\\[1ex]	
\multirow{2}{*}{SM} & $St_{POD}$ & - & - & - & - & 0.0086 & 0.0082 & 0.0083\\
	&  $E_k$ & - & - & - & - & 13.5899 & 0.0983 & 0.8054\\[1ex]	
  \end{tabular}
  \caption{Change of POD modes with sweep angle for $sAR = 2$ wing showing Strouhal numbers according to POD coefficients and percentage of kinetic energy.}{\label{tab:ar2-sweep}}
 \end{center}
\end{table}

\begin{figure}		% POD DMD symmary sweep and AoA
    \begin{subfigure}[b]{0.49\textwidth}
        \begin{overpic}[width=1\textwidth]{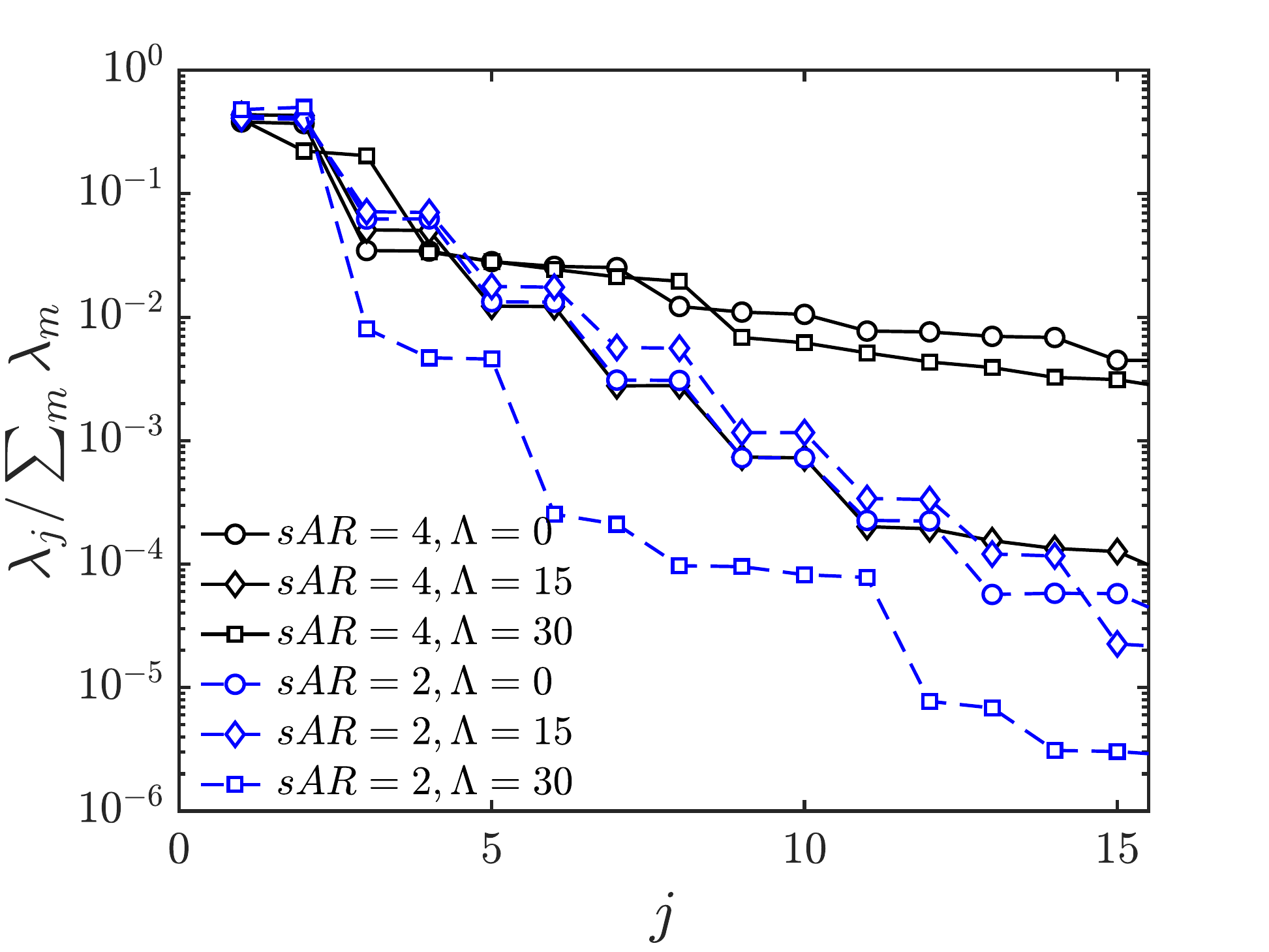}
		\put(-1, 68) {$(a)$}
		\end{overpic}
    \end{subfigure}
    ~ %
    \begin{subfigure}[b]{0.49\textwidth}
    		\begin{overpic}[width=1\textwidth]{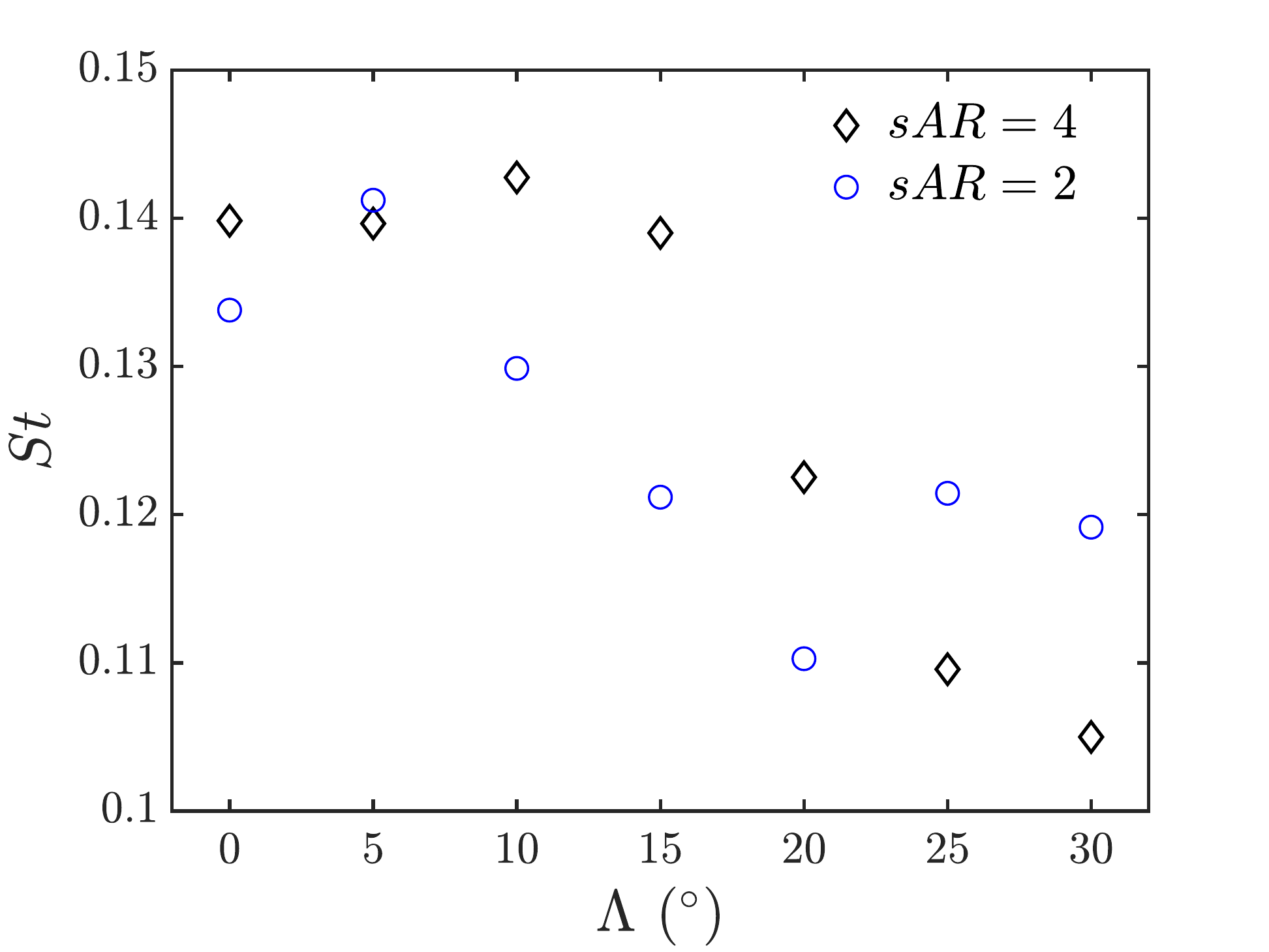}
		\put(-1, 68) {$(b)$}
		\end{overpic}
    \end{subfigure}
    \caption{Effects of sweep at $\alpha=22^\circ$: ($a$) variation of the POD spectrum with increasing $\Lambda$ for $sAR=4$ and 2 ($b$) variation of the dominant interaction mode frequency with sweep angle and aspect ratio.}
    \label{fig:sweep-summary}
\end{figure}

The power spectral density of the IM and the SM modes on the shorter $sAR=2$ wing can be compared with that on the larger wing in figure \ref{fig:pod_frequency}($b$). The dependence of the POD eigenvalues with $\Lambda$ at the two aspect ratios is shown in figure \ref{fig:sweep-summary}($a$). The spectrum at $sAR=2$ is more ordered with each pair of modes representing significantly less energy compared to the larger wing. Interestingly, at $sAR=4, \Lambda=15^\circ$ the POD eigenvalues show a pattern close to the $sAR=2$ behaviour.

The different behaviour of the dominant interaction mode with sweep for the two aspect ratios is summarised in figure \ref{fig:sweep-summary}(b). For the larger $sAR=4$ wing the frequency of the dominant mode does not change significantly with sweep at first, staying within $2\%$ of the unswept wing value in the range $0^\circ \leqslant \Lambda < 15^\circ$, but from $\Lambda=20^\circ$ onward it decreases systematically, dropping by approximately $25\%$ at the extreme end of the sweep scale with the rate of decrease seemingly reducing between $25^\circ$ and $30^\circ$. In the case of the shorter wing, a similar decay of the dominant frequency of IM of about $22\%$ is observed, however this starts at a lower sweep of $\Lambda=5^\circ$ with a constant gradient maintained up to $\Lambda=20^\circ$; after this sweep angle there is an increase of frequency at to $\Lambda=25^\circ$. Both $sAR$ cases show a small increase in the frequency of IM before the start of the decrease stage, that appears at $\Lambda=5^\circ$ and $10^\circ$ for $sAR=4$ and $sAR=2$ respectively. However, the increase is more pronounced in the $sAR=2$ case.

%% file: sections/appendix.tex
\appendix
%=======================================================================
\section{Dynamic mode decompostion}\label{appA}
%=======================================================================
Dynamic mode decomposition (DMD) can identify temporal and spatial coherent structures purely from data. It is based on the best-fit linear operator and allows for the time-resolved flow fileds to be decomposed into modes, each having a single characteristic frequency and growth rate \citep{Schmid_2010}. The method is related to the \citet{Koopman_1931} operator, as discussed by \citet{Rowley_2009} and was introduced into fluid dynamics by \citet{dmd_aps} and and \citet{Schmid_2010}. Several formulations and extensions of the method have been created over the past years \citep[see][]{Brunton_dmd_book} and the algorithm used here is briefly introduced below.

In a manner analogous to POD, snapshots of data are stacked and arranged into two matrices,
\begin{equation}
	\mathsfbi{X}=[\bm{\chi}(t_1) \quad \bm{\chi}(t_2) \quad ... \quad \bm{\chi}(t_m)]  \in \mathbb{R}^{n \times m},
	\label{eq:X}
\end{equation}
\begin{equation}
	\mathsfbi{X}^*=[\bm{\chi}(t_2) \quad \bm{\chi}(t_3) \quad ... \quad \bm{\chi}(t_{m+1})]  \in \mathbb{R}^{n \times m}.
	\label{eq:X*}
\end{equation}
A reduced singular value decomposition (SVD) is performed on the data matrix $\mathsfbi{X} = \mathsfbi{U} \Sigma \mathsfbi{V}^T$ where $T$ denotes conjugate transpose. Optionally the SVD can be truncated by only considering the first $r$ columns of $\mathsfbi{U}$ and $\mathsfbi{V}$ and the first $r$ rows and columns of $\Sigma$ to obtain $\mathsfbi{U}_r$, $\Sigma_r$, and $\mathsfbi{V}_r$. The relationship between the snapshots is approximated in a linear manner, such that
\begin{equation}
	\mathsfbi{X}^* = \mathsfbi{A} \mathsfbi{X} = \mathsfbi{A}\mathsfbi{U}_r \Sigma_r\mathsfbi{V}_r^T,
	\label{eq:DMD-evp}
\end{equation}
where $\mathsfbi{A}=\mathsfbi{X}^* \mathsfbi{X}^+$ with $\mathsfbi{X}^+$ denoting the pseudoinverse of $\mathsfbi{X}$. Multiplying both side by $\mathsfbi{U}^T$ and rearranging while making use of the fact that $\mathsfbi{V}_r^T=\mathsfbi{V}_r^{-1}$ gives
\begin{equation}
	\tilde{\mathsfbi{A}}=\mathsfbi{U}_r^T \mathsfbi{X}^* \mathsfbi{V}_r \Sigma_r^{-1} \quad \in \mathbb{R}^{r \times r},
	\label{eq:A}
\end{equation}
where $\tilde{\mathsfbi{A}}=\mathsfbi{U}_r^T \mathsfbi{A} \mathsfbi{U}$. The eigenvectors $\tilde{\bm{\nu}}_j$ and eigenvalues $\mu_j$ can be found with:
\begin{equation}
	\tilde{\mathsfbi{A}} \tilde{\bm{\nu}}_j = \mu_j \tilde{\bm{\nu}}_j.
	\label{eq:DMD-evp}
\end{equation}
For a DMD eigenvalue (every nonzero $\mu_j$) the corresponding DMD mode $\bm{\nu}_j$ is found by
\begin{equation}
	\bm{\nu}_j = {\mu_j}^{-1} \mathsfbi{X}^* \mathsfbi{V}_r {\Sigma_r}^{-1} \tilde{\bm{\nu}}_j.
	\label{eq:DMD-mode}
\end{equation}
Unlike with POD where the ordering of the modes already represents the energy contribution, it is more difficult to identify the physically relevant modes with DMD. A number of approaches to address this exist. In general, these include ranking using the norm of each mode \citep{Rowley_2009,Wan_2015} that can also be weighted by the magnitude of the DMD eigenvalue to reduce the significance of spurious modes with large norms but large decay rates \citep{Tu_2014}, and using amplitude \citep{Schmid_2012,Sayadi_2014}. All of these mode ranking criteria are effective at identifying dominant modes of periodic flows such as the wake behind a stalled wing considered here.
Mode amplitudes can be calculated using only the first snapshot \citep{Sayadi_2014} or from all snapshots using the Vandermonde matrix constructed from the eigenvalues. The former approach is used here with the amplitude of a mode defined as
\begin{equation}
	\bm{b}_j = {\mathsfbi{W}}^+ \bm{\chi}(t_1),
\end{equation}
where $\mathsfbi{W}$ is a matrix whose columns contain the DMD eigenvectors $\bm{\nu}_j$, $\bm{\chi}(t_1)$ is the first snapshot and $+$ denotes the Moore-Penrose inverse. This approach requires the inverse of $\mathsfbi{W}$ which is a complex $n \times m$ matrix. Singular value decomposition is used for this
\begin{equation}
	{\mathsfbi{W}}^+ = \mathsfbi{V} \Sigma \mathsfbi{U}^T,
\end{equation}
where $\Sigma$ is a diagonal matrix containing the singular values and $\mathsfbi{U}$ and $\mathsfbi{V}$ are matrices containing left and right singular vectors of $\mathsfbi{W}$, respectively. Finally, the oscillation frequency $f_j$ and the growth rate $g_j$ of a DMD mode can be determined from the DMD eigenvalue $\lambda_j$ using
\begin{equation}
	f_j = \angle \lambda_j/(2 \pi \delta t),
\end{equation}
\begin{equation}
	g_j = \log|\lambda_j|/\delta t,
\end{equation}
where $\delta t$ is the uniform sampling increment between snapshots.

%=======================================================================
\section{data-driven analysis validation}\label{appB}

%=======================================================================
\subsection{Code validation}
\label{sec:appB-1}
%=======================================================================

\begin{figure}		%POD sensitivity
 \begin{center}
    \includegraphics[width=0.5\textwidth]{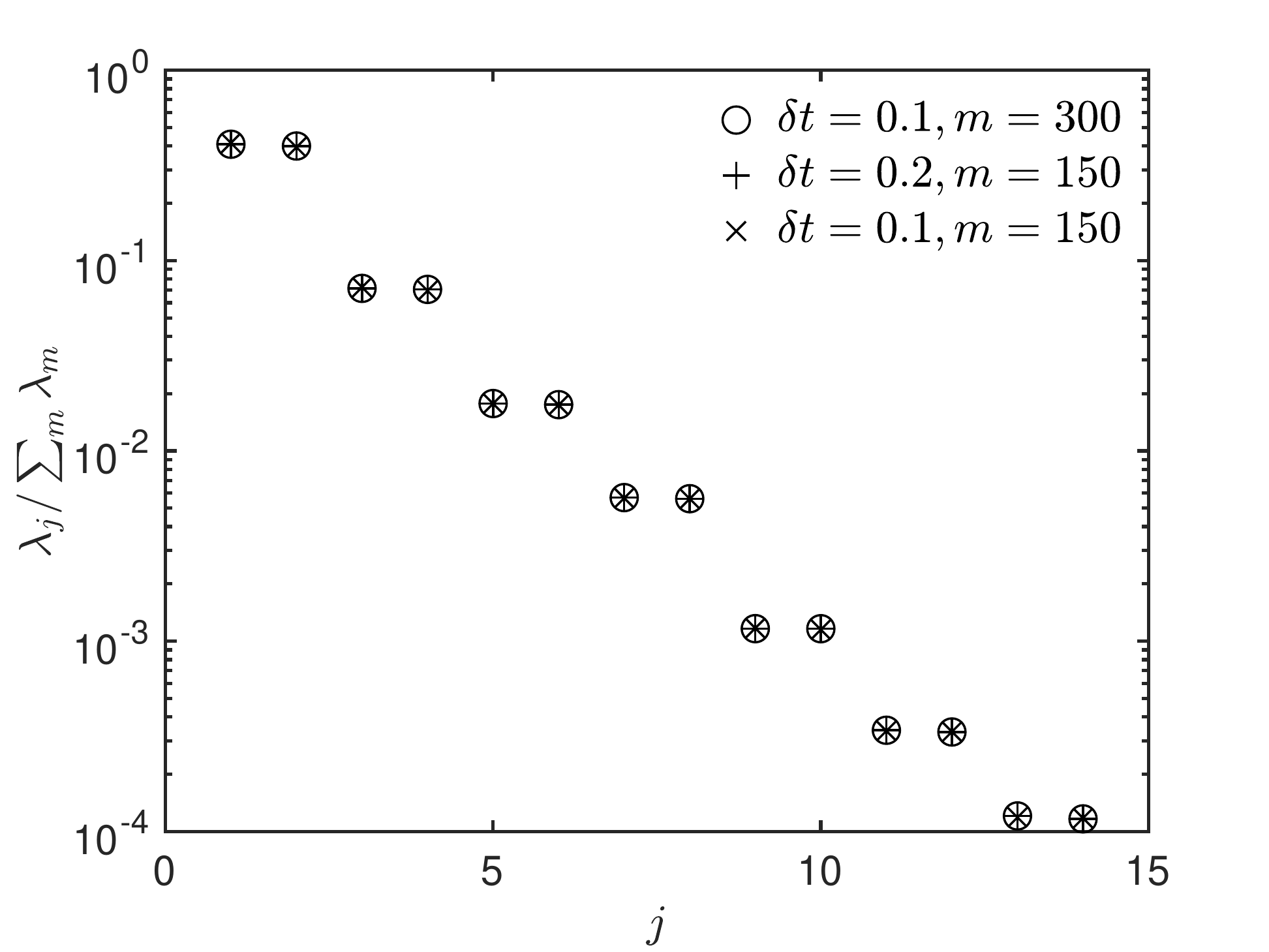}
    \caption{Sensitivity of POD to number of snapshots $m$ and time interval $\delta t$.}
    \label{fig:pod_sensitivity}
\end{center}
\end{figure}

\begin{figure}			% DMD sensitivity
  \begin{center}
    \begin{subfigure}[b]{0.49\textwidth}
        \begin{overpic}[width=1\textwidth]{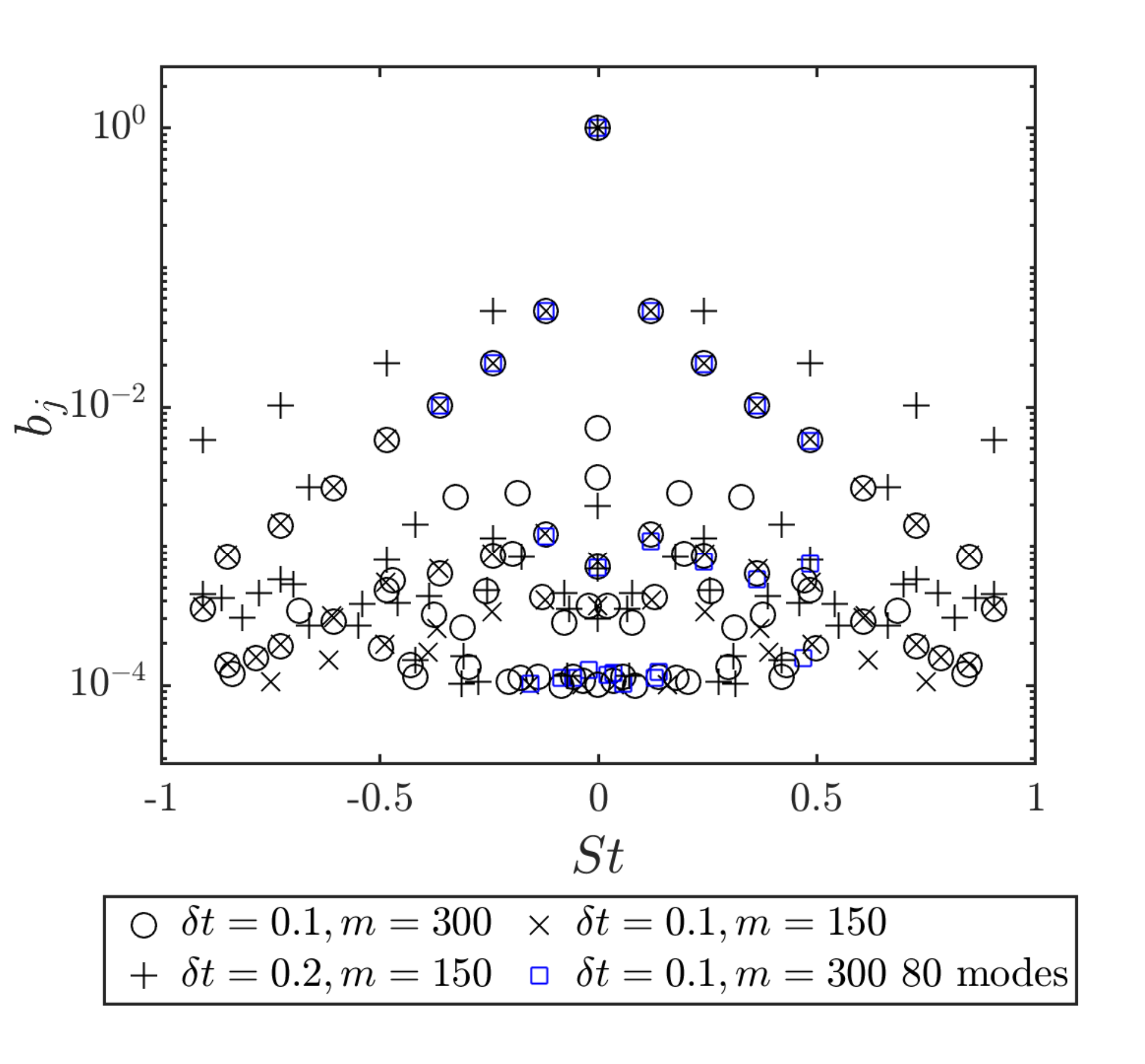}
		\put(0, 85) {$(a)$}
		\end{overpic}
    \end{subfigure}
    %~ 
    \begin{subfigure}[b]{0.49\textwidth}
        \begin{overpic}[width=1\textwidth]{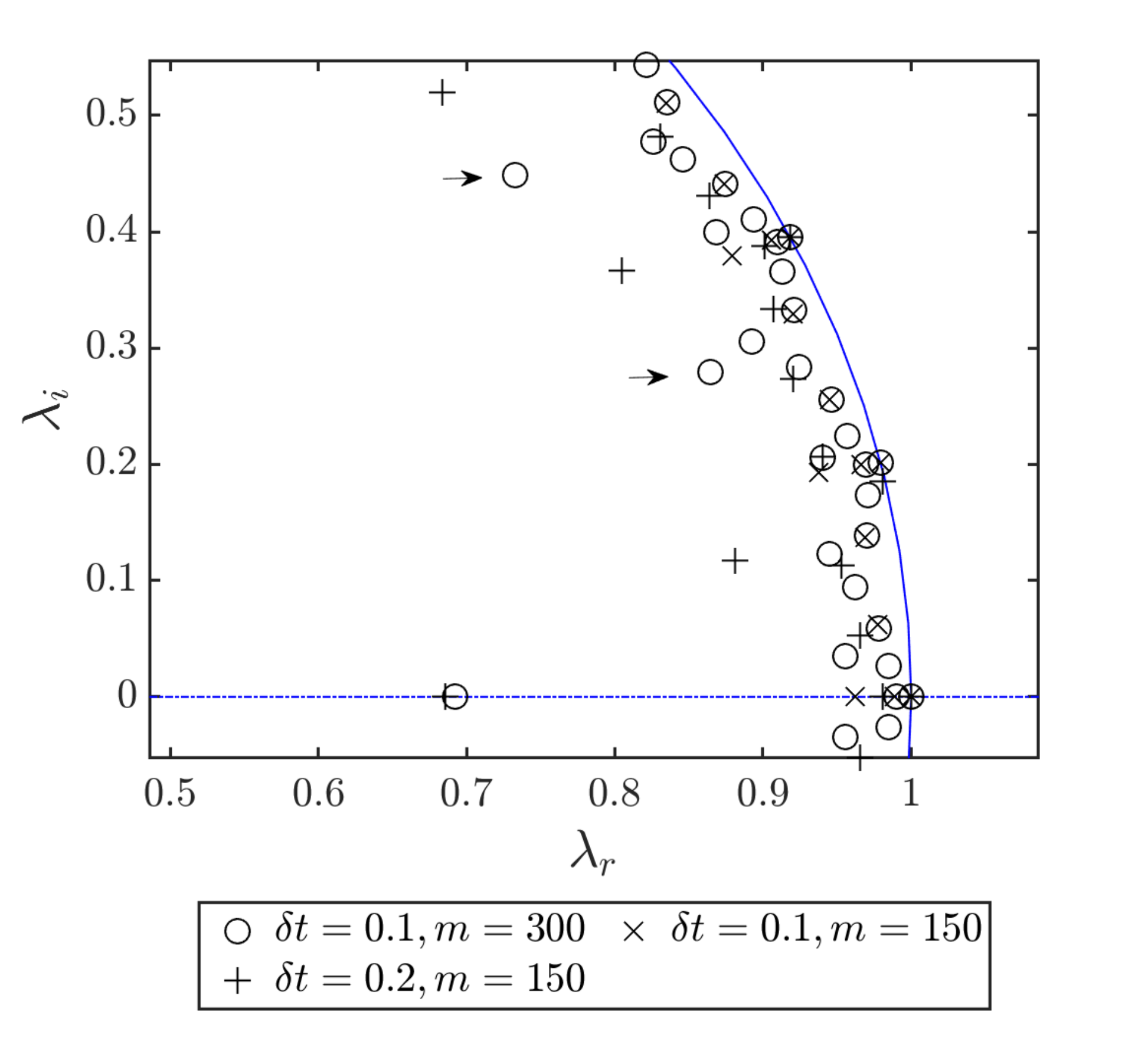}
		\put(0, 85) {$(b)$}
		\end{overpic}
    \end{subfigure}
    \caption{Sensitivity of DMD to number of snapshots $m$ and the interval between them $\delta t$ ($a$) frequencies and normalised initial amplitudes showing the effect of recovering less  than $m$ modes, for clarity modes with $b_j < 10^{-4}$ are omitted ($b$) DMD eigenvalues on the unit circle.}
    \label{fig:dmd-sensitivity}
\end{center}
\end{figure}

In order to verify that the modal results are independent of the number of snapshots $m$ and the interval between them $\delta t$, a sensitivity analysis was performed on a single case $(sAR,\Lambda,\alpha)=(2,15^\circ,22^\circ)$. Three datasets were considered: 300 snapshots in the interval of $40 \leqslant t \leqslant 70$ with $\delta t=0.1$ that cover 10 periods of vortex shedding in the base flow, 150 snapshots in the same interval but with $\delta t=0.2$, and 150 snapshots in the interval of $40 \leqslant t \leqslant 55$ with $\delta t=0.1$ that cover 5 shedding periods. Figure \ref{fig:pod_sensitivity} shows that the POD eigenvalues generated using these datasets are identical for the first 15 modes, which collectively represent 99.995\% of flow kinetic energy. It is therefore concluded that $m=300$ $\delta=0.1$ is more than adequate to resolve the POD modes for the current problem. Similarly, DMD was computed with the same three datasets. The highest frequency DMD can resolve is dictated by the Nyquist sampling theorem while the lowest frequency is limited by the time interval that is spanned by the data. It can be seen in \ref{fig:dmd-sensitivity}($a$) that increasing $\delta t$ to 0.2 ($+$ markers) leads to DMD loosing every other of the high initial amplitude modes. Note that the difference in the initial amplitudes ($b_j$) themselves is not a cause for concern since only the relative difference in $b_j$ is used to differentiate the dominant modes. Reducing the time span of the data by half ($\times$ markers denoting $m=150$, $\delta t=0.1$) recovers dominant modes well but loses some accuracy at lower frequencies of $St < 0.5$. The two nonzero frequency modes at $b_j \approx 2\times 10^{-3}$ (marked with arrows) that are not recovered by the shorter dataset are not physically significant as they lie far from the unit circle as shown in \ref{fig:dmd-sensitivity}($b$). Finally, the effect of computing initial amplitudes based on recovering less than $m$ DMD modes is considered. Once the DMD eigenvalues are found, corresponding modes are computed using equation \ref{eq:DMD-mode}, which is a computationally expensive step. Red square markers in \ref{fig:dmd-sensitivity}($a$) show that $b_j$ computed based on 80 modes is accurate for the first 4 dominant modes. Note that this does not affect the DMD eigenvalues in \ref{fig:dmd-sensitivity}($b$). Therefore, the $m=300$ $\delta=0.1$ dataset with recovering 80 modes was used for DMD analysis.

%=======================================================================
\subsection{Relationship between global stability analysis and data-driven methods}
\label{sec:appB-2}
%=======================================================================
\begin{table}{}		% 2D cylinder comparisson
\begin{center}
\begin{tabular}{ c c c}
	& T1 & S1\\
	BiGlobal EVP (Nektar++) & $\pm 0.11926 + 0.04762i$ & $0 - 0.05145i$\\
	POD, DMD (linear) & $\pm 0.11926 + 0.04761i$ & $0 - 0.05371i$\\	
% 	POD (linear) & 0.11961 + 0.0493i & 
	POD, DMD (nonlinear) & $0.13613 + 0i$ & $0 - 0.05048i$\\
  \end{tabular}
  \caption{Leading travelling and stationary modes for a reference case of two-dimensional cylinder flow at $Re_D=60$.}{\label{tab:2d-cyl}}
 \end{center}
\end{table}

In order to illustrate the relationship between operator based global analysis and data-driven POD and DMD consider a reference problem of a two-dimensional flow over a cylinder at $Re_D=60$. The leading travelling and stationary modes obtained by BiGlobal stability analysis are compared to DMD performed on snapshots recorded in the linear growth regime and in the nonlinear saturated regime as shown in table \ref{tab:2d-cyl}. data-driven analysis shows good agreement with global stability in the linear regime but recovers very different travelling mode when performed on the data taken in saturated regime. It is worth pointing out that POD gives identical results if the frequency and the decay rate of the temporal coefficients.